# Machine learning bridging battery field data and laboratory data

Yanbin Zhao[†1], Hao Liu[†*1], Zhihua Deng[2], Tong Li[1], Haoyi Jiang[1], Zhenfei Ling[1], Xingkai Wang[3], Lei Zhang[4], Xiaoping Ouyang[1]

[1] State Key Laboratory of Fluid Power and Mechatronic Systems, School of Mechanical Engineering, Zhejiang University, Hangzhou, China.
[2] Energy Research Institute at NTU (ERI@N), Nanyang Technological University, Singapore.
[3] School of Materials Science and Engineering, Tianjin University, Tianjin, China.
[4] Wanxiang A123 Systems Corp., Hangzhou, China.
[†] These authors contributed equally to this work: Yanbin Zhao, Hao Liu.
[*] Corresponding authors' email: haoliu7850052@zju.edu.cn (Hao Liu).

## Abstract

Aiming at the dilemma that most laboratory data-driven diagnostic and prognostic methods cannot be applied to field batteries in passenger cars and energy storage systems, this paper proposes a method to bridge field data and laboratory data using machine learning. Only two field real impedances corresponding to a medium frequency and a high frequency are needed to predict laboratory real impedance curve, laboratory charge/discharge curve, and laboratory relaxation curve. Based on the predicted laboratory data, laboratory data-driven methods can be used for field battery diagnosis and prognosis. Compared with the field data-driven methods based on massive historical field data, the proposed method has the advantages of higher accuracy, lower cost, faster speed, readily available, and no use of private data. The proposed method is tested using two open-source datasets containing 249 NMC cells. For a test set containing 76 cells, the mean absolute percentage errors of laboratory real impedance curve, charge curve, and discharge curve prediction results are 0.85%, 4.72%, and 2.69%, respectively. This work fills the gap between laboratory data-driven diagnostic and prognostic methods and field battery applications, making all laboratory data-driven methods applicable to field battery diagnosis and prognosis. Furthermore, this work overturns the fixed path of developing field battery diagnostic and prognostic methods based on massive field historical data, opening up new research and breakthrough directions for field battery diagnosis and prognosis.

## Introduction

With the advantages of low-cost [1], high-power density [2, 3], and long cycle life [4], lithium-ion (Li-ion) batteries have achieved large-scale commercial applications in passenger cars and energy storage systems, and have brought huge carbon emission reduction and environmental benefits. However, Li-ion batteries are currently facing challenges in safety [5]. Frequent safety accidents (such as fires and explosions) have caused a large amount of direct and indirect economic losses [6], and some accidents have even caused casualties. Under strong market demand and strict supervision, improving battery safety is urgent. Battery safety involves almost all processes in design, production, and application [7]. Problems in any link may lead to safety accidents in applications. Battery safety must be improved through the joint improvement of materials, structures, manufacturing processes, and prognostics and health management (PHM). In addition to improving battery safety by optimizing materials [8], structures [9], and manufacturing processes [6] in design and production, PHM in application [10, 11] is also crucial to improving battery safety. Accurate diagnostic and prognostic information can provide a reliable basis for predictive maintenance and optimized management to ensure the safe operation of filed batteries.

Numerous battery diagnostic and prognostic methods have been proposed [10-15], including model-based methods, data-driven methods, and hybrid methods. Recently, thanks to the rapid development of artificial intelligence [16, 17] and the open source of numerous battery datasets [18-22], data-driven methods combining machine learning (ML) algorithms and feature engineering have achieved superior performance in battery diagnosis and prognosis, showing great application potential in battery material development [23], manufacturing processes improvement [24, 25], aging mechanism revelation [26], fast-charging protocol selection [27], management strategy optimization [28, 29], recycling [30, 31], etc. However, most data-driven methods and their feature engineering rely on laboratory data at specific stages under specific conditions, such as charge/discharge data [32, 33], impedance data [34, 35], and relaxation data [36]. Unlike laboratory data with stable test environment and conditions, field data is random due to the coupling of multiple random factors such as application scenarios, operating environment, and user habits [28, 37]. Since the formats of laboratory data are vastly different from those of field data, most existing laboratory data-driven methods cannot be directly applied to field batteries.

Recently, some battery diagnostic and prognostic methods based on field data have also been proposed [28, 38-40]. The main idea of this type of method is to develop data-driven methods based on the massive field historical data generated by field battery operation [41, 42]. As shown in **Figure 1**, based on the needs of field batteries for predictive maintenance and optimized management, we summarize the five main characteristics that ideal diagnostic and prognostic methods should have: (1) High accuracy. The higher the accuracy of the diagnostic and prognostic method, the better the effect of predictive maintenance and optimization management, and the more conducive to the safe operation of field batteries. On the contrary, inaccurate diagnostic and prognostic information may have a negative impact on the safe operation of field batteries. (2) Low cost. The ideal diagnostic and prognostic method should have low development, deployment, and usage costs, which is conducive to large-scale commercialization. (3) Readily available. The ideal diagnostic and prognostic method can perform field battery diagnosis and prognosis anytime and anywhere according to user needs. (4) Fast. The ideal diagnostic and prognostic method can quickly complete the diagnosis and prognosis without negatively affecting the normal operation of field batteries. (5) No use of private data. The ideal diagnostic and prognostic method should not use data involving user privacy, such as historical operating data, which may include historical driving data, historical charging data, user habit data, etc.

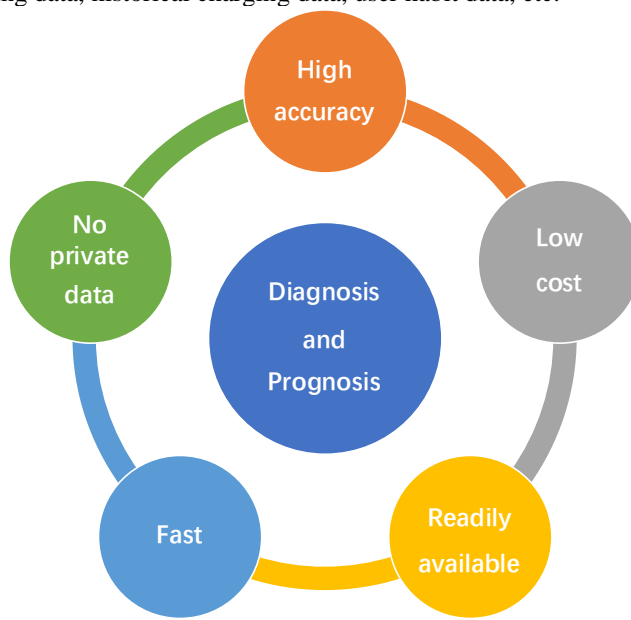

**Figure 1 Five main characteristics of ideal diagnostic and prognostic methods.**

Compared with the above five characteristics that an ideal field battery diagnostic and prognostic method should have, the existing field data-driven methods developed based on massive field historical data are not ideal in all characteristics. In terms of accuracy, when the data volume is the same, the accuracy of the field data-driven method is not as good as that of the laboratory data-driven method. The main reason is that the quality of laboratory data is higher than that of field data. In terms of cost, the field data-driven method needs to collect massive field historical data, and the development cost is high. The field data-driven method needs to adopt large ML models, and some models even need to be deployed on cloud servers, which usually have high deployment and usage costs. In terms of readily available, most of the field data-driven methods need to collect complete or specific stage field battery operation data, which cannot be used anytime and anywhere, and the speed is also slow. In terms of privacy, the field data-driven method needs to use massive field historical data to train the ML model, which may involve the user's privacy data. In addition, most field data-driven methods need to be redeveloped, and the existing results of many laboratory data-driven methods are not fully utilized, resulting in a waste of research resources. Based on first principles thinking, low-cost development of field battery diagnostic and prognostic methods should utilize existing laboratory data-driven methods rather than reinventing them.

Some researchers have begun to work on applying the achievements of laboratory data-driven methods to field batteries. Steininger et al. [43] augmented the field data by transforming the laboratory data into the format of field data, improving the prognostic accuracy of field batteries. Lyu et al. [44] used cumulative utilization lifetime instead of cycle number to reduce the difference in battery capacity aging trajectories between laboratory data and field data, making the battery capacity aging trajectory in field data smooth and easy to predict, just like laboratory data. However, most field battery diagnostic and prognostic methods have not yet gotten rid of their reliance on field historical data.

To completely get rid of the reliance on historical filed data in the process of field battery diagnosis and prognosis, it is necessary to propose a method that can bridge the field data with the laboratory data, and then use the laboratory data-driven method to perform field battery diagnosis and prognosis. The first question we face is what battery data to choose as a bridge? By comparing charge/discharge data, impedance data, and relaxation data, we found that the impedance data is most suitable for use as a bridge. There are three main reasons. First, the studies of Wang et al. [45] and Barcellona et al. [46] showed that when the frequency and aging state are the same, the real impedance (Re) at a specific state-of-charge (SOC) and operating temperature can be predicted by the Re at any other SOCs and operating temperatures, which lays the foundation for bridging field impedance data with laboratory impedance data. Second, researchers have achieved accurate prediction of EIS data based on charge capacity vs. voltage (Q/V) curve [47, 48] and relaxation voltage vs. time (V/t) curve [49]. Guo et al. [50] explored the internal electrochemical relationship between charge Q/V curve and EIS. The experimental results show that the evolution trends of charge Q/V curve and EIS with cycles contain the same battery aging mode information, namely loss of lithium inventory (LLI) and loss of active material (LAM). This study provides mechanistic evidence that charge Q/V curve and EIS can predict each other. Given that the evolution trend of relaxation V/t curve with cycles also contains rich information about battery aging mode [51, 52], we boldly speculate that EIS can also be used to accurately predict charge/discharge Q/V curve and relaxation V/t curve. Third, charge/discharge data and relaxation data are both random passive data generated during battery operation, while impedance data is user-controlled active data that can be collected anytime and anywhere. Therefore, impedance data can meet the readily available needs of ideal diagnostic and prognostic methods. At the same time, the time spent on medium and high (mid-high) frequency impedance measurement is very short (in milliseconds) [53]. At present, some DC/DCs have already integrated impedance measurement functions [54, 55], and the actual deployment and usage costs are relatively low. Therefore, we choose impedance data to bridge field data and laboratory data.

Fortunately, existing studies have solved the problem of directly measuring steady-state EIS during dynamic operation (C rate is not 0) of batteries. Zhu et al. [56] showed that steady-state EIS can be directly generated online based on arbitrary dynamic load profiles and is applicable to different battery chemistries, aging states, SOCs, and operating temperatures. Sihvo et al. [57] proposed a real-time battery steady-state EIS monitoring method based on DST pseudo-random sequence perturbation signal, which is fast, accurate, and easy to implement. The DST sequence and its DFT eigenvector properties can quickly generate valid and consistent steady-state EIS during the dynamic operation of batteries. This method has been shown to perform under various different battery chemistries, current profiles, and operating conditions (various charge, temperature, and dynamic current). On this basis, we propose a method to bridge field data with laboratory data, which can accurately predict laboratory Re vs. frequency (Re/f) curve, charge/discharge Q/V curve, and relaxation V/t curve using only two field Res corresponding to a medium frequency and a high frequency. Specifically, the proposed bridging method first converts the two field Res measured at any SOC and temperature into two laboratory Res at a specific SOC and temperature through ML models, and then predicts the laboratory Re/f curve in the mid-high frequency range (1~1k Hz) based on the ML model and two laboratory Res. Finally, the laboratory charging Q/V curve, discharge Q/V curve, and relaxation V/t curve are predicted based on ML models and the laboratory Re/f curve. After obtaining the predicted laboratory data, the existing laboratory data-driven methods can be directly used for battery diagnosis and prognosis. This work has achieved innovations and breakthroughs in the following four aspects.

(1) A field battery diagnostic and prognostic method based on two field Res is proposed. Compared with the existing field data-driven methods based on massive field historical data, the proposed method has the advantages of high accuracy, low cost, fast, readily available, and no use of private data. ML models are utilized to bridge field data and laboratory data, filling the gap between laboratory data-driven methods and field battery applications, and enabling all laboratory data-driven methods can be applied to field batteries.

(2) Compared with health indicators extracted from random field data, health indicators extracted from regular laboratory data, such as remaining capacity and remaining cycles, can provide more accurate and reliable battery health status results. Compared with health indicators extracted from a single laboratory data, using health indicators extracted from different laboratory data can obtain more accurate and reliable diagnostic and prognostic results.

(3) This work overturns the fixed path of developing field battery diagnostic and prognostic methods based on massive field historical data, and opens up new research and breakthrough directions for field battery diagnosis and prognosis.

(4) The proposed bridging method can be utilized to predict offline performance characterization data that need to be measured regularly during battery life testing, such as EIS, charge/discharge Q/V curve, and relaxation V/t curve, which can reduce the cost and difficulty of battery life testing.

## Results

### Datasets

This paper uses two open-source datasets to test the proposed method of bridging field data and laboratory data, namely Dataset 1 and Dataset 2. Dataset 1 was open sourced by Luh et al. [20] in 2024 and contains 228 NMC cells with a rated capacity of 3 Ah (the manufacturer is LG and the model No. is INR18650HG2). A total of 76 test conditions were designed, including five different depths of discharge (DODs), four different operating temperatures, and three different charge/discharge rates (C-rates), and covering three different aging tests, namely calendar aging, cycle aging, and profile aging. Three cells are assigned to each test condition. We assign the first 2 cells to the training set and the remaining 1 cell to the test set. Then, the training set contains 152 cells and the test set contains 76 cells. For each cell, a reference performance test (PRT) is performed on average every 3 weeks after the first (week 0) and second (week 1) RPTs. The RPT includes the measurement of remaining capacity, steady-state EIS, and resistance, where steady-state EIS and resistance are measured at five different SOC (namely 10%, 30%, 50%, 70%, and 90%) during charging and discharging. The operating temperature of steady-state EIS and resistance measurement during charging is set to 25°C, while the operating temperature of steady-state EIS and resistance measurement during discharging is set to the corresponding test condition temperature. In this paper, we use the steady-state EIS measured at T=25°C and all SOC during charging as the laboratory data, and the steady-state EIS measured at all SOC and operating temperatures during discharging as the field data. At the same time, we use the constant current charge/discharge Q/V data in the remaining capacity measurement as the laboratory data. Since the relaxation time of each cell after full charge and discharge is too short (only 5 min), only the prediction of laboratory charge/discharge Q/V data can be tested. In Dataset 1, the RPT data of some cells are incomplete, and we have removed these incomplete data. After removing the incomplete RPT data, for the laboratory data corresponding to SOC=90%, there are 1627 field EIS samples, 1627 laboratory EIS samples, and 1627 laboratory charge/discharge Q/V curve samples available in the training set of Dataset 1, and there are 805 field EIS samples, 805 laboratory EIS samples, and 805 laboratory charge/discharge Q/V curve samples available in the test set of Dataset 1. For the laboratory data corresponding to SOC=70%, there are 1300 field EIS samples, 1300 laboratory EIS samples, and 1300 laboratory charge/discharge Q/V curve samples available in the training set of Dataset 1, and there are 647 field EIS samples, 647 laboratory EIS samples, and 647 laboratory charge/discharge Q/V curve samples available in the test set of Dataset 1. For the laboratory data corresponding to SOC=50%, there are 1326 field EIS samples, 1326 laboratory EIS samples, and 1326 laboratory charge/discharge Q/V curve samples available in the training set of Dataset 1, and there are 658 field EIS samples, 658 laboratory EIS samples, and 658 laboratory charge/discharge Q/V curve samples available in the test set of Dataset 1. For the laboratory data corresponding to SOC=30%, there are 1289 field EIS samples, 1289 laboratory EIS samples, and 1289 laboratory charge/discharge Q/V curve samples available in the training set of Dataset 1, and there are 645 field EIS samples, 645 laboratory EIS samples, and 645 laboratory charge/discharge Q/V curve samples available in the test set of Dataset 1. For the laboratory data corresponding to SOC=10%, there are 1245 field EIS samples, 1245 laboratory EIS samples, and 1245 laboratory charge/discharge Q/V curve samples available in the training set of Dataset 1, and there are 619 field EIS samples, 619 laboratory EIS samples, and 619 laboratory charge/discharge Q/V curve samples available in the test set of Dataset 1. For more details of Dataset 1, please refer to [20].

Dataset 2 was open-sourced by Mohtat et al. [58] in 2021 and contains 21 NMC cells with a rated capacity of 5 Ah (the manufacturer is the University of Michigan Battery Lab). A total of 7 test conditions were designed, including two different DODs, three different operating temperatures, and four different charge/discharge C-rates, and covering two different aging tests, namely cycle aging and profile aging. Three cells are assigned to each test condition. We select odd-numbered cells to be assigned to the training set and the remaining even-numbered cells to be assigned to the test set. Then, the training set contains 11 cells and the test set contains 10 cells. For each cell, an RPT is performed once at an average capacity loss interval of 5%. The RPT includes the measurement of remaining capacity, steady-state EIS, and resistance, where steady-state EIS and resistance are measured at 10 different SOCs (namely, 0%, 10%, 20%, 30%, 40%, 50%, 60%, 70%, 80%, and 90%) and T=25°C during discharging. During the remaining capacity measurement in the RPT, each cell was rested for one hour after being fully charged, which can be used to verify the predictions of the laboratory relaxation V/t data in the proposed method. Since Dataset 2 only measured steady-state EIS during discharge and at a single temperature, we use the steady-state EIS measured at SOC=90% and T=25°C during discharging as the laboratory data, and the steady-state EIS measured at other SOC and T=25°C during discharging as the field data. At the same time, we use the constant current charge/discharge Q/V curve in the remaining capacity measurement and the relaxation V/t data after full charge as the laboratory data. In Dataset 2, the RPT data of some cells are incomplete, and we have removed these incomplete data. After removing the incomplete RPT data, there are 122 field EIS samples, 122 laboratory EIS samples, 122 laboratory charge/discharge Q/V curve samples, 122 laboratory relaxation V/t curve after full charge samples available in the training set of Dataset 2. In the test set of Dataset 2, there are 118 field EIS samples, 118 laboratory EIS samples, 118 laboratory charge/discharge Q/V curve samples, 118 laboratory relaxation V/t curve after full charge samples. For more details of Dataset 2, please refer to [58].

This paper uses remaining capacity to characterize the battery state-of-health (SoH). Meanwhile, the remaining cycles or days corresponding to the full discharge capacity decay to 80% of the initial capacity is used as the battery life.

## Bridging Method

As shown in **Figure 2**, the proposed method for bridging field data and laboratory data mainly includes five steps, namely field Res collection (**Step 1**), laboratory Res prediction (**Step 2**), laboratory Re/f curve prediction (**Step 3**), other laboratory data prediction (**Step 4**), and diagnosis and prognosis (**Step 5**). The details of each step are as follows.

**Step 1: Field Res collection.**

Collect the Res corresponding to two preset frequencies of the field battery online. The key to this step is to find the two preset frequencies that need to be obtained. In this paper, we use the k-means clustering algorithm to select two frequencies as the preset frequencies from the medium frequency range (1 Hz~100 Hz) and high frequency range (100 Hz~1k Hz) of the Re/f curve. We named the frequency extracted from the medium frequency range as $f_1$, and the frequency extracted from the high frequency range as $f_2$. The field operating temperature and field SOC are named as $T^F$ and $SOC^F$, respectively. The Res corresponding to $f_1$ and $f_2$ are named as $Re_1^F$ and $Re_2^F$, respectively. For more details on the selection method of two preset frequencies $f_1$ and $f_2$, please refer to **Supplementary Note 1**.

In Dataset 1, based on the laboratory Re/f curves in the training set, the k-means clustering algorithm is used to obtain the clustering results in the medium frequency range (1 Hz~100 Hz) and the high frequency range (100 Hz~1k Hz), as shown in **Figure 3 (a)**. The extracted two preset frequencies are $f_1$=10 Hz and $f_2$=312.5 Hz. The positions of the two preset frequencies on the field EIS and laboratory EIS are shown in **Figure 3 (b)**.

In Dataset 2, based on the laboratory Re/f curves in the training set, the k-means clustering algorithm is used to obtain the clustering results in the medium frequency range (1 Hz~100 Hz) and the high frequency range (100 Hz~1k Hz), as shown in **Figure 4 (a)**. The extracted two preset frequencies are $f_1$=5.53 Hz and $f_2$=193.03 Hz. The positions of the two preset frequencies on the field EIS and laboratory EIS are shown in **Figure 4 (b)**.

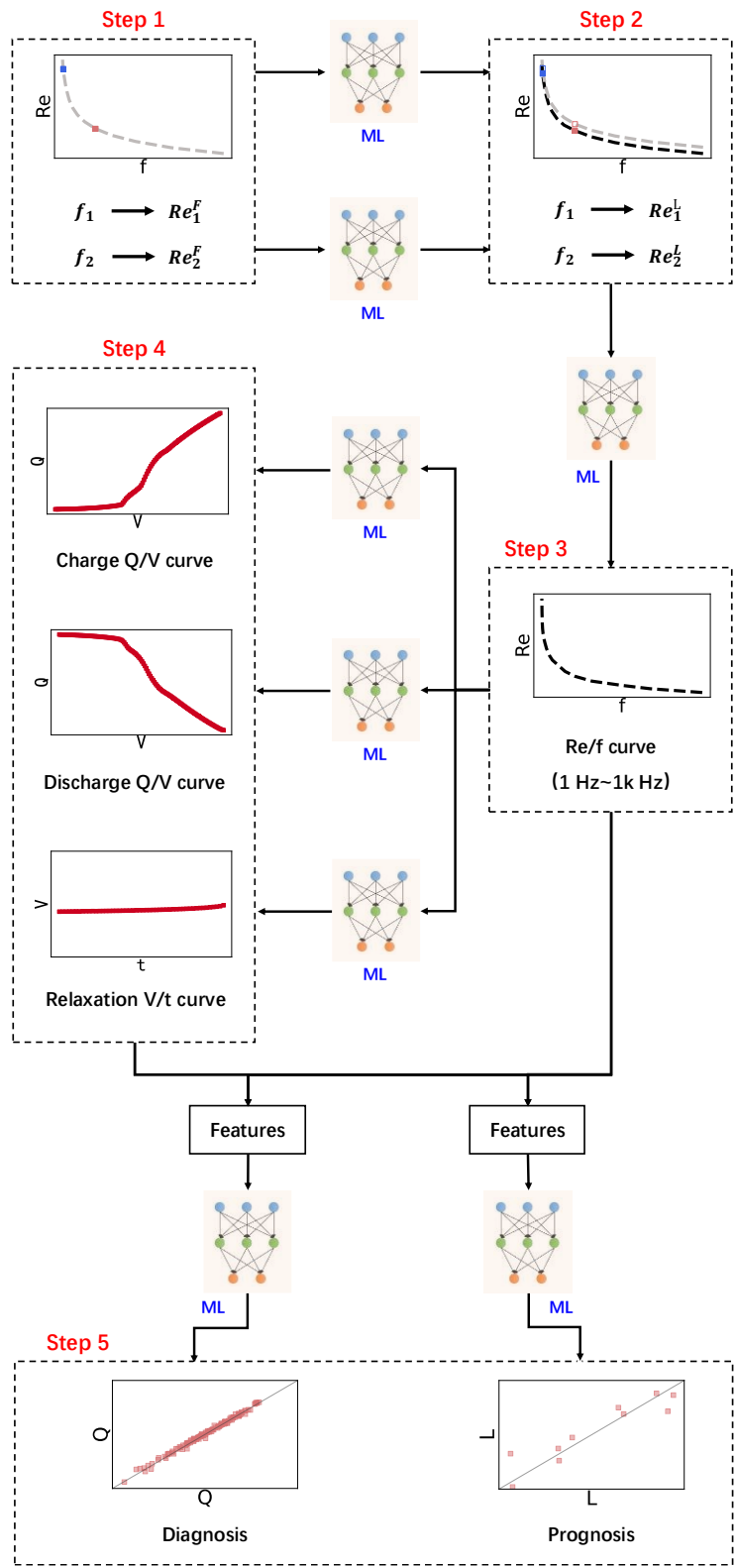

**Figure 2 Steps of the bridging method.**

**Step 2: Laboratory Res prediction.**

The laboratory operating temperature and laboratory SOC are named as $T^L$ and $SOC^L$, respectively. Input $Re_1^F$, $SOC^F$, and $T^F$ into the ML model, and predict the value of the laboratory $Re_1^L$ corresponding to $f_1$ and $SOC^L$. Input $Re_2^F$, $SOC^F$, and $T^F$ into the ML model, and predict the value of the laboratory $Re_2^L$ corresponding to $f_2$ and $SOC^L$. For more details on the prediction method of $Re_1^L$ and $Re_2^L$, please refer to **Supplementary Note 2**.

Taking $SOC^L$=90% as an example, in the test set of Dataset 1, the $Re_1^F$, $SOC^F$ ∈[90%, 100%], and $T^F$ =0°C, 10°C, 25°C, or 40°C are input into the trained ML model to predict $Re_1^L$ corresponding to $f_1$=10 Hz and $SOC^L$=90%, and the prediction results are shown in **Figure 5 (a)**. The mean absolute error (MEA), root mean square error (RMSE), and mean absolute percentage error (MAPE) of the $Re_1^L$ prediction results are 0.24 mΩ, 0.37, and 1.19%, respectively. Similarly, the prediction results of $Re_2^L$ corresponding to $f_2$=312.5 Hz and $SOC^L$=90% are shown in **Figure 5 (b)**. The MEA, RMSE, and MAPE of the $Re_2^L$ prediction result are 0.12 mΩ, 0.19, and 0.68%, respectively.

In the test set of Dataset 2, the $Re_1^F$, $SOC^F$ ∈[0%, 90%), and $T^F$ =25°C are input into the trained ML model to predict $Re_1^L$ corresponding to $f_1$=5.53 Hz and $SOC^L$=90%, and the prediction results are shown in **Figure 6 (a)**. The MEA, RMSE, and MAPE of the $Re_1^L$ prediction result are 0.09 mΩ, 0.14, and 1.09%, respectively. Similarly, the prediction result of $Re_2^L$ corresponding to $f_2$=193.03 Hz and $SOC^L$=90% is shown in **Figure 6 (b)**. The MEA, RMSE, and MAPE of the $Re_2^L$ prediction result are 0.05 mΩ, 0.09, and 0.81%, respectively.

**Step 3: Laboratory Re/f curve prediction.**

Based on the predicted laboratory $Re_1^L$ and $Re_2^L$ data in **Step 2**, the laboratory Re/f curve in the mid-high frequency range (1 Hz~1k Hz) is predicted. The input of the ML model is $Re_1^L$ and $Re_2^L$, and the output is the laboratory Re/f curve in the mid-high frequency range. For more details on the prediction method of the laboratory Re/f curve, please refer to **Supplementary Note 3**.

Taking $SOC^L$=90% as an example, in the test set of Dataset 1, the prediction results of the laboratory Re/f curve in the mid-high frequency range are shown in **Figure 7 (a)**. The MEA, RMSE, and MAPE of the predicted laboratory Re/f curve on the test set are 0.18 mΩ, 0.32, and 0.85%, respectively. To further verify the accuracy of the predicted laboratory Re/f curve, we combine the predicted laboratory Re/f curve with the measured laboratory imaginary impedance (Im) vs. frequency (Im/f) curve to plot the predicted EIS, as shown in **Figure 7 (b)**. Furthermore, the same distribution of relaxation times (DRT) calculation is performed on the predicted EIS and the measured EIS, and the obtained DRT curves are shown in **Figure 7 (c)**. The MEA, RMSE, and MAPE on the test set are 0.07 mΩ, 0.01, and 0.19%, respectively. **Figures 7 (b)** and **7 (c)** indirectly verify that the prediction of the laboratory Re/f curve in the mid-high frequency range is accurate.

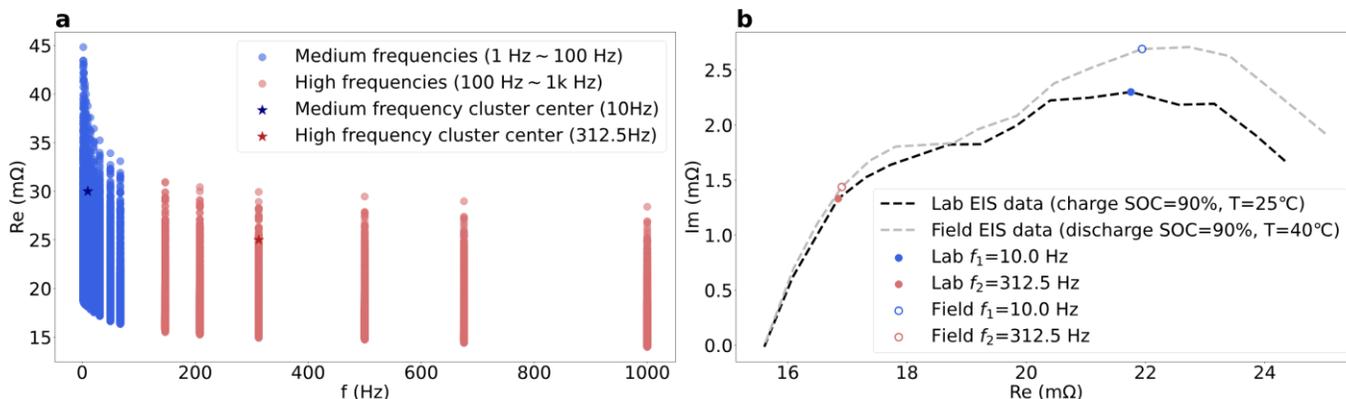

**Figure 3** Test results of Step 1 on the training set of Dataset 1. **a.** The clustering results in the medium frequency range (1 Hz~100 Hz) and the high frequency range (100 Hz~1k Hz) of laboratory Re/f curve. **b.** The positions of the two preset frequencies $f_1$ and $f_2$ on the field EIS and laboratory EIS of the representative cell (numbered as P070_1_S06_C02 in the training set of Dataset 1).

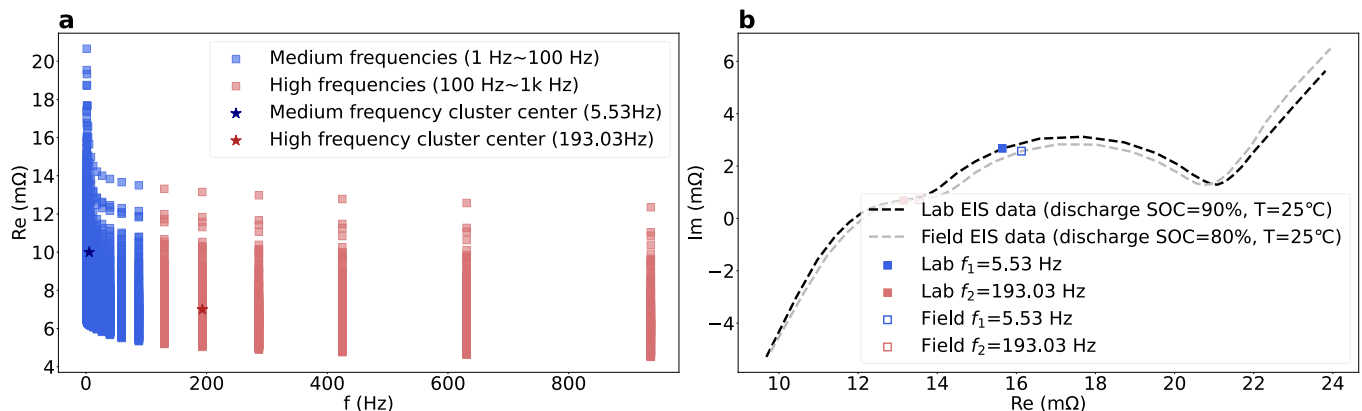

**Figure 4** Test results of Step 1 on the training set of Dataset 2. **a.** The clustering results in the medium frequency range (1 Hz~100 Hz) and the high frequency range (100 Hz~1k Hz) of laboratory Re/f curve. **b.** The positions of the two preset frequencies $f_1$ and $f_2$ on the field EIS and laboratory EIS of the representative cell (numbered as 21 in the training set of Dataset 2).

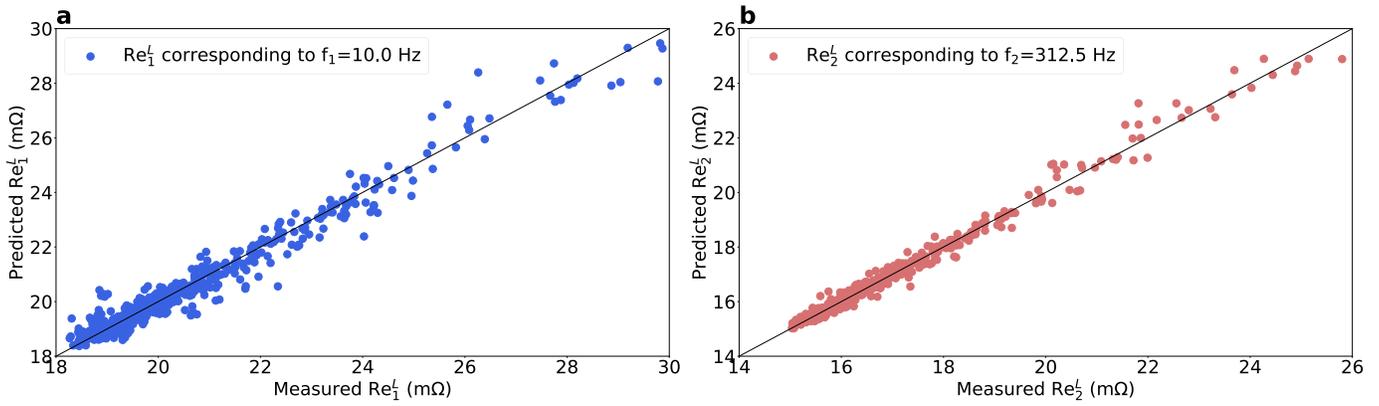

**Figure 5** Test results of Step 2 on the test set of Dataset 1. a. The prediction results of $Re_1^L$ corresponding to $f_1$=10 Hz and $SOC^L$=90%. b. The prediction results of $Re_2^L$ corresponding to $f_2$=312.5 Hz and $SOC^L$=90%.

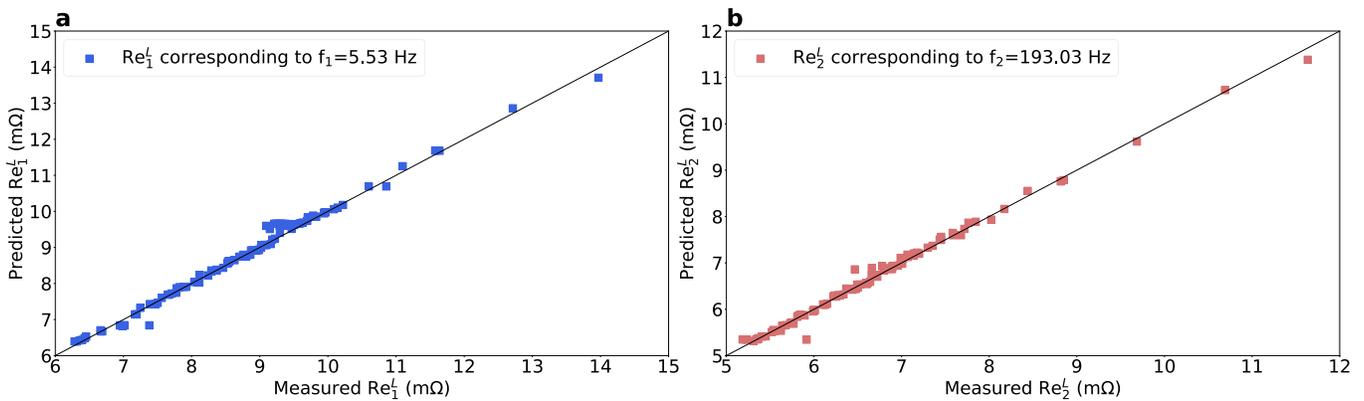

**Figure 6** Test results of step 2 on the test set of Dataset 2. a. The prediction results of $Re_1^L$. b. The prediction results of $Re_2^L$.

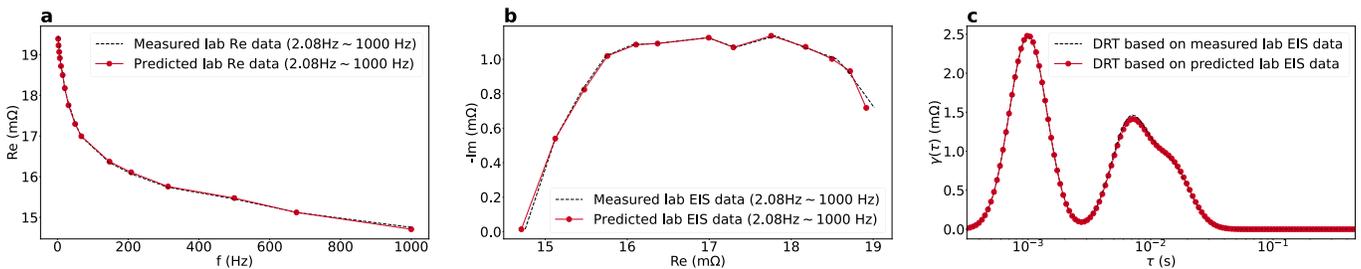

**Figure 7** Test results of Step 3 on the representative cell (numbered as P013_3_S19_C09 in the test set of Dataset 1). a. The prediction results of the laboratory Re/f curve in the mid-high frequency range. b. The predicted EIS and the measured EIS. c. The DRT curves based on the predicted EIS and measured EIS.

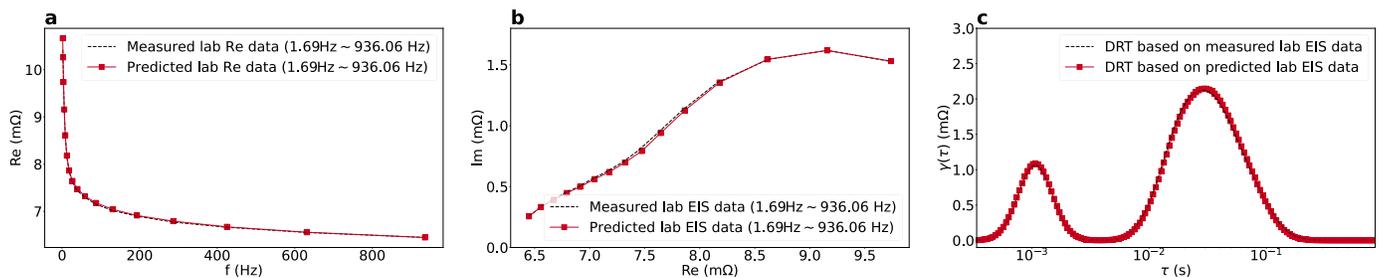

**Figure 8** Test results of Step 3 on the representative cell (numbered as 20 in the test set of Dataset 2). a. The prediction results of the laboratory Re/f curve in the mid-high frequency range. b. The predicted EIS and the measured EIS. c. The DRT curves based on the predicted EIS and measured EIS.

In the test set of Dataset 2, the prediction results of the laboratory Re/f curve in the mid-high frequency range are shown in **Figure 8 (a)**. The MEA, RMSE, and MAPE of the predicted laboratory Re/f curve on the test set are 0.09 mΩ, 0.14, and 0.88%, respectively. To further verify the accuracy of the predicted laboratory Re/f curve, we combine the predicted laboratory Re/f curve with the measured laboratory Im/f curve to plot the predicted EIS, as shown in **Figure 8 (b)**. Furthermore, the same distribution of relaxation times (DRT) calculation is performed on the predicted EIS and the measured EIS, and the obtained DRT curves are shown in **Figure 8 (c)**. The MEA, RMSE, and MAPE on the test set are 0.01 mΩ, 0.02, and 1.11%, respectively. **Figures 8 (b)** and **8 (c)** indirectly verify that the prediction of the laboratory Re/f curve in the mid-high frequency range is accurate.

**Step 4: Other laboratory data prediction.**

Based on the predicted laboratory Re/f curve in **Step 3**, the laboratory charge Q/V curve, discharge Q/V curve, and relaxation V/t curve after full charge are predicted. The inputs of the ML models used to predict laboratory charge Q/V curve, discharge Q/V curve, and relaxation V/t curve after full charge are all laboratory Re/f curve. For more details of the prediction method, please refer to **Supplementary Note 4**.

In the test set of Dataset 1, taking $SOC^L$=90% as an example, the prediction results of the laboratory charge Q/V curve are shown in **Figure 9 (a)**. The MEA, RMSE, and MAPE of the laboratory charge Q/V curve prediction results on the test set are 0.02 Ah, 0.03, and 4.72%, respectively. To further verify the accuracy of the prediction results, the predicted and measured laboratory charge Q/V curve are derived, and the calculated charge incremental capacity (IC) curve and charge differential voltage (DV) curve are shown in **Figures 9 (b)** and **9 (c)**, respectively. The MEA, RMSE, and MAPE of the laboratory charge IC curve prediction results on the test set are 0.05 Ah/V, 0.12, and 10.55%, respectively. The MEA, RMSE, and MAPE of the laboratory charge DV curve prediction results on the test set are 0.86 V/Ah, 1.65, and 9.42%, respectively. **Figures 9 (b)** and **9 (c)** indirectly verify that the prediction of the laboratory charge Q/V curve is accurate.

In the test set of Dataset 1, taking $SOC^L$=90% as an example, the prediction results of the laboratory discharge Q/V curve are shown in **Figure 9 (d)**. The MEA, RMSE, and MAPE of the laboratory discharge Q/V curve prediction results on the test set are 0.03 Ah, 0.04, and 2.69%, respectively. To further verify the accuracy of the prediction results, the predicted and measured laboratory discharge Q/V curve are derived, and the calculated discharge IC curve and discharge DV curve are shown in **Figures 9 (e)** and **9 (f)**, respectively. The MEA, RMSE, and MAPE of the laboratory discharge IC curve prediction results on the test set are 0.06 Ah/V, 0.13, and 9.83%, respectively. The MEA, RMSE, and MAPE of the laboratory discharge DV curve prediction results on the test set are 0.51 V/Ah, 1.03, and 8.80%, respectively. **Figures 9 (e)** and **9 (f)** indirectly verify that the prediction of the laboratory discharge Q/V curve is accurate.

In the test set of Dataset 2, the prediction results of the laboratory charge Q/V curve are shown in **Figure 10 (a)**. The MEA, RMSE, and MAPE of the laboratory charge Q/V curve prediction results on the test set are 0.07 Ah, 0.14, and 8.15%, respectively. To further verify the accuracy of the prediction results, the predicted and measured laboratory charge Q/V curve are derived, and the calculated charge IC curve and charge DV curve are shown in **Figures 10 (b)** and **10 (c)**. The MEA, RMSE, and MAPE of the laboratory charge IC curve prediction results on the test set are 0.26 Ah/V, 0.68, and 9.07%, respectively. The MEA, RMSE, and MAPE of the laboratory charge DV curve prediction results on the test set are 0.32 V/Ah, 11.81, and 7.76%, respectively. **Figures 10 (b)** and **10 (c)** indirectly verify that the prediction of the laboratory charge Q/V curve is accurate.

In the test set of Dataset 2, the prediction results of the laboratory discharge Q/V curve are shown in **Figure 10 (d)**. The MEA, RMSE, and MAPE of the laboratory discharge Q/V curve prediction results on the test set are 0.07 Ah, 0.11, and 2.51%, respectively. To further verify the accuracy of the prediction results, the predicted and measured laboratory discharge Q/V curve are derived, and the calculated discharge IC curve and discharge DV curve are shown in **Figures 10 (e)** and **10 (f)**. The MEA, RMSE, and MAPE of the laboratory discharge IC curve prediction results on the test set are 0.25 Ah/V, 0.56, and 6.86%, respectively. The MEA, RMSE, and MAPE of the laboratory discharge DV curve prediction results on the test set are 0.06 V/Ah, 0.18, and 6.11%, respectively. **Figures 10 (e)** and **10 (f)** indirectly verify that the prediction of the laboratory discharge Q/V curve is accurate.

In the test set of Dataset 2, the prediction results of the laboratory relaxation V/t curve after full charge are shown in **Figure 10 (g)**. The MEA, RMSE, and MAPE of the laboratory relaxation V/t curve prediction results on the test set are 2.82e-4 V, 3.84e-4, and 6.72e-3%, respectively. To further verify the accuracy of the prediction results, the predicted and actually measured laboratory relaxation V/t curve are derived, and the calculated laboratory relaxation dV/dt curve are shown in **Figures 10 (h)**. The MEA, RMSE, and MAPE of the laboratory relaxation dV/dt curve prediction results on the test set are 6.11e-7 V/s, 7.80e-7, and 2.93e-4%, respectively. **Figure 10 (h)** indirectly verify that the prediction of the laboratory relaxation V/t curve is accurate.

**Step 5: Diagnosis and prognosis**

Extract diagnostic and prognostic features first from the laboratory data predicted in **Step 3** and **Step 4**, respectively. Then, input the extracted diagnostic features into the ML model to predict the remaining capacity, and input the extracted prognostic features into the ML model to predict the remaining cycles or days. In this paper, the features extracted from the predicted laboratory data are the best two-point features (BTPFs). The extraction method of BTPF is detailed in [59]. For more details on the diagnostic and prognostic methods, please refer to **Supplementary Note 5**.

Taking $SOC^L$=90% as an example, in the training set of Dataset 1, BTPFs for diagnosis are extracted from the laboratory Re/f curve in the mid-high frequency range, charge Q/V curve, charge IC curve, charge DV curve, discharge Q/V curve, discharge IC curve, and discharge DV curve, as shown in **Supplementary Figures 6 and 7**. The extracted BTPFs for diagnosis are input into the trained ML model to predict the remaining capacity. In the test set of Dataset 1, the BTPFs extracted from laboratory Re/f curve, charge Q/V curve, charge IC curve, charge DV curve, discharge Q/V curve, discharge IC curve, and discharge DV curve have 805 samples, respectively, and the remaining capacity has 805 samples. The predicted remaining capacity of all cells in the test set of Dataset 1 is shown in **Figures 11 (a)** and **11 (b)**. Based the measured laboratory data, the MEA, RMSE, and MAPE of the remaining capacity prediction results on the test set are 0.03 Ah, 0.04, and 1.17%, respectively. Based the predicted laboratory data, the MEA, RMSE, and MAPE of the remaining capacity prediction results on the test set are 0.05 Ah, 0.06, and 1.89%, respectively.

Taking $SOC^L$=90% as an example, in the training set of Dataset 1, BTPFs for prognosis are extracted from the laboratory Re/f curve in the mid-high frequency range, charge Q/V curve, charge IC curve, charge DV curve, discharge Q/V curve, discharge IC curve, and discharge DV curve, as shown in **Supplementary Figures 6 and 7**. The extracted BTPFs for prognosis are input into the trained ML model to predict the remaining days. In the test set of Dataset 1, the BTPFs extracted from laboratory Re/f curve, charge Q/V curve, charge IC curve, charge DV curve, discharge Q/V curve, discharge IC curve, and discharge DV curve have 46 samples, respectively, and the remaining days has 46 samples. The predicted remaining cycles of all cells in the test set of Dataset 1 is shown in **Figures 11 (c)** and **11 (d)**. Based the measured laboratory data, the MEA, RMSE, and MAPE of the remaining cycles prediction results on the test set are 60.21 Ah, 81.53, and 23.33%, respectively. Based the predicted laboratory data, the MEA, RMSE, and MAPE of the remaining cycles prediction results on the test set are 85.87 Ah, 100.87, and 30.61%, respectively.

In the training set of Dataset 2, BTPFs for diagnosis are extracted from the laboratory Re/f curve in the mid-high frequency range, charge Q/V curve, charge IC curve, charge DV curve, discharge Q/V curve, discharge IC curve, discharge DV curve, and relaxation V/t curve after full charge, as shown in **Supplementary Figures 8 and 9**. The extracted BTPFs for diagnosis are input into the trained ML model to predict the remaining capacity. In the test set of Dataset 2, the laboratory Re/f curve, charge Q/V curve, charge IC curve, charge DV curve, discharge Q/V curve, discharge IC curve, discharge DV curve, and relaxation V/t curve after full charge have 118 samples, and the remaining capacity has 118 samples. The predicted remaining capacity of all cells in the test set of Dataset 2 is shown in **Figures 12 (a)** and **12 (b)**. Based the measured laboratory data, the MEA, RMSE, and MAPE of the remaining capacity prediction results on the test set are 0.03 Ah, 0.14, and 1.71%, respectively. Based the predicted laboratory data, the MEA, RMSE, and MAPE of the remaining capacity prediction results on the test set are 0.12 Ah, 0.16, and 3.08%, respectively.

In the training set of Dataset 2, BTPFs for prognosis are extracted from the laboratory Re/f curve in the mid-high frequency range, charge Q/V curve, charge IC curve, charge DV curve, discharge Q/V curve, discharge IC curve, discharge DV curve, and relaxation V/t curve after full charge, as shown in **Supplementary Figures 8 and 9**. The extracted BTPFs for prognosis are input into the trained ML model to predict the remaining cycles. In the test set of Dataset 2, the laboratory Re/f curve, charge Q/V curve, charge IC curve, charge DV curve, discharge Q/V curve, discharge IC curve, discharge DV curve, and relaxation V/t curve after full charge have 10 samples, and the remaining cycle data has 10 samples. The predicted remaining cycles of all cells in the test set of Dataset 2 are shown in **Figures 12 (c)** and **12 (d)**. Based the measured laboratory data, the MEA, RMSE, and MAPE of the remaining cycles prediction results on the test set are 42.28 Ah, 53.69, and 12.21%, respectively. Based the predicted laboratory data, the MEA, RMSE, and MAPE of the remaining cycles prediction results on the test set are 63.86 Ah, 91.40, and 16.91%, respectively.

For the test set of Dataset 1, when $SOC^L$=10%, 30%, 50%, 70%, or 90%, the laboratory data prediction results of **Step 2** to **Step 4** are shown in **Supplementary Tables 1** to **5**. From **Step 2** to **Step 5**, all ML models are random forest (RF) regression models [60]. The hyperparameter settings and training process of RF models are detailed in **Supplementary Note 6**. The evaluation metrics for diagnostic and prognostic results are detailed in **Supplementary Note 7**.

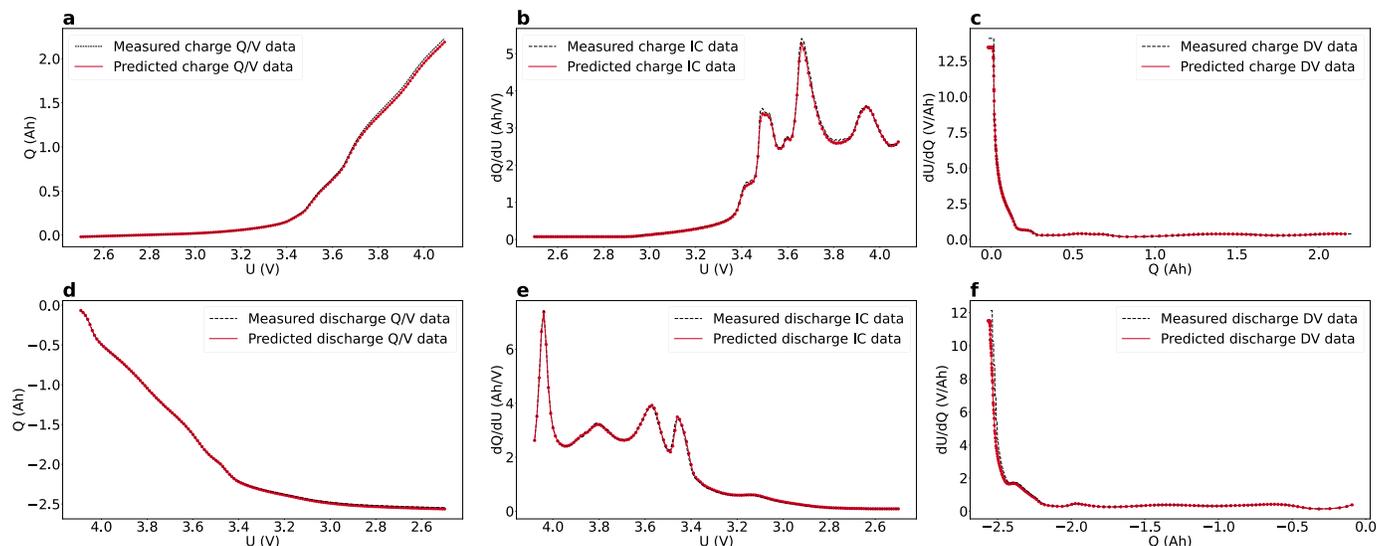

**Figure 9 Test results of Step 4 on the representative cell (numbered as P013_3_S19_C09 in the test set of Dataset 1). a. The prediction results of the laboratory charge Q/V curve. b. The prediction results of the laboratory charge IC curve. c. The prediction results of the laboratory charge DV curve. d. The prediction results of the laboratory discharge Q/V curve. e. The prediction results of the laboratory discharge IC curve. f. The prediction results of the laboratory discharge DV curve.**

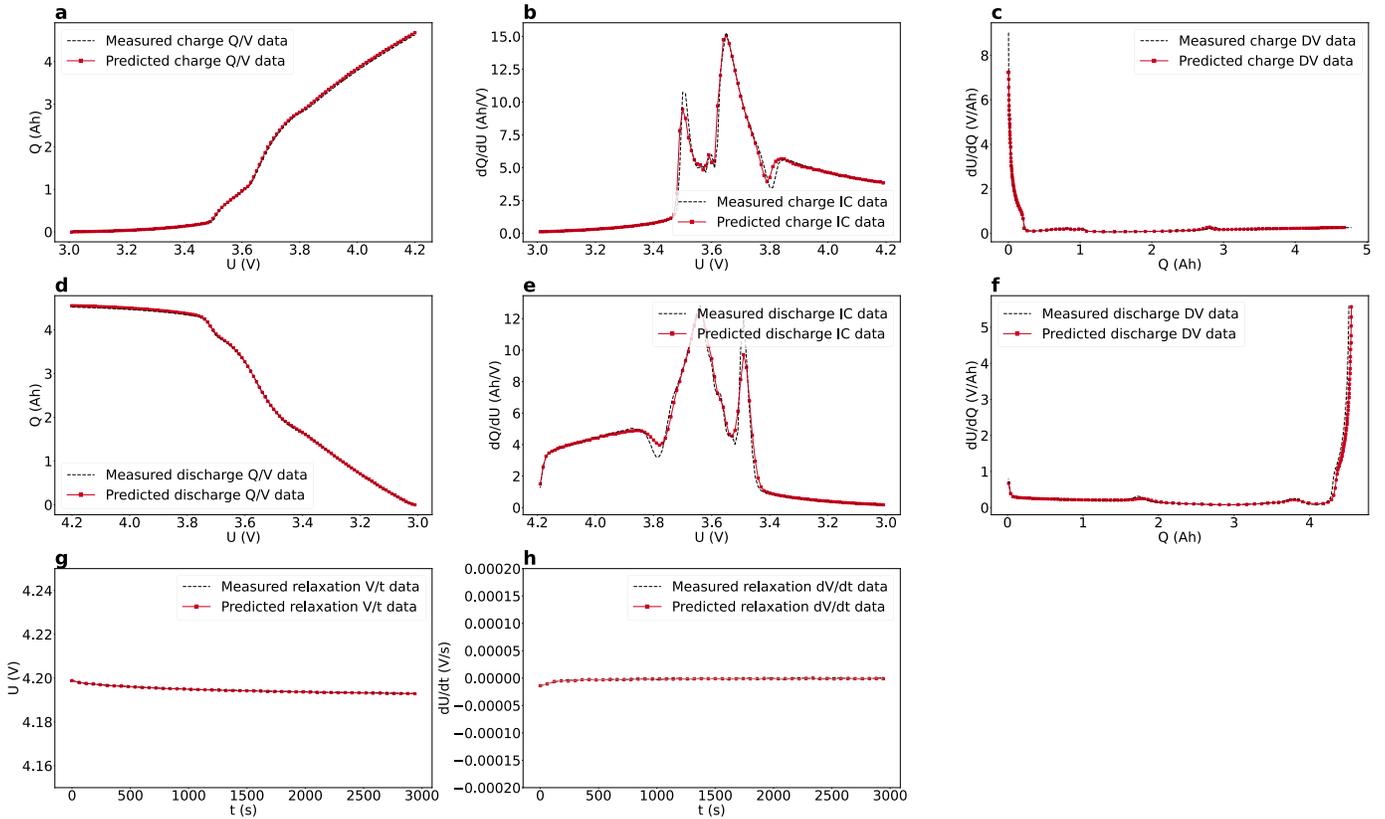

Figure 10 Test results of Step 4 on the representative cell (numbered as 20 in the test set of Dataset 2). a. The prediction results of the laboratory charge Q/V curve. b. The prediction results of the laboratory charge IC curve. c. The prediction results of the laboratory charge DV curve. d. The prediction results of the laboratory discharge Q/V curve. e. The prediction results of the laboratory discharge IC curve. f. The prediction results of the laboratory discharge DV curve. g. The prediction results of the laboratory relaxation V/t curve after full charge. h. The prediction results of the laboratory relaxation dV/dt curve after full charge.

## Discussions

The proposed method of bridging field data and laboratory data has the advantages of high accuracy, low cost, fast speed, readily available, and no use of private data.

(1) In terms of accuracy, the proposed method uses laboratory data for battery diagnosis and prognosis, and can achieve higher accuracy than the field data-driven method when using the same amount of data. The higher the accuracy of the diagnostic and prognostic method, the better the effect of predictive maintenance and optimization management, and the more beneficial it is to the safe operation of field batteries.

(2) In terms of cost, compared with the field data-driven method that uses massive field historical data and needs to be redeveloped, the proposed method requires less data (most of the data is laboratory data, which is cheaper and easier to collect) and does not require redevelopment (the existing laboratory data-driven method can be directly used based on the predicted laboratory data), and the development cost is low. At present, some DC/DCs have integrated impedance measurement functions, and the actual deployment cost of the proposed method is also low. Compared with the large ML models (such as deep learning models) commonly used in field data-driven methods, the ML models used in the proposed method are smaller and have lower use costs. In summary, compared with the field data-driven method, the proposed method has lower development cost, deployment cost, and usage cost, which is more conducive to large-scale commercialization.

(3) In terms of speed, compared with the field data-driven method that requires the input of battery operation data within a specific stage, the proposed method only needs to collect two Res corresponding to a medium frequency and a high frequency, which takes a very short time (in milliseconds) and basically does not affect the normal operation of field batteries.

(4) In terms of readily available, compared with the field data-driven method that relies on random passive data generated during battery operation, the two field Res used in the proposed method are user-controlled active data that can be collected anytime and anywhere. Therefore, based on the proposed method, field battery diagnosis and prognosis can be performed anytime and anywhere according to user needs.

(5) In terms of privacy, compared with the field data-driven method that uses massive historical field data which may involve user privacy, the proposed method only uses two Res and does not use any data involving user privacy.

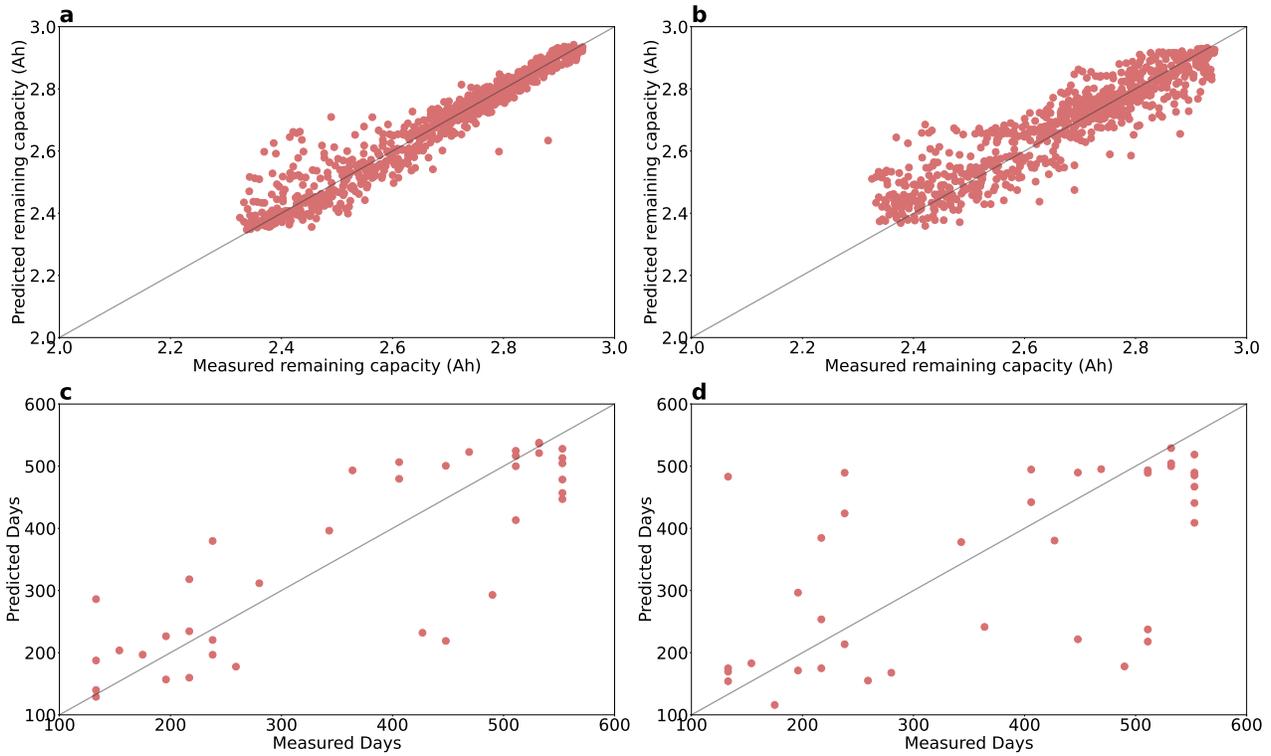

**Figure 11** Test results of Step 5 on the test set of Dataset 1. a. The predicted remaining capacity of all 76 cells in the test set based the measured laboratory data. b. The predicted remaining capacity of all 76 cells in the test set based the predicted laboratory data. c. The predicted life of all 46 cells in the test set based the measured laboratory data. e. The predicted life of all 46 cells in the test set based the predicted laboratory data.

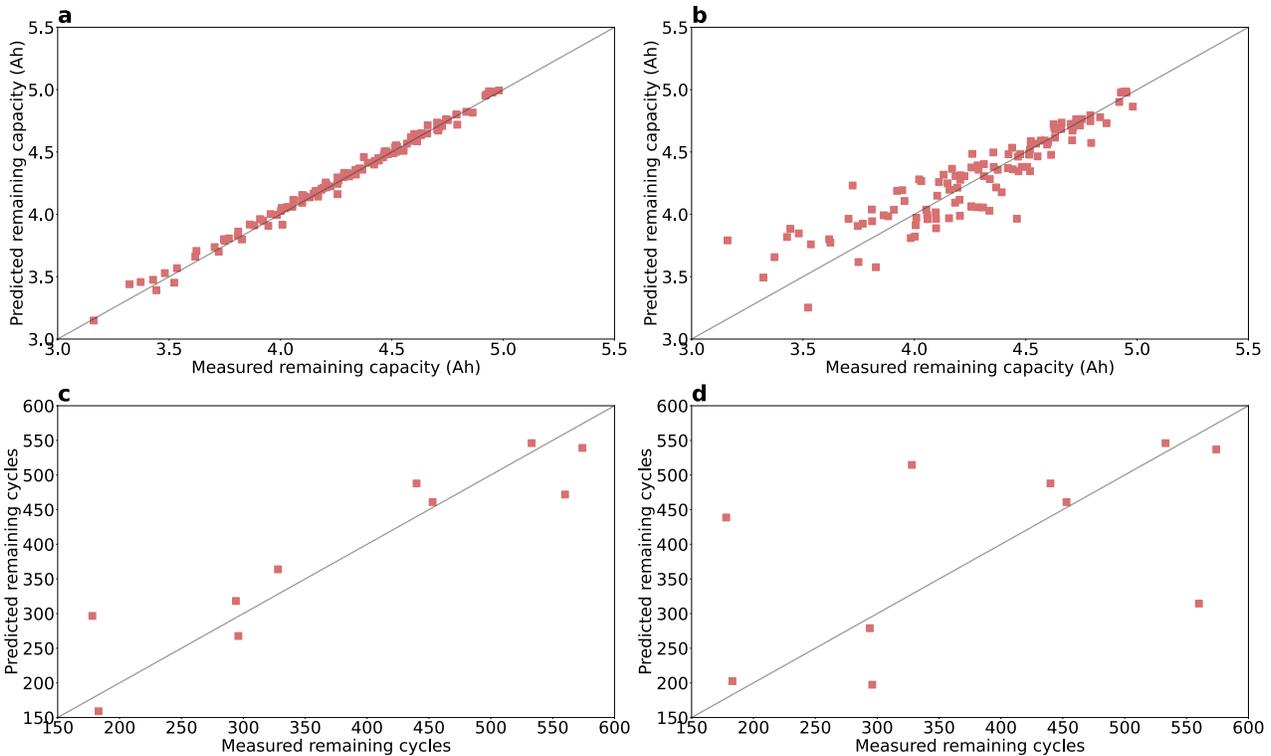

**Figure 12** Test results of Step 5 on the test set of Dataset 2. a. The predicted remaining capacity of all 10 cells in the test set based the measured laboratory data. b. The predicted remaining capacity of all 10 cells in the test set based the predicted laboratory data. c. The predicted life of all 10 cells in the test set based the measured laboratory data. e. The predicted life of all 10 cells in the test set based the predicted laboratory data.

## Conclusions

Accurate health diagnostic and prognostic information can provide a reliable basis for predictive maintenance and optimized management, thereby ensuring the safe operation of equipment. In current practical applications, such as passenger cars and energy storage systems, there is still a lack of high accuracy, low cost, and readily available diagnostic and prognostic methods for field batteries. The main reason is that the formats of laboratory data and field data are different, making it impossible for laboratory data-driven methods to be directly applied to field batteries. This work proposes a bridging method that uses ML to bridge field data and laboratory data. Only two Res collected under field conditions can accurately predict the impedance data, charge/discharge data, and relaxation data under specific laboratory conditions. Based on the predicted laboratory data, the laboratory data-driven method can be used for field battery diagnosis and prognosis. This paper uses two open-source datasets (a total of 249 NMC cells) to test the proposed method. Test results show that the predicted laboratory data are accurate, with the MAPEs of 0.85%, 4.72%, and 2.69% for laboratory Re/f curve, charge Q/V curve, and discharge Q/V curve prediction on the test set of Dataset 1. Compared with the field data-driven methods developed based on massive field historical data, the proposed bridging method has the advantages of high accuracy, low cost, fast speed, readily available, and no use of privacy data. This work prioritizes method exploration rather than maximizing prediction accuracy.

## Data availability

All datasets used in this study are publicly accessible and include Dataset 1 (https://www.nature.com/articles/s41597-024-03831-x) and Dataset 2 (https://iopscience.iop.org/article/10.1149/1945-7111/ac2d3e#jesac2d3eapp1). Source data of this paper will be available after publication.

## Code availability

The code needed to replicate the results and figures in this paper will be available after publication.

## References

[1] Steffen Link, Annegret Stephan, Daniel Speth & Patrick Plötz. Rapidly declining costs of truck batteries and fuel cells enable large-scale road freight electrification, *Nature Energy*, **9**, 1032–1039 (2024).
[2] Zachary P. Cano, Dustin Banham, Siyu Ye, Andreas Hintennach, Jun Lu, Michael Fowler & Zhongwei Chen. Batteries and fuel cells for emerging electric vehicle markets, *Nature Energy*, **3**, 279–289 (2018).
[3] Venkatasubramanian Viswanathan, Alan H. Epstein, Yet-Ming Chiang, Esther Takeuchi, Marty Bradley, John Langford & Michael Winter. The challenges and opportunities of battery-powered flight, *Nature*, **601**, 519–525 (2022).
[4] Kristen A. Severson, Peter M. Attia, Norman Jin, Nicholas Perkins, Benben Jiang, Zi Yang, Michael H. Chen, Muratahan Aykol, Patrick K. Herring, Dimitrios Fraggedakis, Martin Z. Bazant, Stephen J. Harris, William C. Chueh & Richard D. Braatz. Data-driven prediction of battery cycle life before capacity degradation, *Nature Energy*, **4**, 383–391 (2019).
[5] Jingyuan Zhao, Zhilong Lv, Di Li, Xuning Feng, Zhenghong Wang, Yuyan Wu, Dapai Shi, Michael Fowler & Andrew F. Burke. Battery engineering safety technologies (BEST): M5 framework of mechanisms, modes, metrics, modeling, and mitigation, *eTransportation*, **22**, 100364 (2024).
[6] Peter M. Attia, Eric Moch & Patrick K. Herring. Perspective: Challenges and opportunities for high-quality battery production at scale, *arXiv*, 2403.01065 (2024).
[7] Xuebing Han, Languang Lu, Yuejiu Zheng, Xuning Feng, Zhe Li, Jianqiu Li & Minggao Ouyang. A review on the key issues of the lithium-ion battery degradation among the whole life cycle, *eTransportation*, **1**, 100005 (2019).
[8] Richard Schmuch, Ralf Wagner, Gerhard Hörpel, Tobias Placke & Martin Winter. Performance and cost of materials for lithium-based rechargeable automotive batteries, *Nature Energy*, **3**, 267–278 (2018).
[9] Chao-Yang Wang, Teng Liu, Xiao-Guang Yang, Shanhai Ge, Nathaniel V. Stanley, Eric S. Rountree, Yongjun Leng & Brian D. McCarthy. Fast charging of energy-dense lithium-ion batteries, *Nature*, **611**, 485–490 (2022).
[10] Xiaosong Hu, Le Xu, Xianke Lin & Michael Pecht. Battery lifetime prognostics, *Joule*, 4(2), 310-346, (2020).
[11] Man-Fai Ng, Jin Zhao, Qingyu Yan, Gareth J. Conduit & Zhi Wei She. Predicting the state of charge and health of batteries using data-driven machine learning, *Nature Machine Intelligence*, **2**, 161–170 (2020).
[12] Adam Thelen, Xun Huan, Noah Paulson, Simona Onori, Zhen Hu & Chao Hu. Probabilistic machine learning for battery health diagnostics and prognostics—review and perspectives, *npj materials sustainability*, **2**, 14 (2024).
[13] Minxing Yang, Xiaofei Sun, Rui Liu, Lingzhi Wang, Fei Zhao & Xuesong Mei. Predict the lifetime of lithium-ion batteries using early cycles: A review, *Applied Energy*, **376**, 124171 (2024).
[14] Muratahan Aykol, Chirranjeevi Balaji Gopal, Abraham Anapolsky, Patrick K. Herring, Bruis van Vlijmen, Marc D. Berliner, Martin Z. Bazant, Richard D. Braatz, William C. Chueh & Brian D. Storey. Perspective-Combining Physics and Machine Learning to Predict Battery Lifetime, *Journal of The Electrochemical Society*, **168**, 030525 (2021).
[15] Han Zhang, Yuqi Li, Shun Zheng, Ziheng Lu, Xiaofan Gui, Wei Xu & Jiang Bian. Battery lifetime prediction across diverse ageing conditions with inter-cell deep learning, *Nature Machine Intelligence*, **7**, 270-277 (2025).

[16] Shuangqi Li, H. Oliver Gao & Fengqi You. AI for science in electrochemical energy storage: A multiscale systems perspective on transportation electrification, *Nexus*, 1(3), 100026 (2024).

[17] Yanbin Ning, Feng Yang, Yan Zhang, Zhuomin Qiang, Geping Yin, Jiajun Wang & Shuaifeng Lou. Bridging multimodal data and battery science with machine learning, *Matter*, **7(6)**, 2011-2032 (2024).

[18] Gabriele Pozzato, Anirudh Allam, Luca Pulvirenti, Gianina Alina Negoita, William A. Paxton & Simona Onori. Analysis and key findings from real-world electric vehicle field data, *Joule*, **7(9)**, 2035-2053 (2023).

[19] Gonçalo dos Reis, Calum Strange, Mohit Yadav & Shawn Li. Lithium-ion battery data and where to find it, *Energy and AI*, **5**, 100081 (2021).

[20] Matthias Luh & Thomas Blank. Comprehensive battery aging dataset capacity and impedance fade measurements of a lithium-ion NMCC-SiO cell, *scientific data*, **11**, 1004 (2024).

[21] Florian Stroebl, Ronny Petersohn, Barbara Schricker, Florian Schaeufl, Oliver Bohlen & Herbert Palm. A multi-stage lithium-ion battery aging dataset using various experimental design methodologies, *scientific data*, **11**, 1020 (2024).

[22] Xiaopeng Tang, Xin Lai, Changfu Zou, Yuanqiang Zhou, Jiajun Zhu, Yuejiu Zheng & Furong Gao. Detecting abnormality of battery lifetime from first-cycle data using few‐shot learning, *Advanced Science*, **11(6)**, 2305315 (2024).

[23] Noah H. Paulson, Joseph Kubal, Logan Ward, Saurabh Saxena, Wenquan Lu & Susan J. Babinec. Feature engineering for machine learning enabled early prediction of battery lifetime, *Journal of Power Sources*, **527**, 231127 (2022).

[24] Andrew Weng, Peyman Mohtat, Peter M. Attia, Valentin Sulzer, Suhak Lee, Greg Less & Anna Stefanopoulou. Predicting the impact of formation protocols on battery lifetime immediately after manufacturing, *Joule*, **5(11)**, 2971-2992 (2021).

[25] Xiao Cui, Stephen Dongmin Kang, Sunny Wang, Justin A. Rose, Huada Lian, Alexis Geslin, Steven B. Torrisi, Martin Z. Bazant, Shijing Sun & William C. Chueh. Data-driven analysis of battery formation reveals the role of electrode utilization in extending cycle life, Joule, 8(11), 3072-3087, (2024).

[26] Bruis van Vlijmen, Vivek Lam, Patrick A. Asinger, Xiao Cui, Devi Ganapathi, Shijing Sun, Patrick K. Herring, Chirranjeevi Balaji Gopal, Natalie Geise, Haitao D. Deng, Henry L. Thaman, Stephen Dongmin Kang, Amalie Trewartha, Abraham Anapolsky, Brian D. Storey, William E. Gent, Richard D. Braatz & William C. Chueh. Interpretable Data-Driven Modeling Reveals Complexity of Battery Aging, *ChemRxiv*, (2023).

[27] Peter M. Attia, Aditya Grover, Norman Jin, Kristen A. Severson, Todor M. Markov, Yang-Hung Liao, Michael H. Chen, Bryan Cheong, Nicholas Perkins, Zi Yang, Patrick K. Herring, Muratahan Aykol, Stephen J. Harris, Richard D. Braatz, Stefano Ermon & William C. Chueh. Closed-loop optimization of fast-charging protocols for batteries with machine learning, *Nature*, **578**, 397–402 (2020).

[28] Jingzhao Zhang, Yanan Wang, Benben Jiang, Haowei He, Shaobo Huang, Chen Wang, Yang Zhang, Xuebing Han, Dongxu Guo, Guannan He & Minggao Ouyang. Realistic fault detection of li-ion battery via dynamical deep learning, *Nature Communications*, **14**, 5940 (2023).

[29] Yujie Wang, Jiaqiang Tian, Zhendong Sun, Li Wang, Ruilong Xu, Mince Li & Zonghai Chen. A comprehensive review of battery modeling and state estimation approaches for advanced battery management systems, *Renewable and Sustainable Energy Reviews*, 131, 110015 (2020).

[30] Shengyu Tao, Haizhou Liu, Chongbo Sun, Haocheng Ji, Guanjun Ji, Zhiyuan Han, Runhua Gao, Jun Ma, Ruifei Ma, Yuou Chen, Shiyi Fu, Yu Wang, Yaojie Sun, Yu Rong, Xuan Zhang, Guangmin Zhou & Hongbin Sun. Collaborative and privacy-preserving retired battery sorting for profitable direct recycling via federated machine learning, *Nature Communications*, **14**, 8032 (2023).

[31] Shengyu Tao, Ruifei Ma, Zixi Zhao, Guangyuan Ma, Lin Su, Heng Chang, Yuou Chen, Haizhou Liu, Zheng Liang, Tingwei Cao, Haocheng Ji, Zhiyuan Han, Minyan Lu, Huixiong Yang, Zongguo Wen, Jianhua Yao, Rong Yu, Guodan Wei, Yang Li, Xuan Zhang, Tingyang Xu & Guangmin Zhou. Generative learning assisted state-of-health estimation for sustainable battery recycling with random retirement conditions, *Nature Communications*, **15**, 10154 (2024).

[32] Darius Roman, Saurabh Saxena, Valentin Robu, Michael Pecht & David Flynn. Machine learning pipeline for battery state-of-health estimation, *Nature Machine Intelligence*, **3**, 447–456 (2021).

[33] Fujin Wang, Zhi Zhai, Zhibin Zhao, Yi Di & Xuefeng Chen. Physics-informed neural network for lithium-ion battery degradation stable modeling and prognosis, *Nature Communications*, **15**, 4332 (2024).

[34] Penelope K. Jones, Ulrich Stimming & Alpha A. Lee. Impedance-based forecasting of lithium-ion battery performance amid uneven usage, *Nature Communications*, **13**, 4806 (2022).

[35] Yunwei Zhang, Qiaochu Tang, Yao Zhang, Jiabin Wang, Ulrich Stimming & Alpha A. Lee. Identifying degradation patterns of lithium ion batteries from impedance spectroscopy using machine learning, *Nature Communications*, **11**, 1706 (2020).

[36] Jiangong Zhu, Yixiu Wang, Yuan Huang, R. Bhushan Gopaluni, Yankai Cao, Michael Heere, Martin J. Mühlbauer, Liuda Mereacre, Haifeng Dai, Xinhua Liu, Anatoliy Senyshyn, Xuezhe Wei, Michael Knapp & Helmut Ehrenberg. Data-driven capacity estimation of commercial lithium-ion batteries from voltage relaxation, *Nature Communications*, **13**, 2261 (2022).

[37] Gabriele Pozzato, Anirudh Allam, Luca Pulvirenti, Gianina Alina Negoita, William A. Paxton & Simona Onori. Analysis and key findings from real-world electric vehicle field data, *Joule*, **7(9)**, 2035-2053 (2023).

[38] Qiushi Wang, Zhenpo Wang, Peng Liu, Lei Zhang, Dirk Uwe Sauer & Weihan Li. Large-scale field data-based battery aging prediction driven by statistical features and machine learning, *Cell Reports Physical Science*, **4(12)**, 101720 (2023).

[39] Antti Aitio & David A. Howey. Predicting battery end of life from solar off-grid system field data using machine learning, *Joule*, **5(12)**, 3204-3220 (2021).
[40] Jan Figgener, Jonas van Ouwerkerk, David Haberschusz, Jakob Bors, Philipp Woerner, Marc Mennekes, Felix Hildenbrand, Christopher Hecht, Kai-Philipp Kairies, Oliver Wessels & Dirk Uwe Sauer. Multi-year field measurements of home storage systems and their use in capacity estimation, *Nature Energy*, **9**, 1438–1447 (2024).
[41] Hongao Liu, Chang Li, Xiaosong Hu, Jinwen Li, Kai Zhang, Yang Xie, Ranglei Wu & Ziyou Song. Multi-modal framework for battery state of health evaluation using open-source electric vehicle data, *Nature Communications*, **16**, 1137 (2025).
[42] Rui Cao, Zhengjie Zhang, Runwu Shi, Jiayi Lu, Yifan Zheng, Yefan Sun, Xinhua Liu & Shichun Yang. Model-constrained deep learning for online fault diagnosis in Li-ion batteries over stochastic conditions, *Nature Communications*, **16**, 1651 (2025).
[43] Valentin Steininger, Katharina Rumpf, Peter Hüsson, Weihan Li & Dirk Uwe Sauer. Automated feature extraction to integrate field and laboratory data for aging diagnosis of automotive lithium-ion batteries, *Cell Reports Physical Science*, **4(10)**, 101596 (2023).
[44] Dongzhen Lyu, Bin Zhang, Enrico Zio & Jiawei Xiang. Battery cumulative lifetime prognostics to bridge laboratory and real-life scenarios, *Cell Reports Physical Science*, **5(9)**, 102164 (2024).
[45] Xueyuan Wang, Xuezhe Wei & Haifeng Dai. Estimation of state of health of lithium-ion batteries based on charge transfer resistance considering different temperature and state of charge, *Journal of Energy Storage*, **21**, 618-631 (2019).
[46] Simone Barcellona, Silvia Colnago, Giovanni Dotelli, Saverio Latorrata & Luigi Piegari. Aging effect on the variation of Li-ion battery resistance as function of temperature and state of charge, *Journal of Energy Storage*, **50**, 104658 (2022).
[47] Yue Sun, Rui Xiong, Xiangfeng Meng, Xuanrou Deng, Hailong Li & Fengchun Sun. Battery degradation evaluation based on impedance spectra using a limited number of voltage-capacity curves, eTransportation, **22**, 100347 (2024).
[48] Yanzhou Duan, Jinpeng Tian, Jiahuan Lu, Chenxu Wang, Weixiang Shen & Rui Xiong. a Deep neural network battery impedance spectra prediction by only using constant-current curve, *Energy Storage Materials*, **41**, 24-31 (2021).
[49] Aihua Tang, Yuchen Xu, Pan Liu, Jinpeng Tian, Zikang Wu, Yuanzhi Hu & Quanqing Yu. Deep learning driven battery voltage-capacity curve prediction utilizing short-term relaxation voltage, *eTransportation*, **22**, 100378 (2024).
[50] Jia Guo, Yunhong Che, Kjeld Pedersen & Daniel-Ioan Stroe. Battery impedance spectrum prediction from partial charging voltage curve by machine learning, *Journal of Energy Chemistry*, **79**, 211-221 (2023).
[51] Bor-Rong Chen, Cody M. Walker, Sangwook Kim, M. Ross Kunz, Tanvir R. Tanim & Eric J. Dufek. Battery aging mode identification across NMC compositions and designs using machine learning, *Joule*, **6(12)**, 2776-2793 (2022).
[52] Bor-Rong Chen, M. Ross Kunz, Tanvir R. Tanim & Eric J. Dufek. A machine learning framework for early detection of lithium plating combining multiple physics-based electrochemical signatures, *Cell Reports Physical Science*, **2(3)**, 100352 (2021).
[53] Chun Chang, Yaliang Pan, Shaojin Wang, Jiuchun Jiang, Aina Tian, Yang Gao, Yan Jiang, & Tiezhou Wu. Fast EIS acquisition method based on SSA-DNN prediction model, *Energy*, **288**, 129768 (2024).
[54] Chiaki Mizutani, Masahiro Shiozawa, Tsuyoshi Maruo & Shinji Aso. On-board control system of water content inside FCV stack by electrochemical impedance spectroscopy, *ECS Transactions*, **80(8)**, 357 (2017).
[55] Takahiko Hasegawa, Hiroyuki Imanishi, Mitsuhiro Nada & Yoshihiro Ikogi. Development of the fuel cell system in the Mirai FCV, *SAE Technical Paper*, 2016-01-1185 (2016).
[56] Jiajun Zhu, Xin Lai, Xiaopeng Tang, Yuejiu Zheng, Hengyun Zhang & Haifeng Dai, Yunfeng Huang. Online multi-scenario impedance spectra generation for batteries based on small-sample learning, *Cell Reports Physical Science*, **5(8)**, 102134 (2024).
[57] J. Sihvo & D-I. Stroe. Real-time impedance monitoring of Li-ion batteries under dynamic operating conditions using the discrete Fourier transform eigenvector approach, *Cell Reports Physical Science*, **6(4)**, 102512 (2025).
[58] Peyman Mohtat, Suhak Lee, Jason B. Siegel & Anna G. Stefanopoulou. Reversible and Irreversible Expansion of Lithium-Ion Batteries Under a Wide Range of Stress Factors, *Journal of The Electrochemical Society*, **168(10)**, 100520 (2021).
[59] Hao Liu, Yanbin Zhao, Huarong Zheng, Xiulin Fan, Zhihua Deng, Mengchi Chen, Xingkai Wang, Zhiyang Liu, Jianguo Lu & Jian Chen. Two points are enough, arXiv, 2408.11872 (2024).
[60] Leo Breiman. Random forests, *Machine learning*, **45**, 5-32 (2001).

## Acknowledgements

This work was funded by the Natural Science Foundation of Zhejiang Province (Grant No. LQ23E050013) and National Natural Science Foundation of China (Grant No. U2141209).

## Author contributions

Y. Zhao: Conceptualization, software, methodology, data curation and analysis, visualization, writing-review and editing.
H. Liu: Conceptualization, resources, software, methodology, data curation and analysis, visualization, writing-review and editing, project administration, supervision, funding acquisition.
Z. Deng: Visualization, writing-review and editing.
T. Li: Visualization, writing-review and editing.
H. Jiang: Visualization, writing-review and editing.

Z. Ling: Visualization, writing-review and editing.
X. Wang: Writing-review and editing.
L. Zhang: Writing-review and editing.
X. Ouyang: Writing-review and editing, project administration, funding acquisition.

## Competing interests

The authors declare no competing interests.

Supplementary Information

# Machine learning bridging battery field data and laboratory data

Yanbin Zhao[†1], Hao Liu[†*1], Zhihua Deng[2], Tong Li[1], Haoyi Jiang[1], Zhenfei Ling[1], Xingkai Wang[3], Lei Zhang[4], Xiaoping Ouyang[1]

[1] State Key Laboratory of Fluid Power and Mechatronic Systems, School of Mechanical Engineering, Zhejiang University, Hangzhou, China.
[2] Energy Research Institute at NTU (ERI@N), Nanyang Technological University, Singapore.
[3] School of Materials Science and Engineering, Tianjin University, Tianjin, China.
[4] Wanxiang A123 Systems Corp., Hangzhou, China.
[†] These authors contributed equally to this work: Yanbin Zhao, Hao Liu.
[*] Corresponding authors' email: haoliu7850052@zju.edu.cn (Hao Liu).

## Supplementary Note 1

As shown in **Supplementary Figure 1**, based on the laboratory Re/f curves in the training set, the k-means clustering algorithm is utilized to extract two preset frequencies from the Re/f curve containing $n$ frequencies. The medium frequency part (1 Hz~100 Hz) and high frequency part (100 Hz~10K Hz) in the Re/f curve are extracted. Each Re/f data point contains a frequency value and a corresponding Re value. First, the number of clusters is set to 2, and two data points on the Re/f curve are randomly selected as the initial cluster centers using the *k-means++* algorithm. Then, the k-means clustering algorithm is used, and two clusters of medium frequency and high frequency are output. Next, the mean of the frequency values in the medium frequency cluster is calculated, and the actual frequency value closest to the frequency mean is selected as the center frequency of the medium frequency cluster. Similarly, the center frequency of the high frequency cluster can be selected. These two center frequencies are used as a center frequency combination. Finally, the number of occurrences of different center frequency combinations in all Re/f curves is counted, and the center frequency combination with the most occurrences is selected as the two preset frequencies. In Dataset 1, $n$ is 16. In Dataset 2, $n$ is 17.

In Dataset 1, $f_1$=10 Hz and $f_2$=312.5 Hz, $SOC^F$=10%, 30%, 50%, 70%, or 90%, and $T^F$=0°C, 10°C, 25°C, or 40°C.

In Dataset 2, $f_1$=5.53 Hz and $f_2$=193.03 Hz, $SOC^F$=10%, 20%, 30%, 40%, 50%, 60%, 70% or 80%, and $T^F$=25°C.

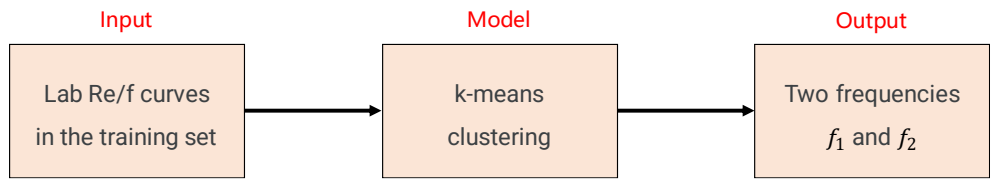

**Supplementary Figure 1** The extraction method of two preset frequencies $f_1$ and $f_2$.

Supplementary Note 2

In Dataset 1, $T^L$=25°C and $SOC^L$=10%, 30%, 50%, 70%, or 90%. The ML models used to predict $Re_1^L$ and $Re_2^L$ are trained using the cell data in the training set. Taking $SOC^L$=90% as an example, there are 1627 samples of field $Re_1^F$ and $Re_2^F$ in the training set, and 1627 samples of laboratory $Re_1^L$ and $Re_2^L$. The inputs of the ML model used to predict $Re_1^L$ are $Re_1^F$, $T^F$, and $SOC^F$, and the output is $Re_1^L$ corresponding to $f_1$=10 Hz and $SOC^L$=10%, 30%, 50%, 70%, or 90%. The inputs of the ML model used to predict $Re_2^L$ are $Re_2^F$, $T^F$, and $SOC^F$, and the output is $Re_2^L$ corresponding to $f_2$=312.5 Hz and $SOC^L$=10%, 30%, 50%, 70%, or 90%.

In Dataset 2, the SOC and operating temperature corresponding to the laboratory condition are 90% and 25°C, respectively. The ML models used to predict $Re_1^L$ and $Re_2^L$ are trained using the cell data in the training set, where there are 122 samples of field $Re_1^F$ and $Re_2^F$ data, and 122 samples of laboratory $Re_1^L$ and $Re_2^L$ data. The inputs of the ML model used to predict $Re_1^L$ are $Re_1^F$, $T^F$, and $SOC^F$, and the output is $Re_1^L$ corresponding to $f_1$=5.53 Hz. The inputs of the ML model used to predict $Re_2^L$ are $Re_2^F$, $T^F$, and $SOC^F$, and the output is $Re_2^L$ corresponding to $f_2$=193.03 Hz and SOC=90%.

Note that, to improve the prediction accuracy of $Re_1^L$ and $Re_2^L$, we first divided the field SOC into multiple independent intervals. Specifically, the value of $SOC^F$ is divided into ten independent intervals, namely [0%, 10%), [10%, 20%), [20%, 30%), [30%, 40%), [40%, 50%), [50%, 60%), [60%, 70%), [70%, 80%), [80%, 90%), and [90%, 100%]. As shown in **Supplementary Figure 2**, when $SOC^F$ ∈[0%, 10%), [10%, 20%), [20%, 30%), [30%, 40%), [40%, 50%), [50%, 60%), [60%, 70%), [70%, 80%), [80%, 90%), or [90%, 100%], the laboratory $Re_1^L$ and $Re_2^L$ output by the ML model corresponds to $SOC^L$=0%, 10%, 20%, 30%, 40%, 50%, 60%, 70%, 80%, or 90%, respectively. Note that in Dataset 2, we use the steady-state EIS measured at SOC=90% and T=25°C during discharging as the laboratory data, and the steady-state EIS measured at other SOC and T=25°C during discharging as the field data. when $SOC^F$ ∈[0%, 10%), [10%, 20%), [20%, 30%), [30%, 40%), [40%, 50%), [50%, 60%), [60%, 70%), [70%, 80%), [80%, 90%), or [90%, 100%], the laboratory $Re_1^L$ and $Re_2^L$ output by the ML model corresponds to $SOC^L$=90%.

Then, we divided the field Re values into multiple independent intervals and trained a ML model for each independent interval.

Specifically, in Dataset 1, according to the numerical range of $Re_1^F$ corresponding to $f_1$=10 Hz in the training set, the value of $Re_1^F$ is divided into four independent intervals, namely [14, 18), [18, 22), [22, 26), and [26, 30]. Therefore, a total of forty ML models are trained for the prediction of $Re_1^L$ corresponding to $f_1$=10 Hz. According to the numerical range of $Re_2^F$ corresponding to $f_2$=312.5 Hz in the training set, the value of $Re_2^F$ is divided into four independent intervals, namely [14, 18), [18, 22), [22, 26), and [26, 30]. Therefore, a total of forty ML models are trained for the prediction of $Re_2^L$ corresponding to $f_2$=312.5 Hz.

In Dataset 2, according to the numerical range of $Re_1^F$ corresponding to $f_1$=5.53 Hz in the training set, the value of $Re_1^F$ is divided into six independent intervals, namely [4, 6), [6, 8), [8, 10), [10, 12), [12, 14), and [14, 16]. Therefore, a total of sixty ML models are trained for the prediction of $Re_1^L$ corresponding to $f_1$=5.53 Hz. According to the numerical range of $Re_2^F$ corresponding to $f_2$=193.03 Hz in the training set, the value of $Re_2^F$ is divided into six independent intervals, namely [4, 6), [6, 8), [8, 10), [10, 12), [12, 14), [14, 16]. Therefore, a total of sixty ML models are trained for the prediction of $Re_2^L$ corresponding to $f_2$=193.03 Hz.

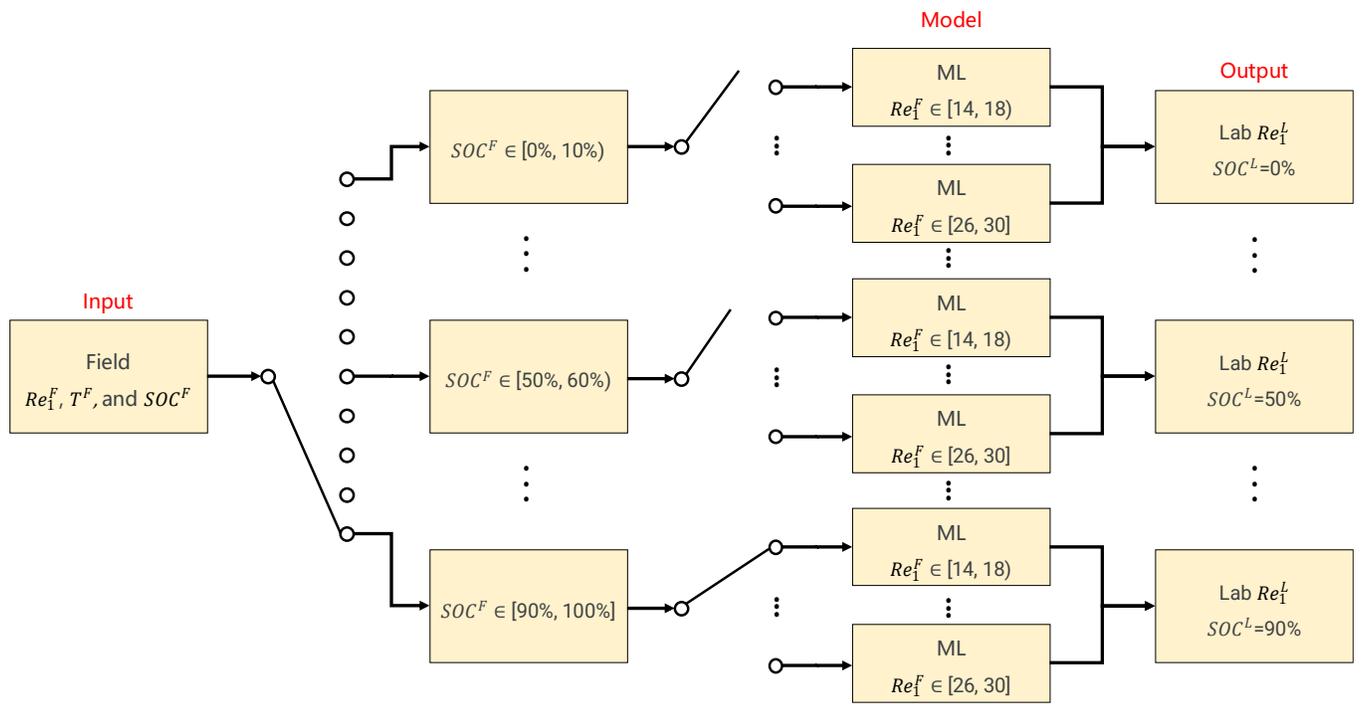

(a)

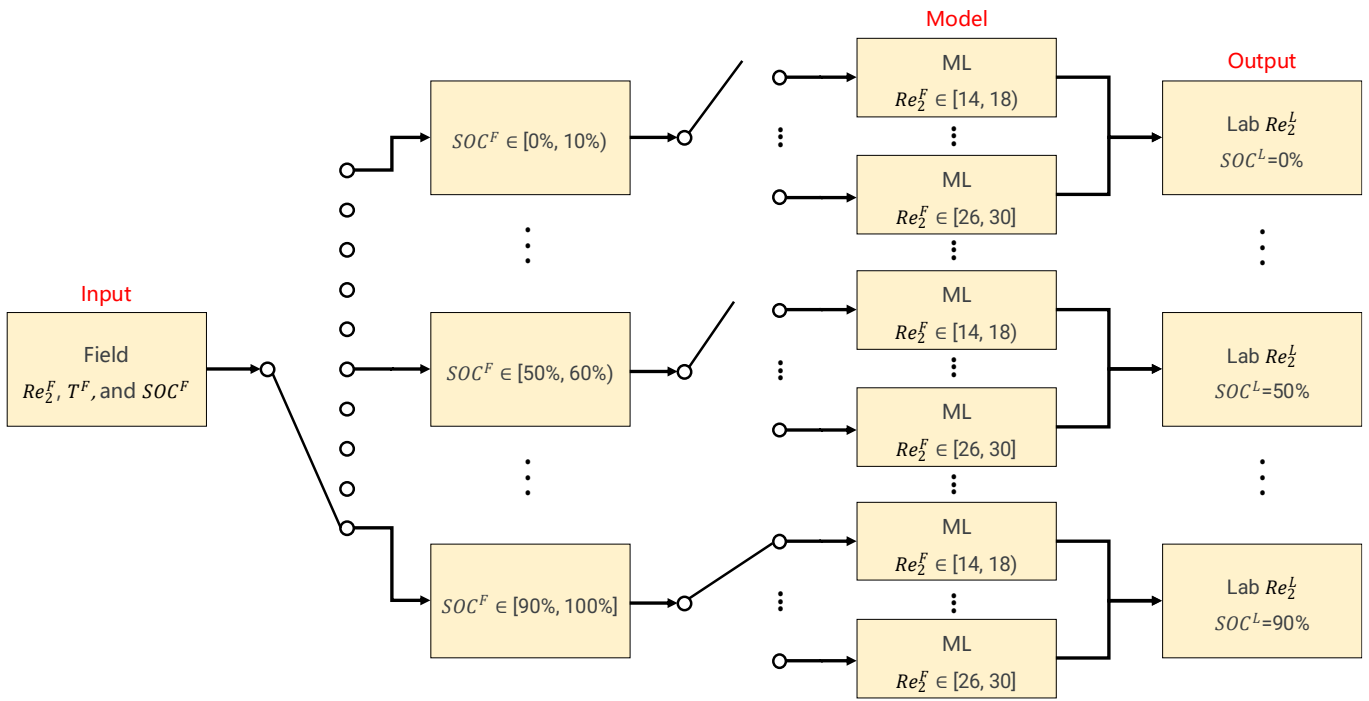

(b)

**Supplementary Figure 2** (a) The prediction method of $Re_1^L$. (b) The prediction method of $Re_2^L$.

Supplementary Note 3

In Dataset 1, the ML model used to predict laboratory Re/f curve in the mid-high frequency range is trained using the cell data in the training set. Taking $SOC^L$=90% as an example, there are 1627 samples of $Re_1^L$ and $Re_2^L$ in the training set, and 1627 samples of laboratory Re/f curve (EIS). As shown in **Supplementary Figure 3**, the inputs of the ML model are $Re_1^L$ and $Re_2^L$, and the outputs are laboratory Re/f curve in the frequency range of [2.08 Hz, 1000 Hz] (with 16 Re/f data).

In Dataset 2, the ML model used to predict laboratory Re/f curve in the mid-high frequency range is trained using the cell data in the training set, where there are 122 samples of $Re_1^L$ and $Re_2^L$ data, and 122 samples of laboratory Re/f curve (EIS). The inputs of the ML model are $Re_1^L$ and $Re_2^L$, and the outputs are laboratory Re/f curve in the frequency range of [1.69 Hz, 936.06 Hz] (with 17 Re/f data).

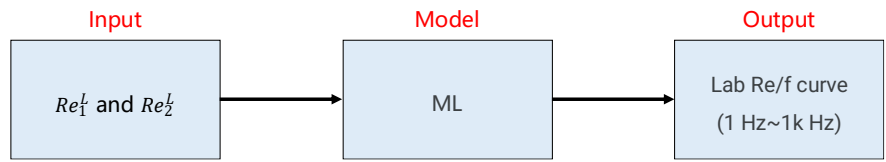

**Supplementary Figure 3** The prediction method of the laboratory Re/f curve in the mid-high frequency range.

## Supplementary Note 4

In Dataset 1, the ML models used for laboratory charge Q/V curve and discharge Q/V curve prediction are trained using the cell data in the training set. Taking $SOC^L$=90% as an example, there are 1627 samples of laboratory Re/f curve in the mid-high frequency range, and 1627 samples of laboratory charge Q/V curve and discharge Q/V curve in the training set. As shown in **Supplementary Figures 4(a)** and **4(b)**, the inputs of the ML models are 16 laboratory Re/f data in the frequency range of [2.08 Hz, 1000 Hz], and the outputs are 160 capacitance values in the charging voltage range of [2.5 V, 4.1 V] (the voltage interval is 0.01 V) or 160 capacitance values in the discharge voltage range of [2.5 V, 4.1 V] (the voltage interval is 0.01 V).

In Dataset 2, the ML models used for laboratory charge Q/V curve, discharge Q/V curve, and relaxation V/t curve after full charge are trained using the cell data in the training set, where there are 122 samples of laboratory Re/f curve in the mid-high frequency range, and 122 samples of laboratory charge Q/V curve, discharge Q/V curve, and relaxation V/t curve after full charge. As shown in **Supplementary Figure 4**, the inputs of the ML models are 17 laboratory Re/f data in the frequency range of [1.69 Hz, 936.06 Hz], and the outputs are 120 capacitance values in the charging voltage range of [3.01 V, 4.20 V] (the voltage interval is 0.01 V), 120 capacitance values in the discharge voltage range of [3.01 V, 4.20 V] (the voltage interval is 0.01 V), or 50 voltage values in the time range of [0 s, 3600 s] (the time interval is 60 s) after full charge.

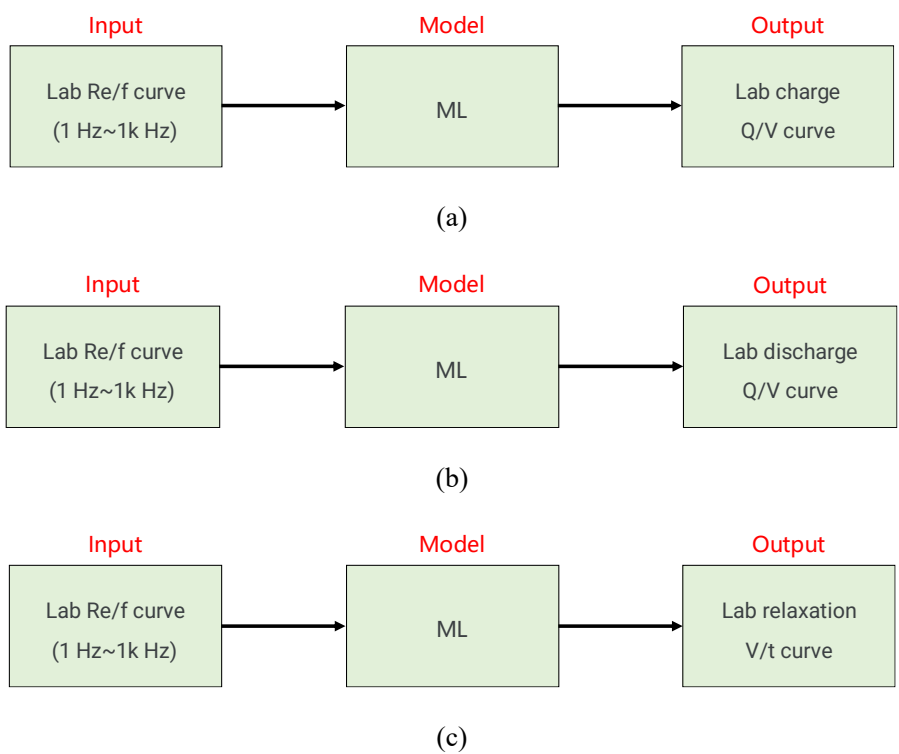

**Supplementary Figure 4** (a) The prediction method of the laboratory charge Q/V curve. (b) The prediction method of the laboratory discharge Q/V curve. (c) The prediction method of the laboratory relaxation V/t curve after full charge.

Supplementary Note 5

Specifically, in Dataset 1, the ML model used for diagnosis is trained using the cell data in the training set. Taking $SOC^L$=90% as an example, the laboratory Re/f curve in the mid-high frequency range, charge Q/V curve, and discharge Q/V curve have 1627 samples in the training set, and the remaining capacity has 1627 samples. The difference data used to extract BTPFs for diagnosis is obtained by subtracting the first RPT data from each RPT data. As shown in **Supplementary Figure 5**, the inputs of the ML model for diagnosis are seven BTPFs [1] extracted from the laboratory Re/f curve in the mid-high frequency range, charge Q/V curve, charge IC curve, charge DV curve, discharge Q/V curve, discharge IC curve, and discharge DV curve, and the output is the remaining capacity. The ML model for prognosis is trained using the cell data in the training set. Taking $SOC^L$=90% as an example, the laboratory Re/f curve in the mid-high frequency range, charge Q/V curve, and discharge Q/V curve have 152 samples in the training set, and the remaining days has 152 samples. The difference data used to extract BTPFs for prognosis is obtained by subtracting the first RPT data from the third RPT data. The inputs of the ML model for prognosis are seven BTPFs extracted from the laboratory Re/f curve in the mid-high frequency range, charge Q/V curve, charge IC curve, charge DV curve, discharge Q/V curve, discharge IC curve, and discharge DV curve, and the output is the remaining days. The selection of BTPFs for diagnosis and prognosis from laboratory data in the training set is shown in **Supplementary Figure 6** and **Supplementary Figure 7**.

In Dataset 2, the ML model used for diagnosis is trained using the cell data in the training set, where the laboratory Re/f curve in the mid-high frequency range, charge Q/V curve, discharge Q/V curve, and relaxation V/t curve after full charge have 122 samples, and the remaining capacity has 122 samples. The difference data used to extract BTPFs for diagnosis is obtained by subtracting the first RPT data from each RPT data. As shown in **Supplementary Figure 5**, the inputs of the ML model for diagnosis are eight BTPFs extracted from the laboratory Re/f curve in the mid-high frequency range, charge Q/V curve, charge IC curve, charge DV curve, discharge Q/V curve, discharge IC curve, discharge DV curve, and relaxation V/t data after full charge, and the output is the remaining capacity. The ML model for prognosis is trained using the cell data in the training set, where the laboratory Re/f curve in the mid-high frequency range, charge Q/V curve, charge IC curve, charge DV curve, discharge Q/V curve, discharge IC curve, discharge DV curve, and relaxation V/t data after full charge have 11 samples, and the remaining cycles has 11 samples. The difference data used to extract BTPFs for prognosis is obtained by subtracting the first RPT data from the second RPT data. The inputs of the ML model for prognosis are eight BTPFs extracted from the laboratory Re/f curve in the mid-high frequency range, charge Q/V curve, charge IC curve, charge DV curve, discharge Q/V curve, discharge IC curve, discharge DV curve, and relaxation V/t data after full charge, and the output is the remaining cycles. The selection of BTPFs for diagnosis and prognosis from laboratory data in the training set is shown in **Supplementary Figure 8** and **Supplementary Figure 9**.

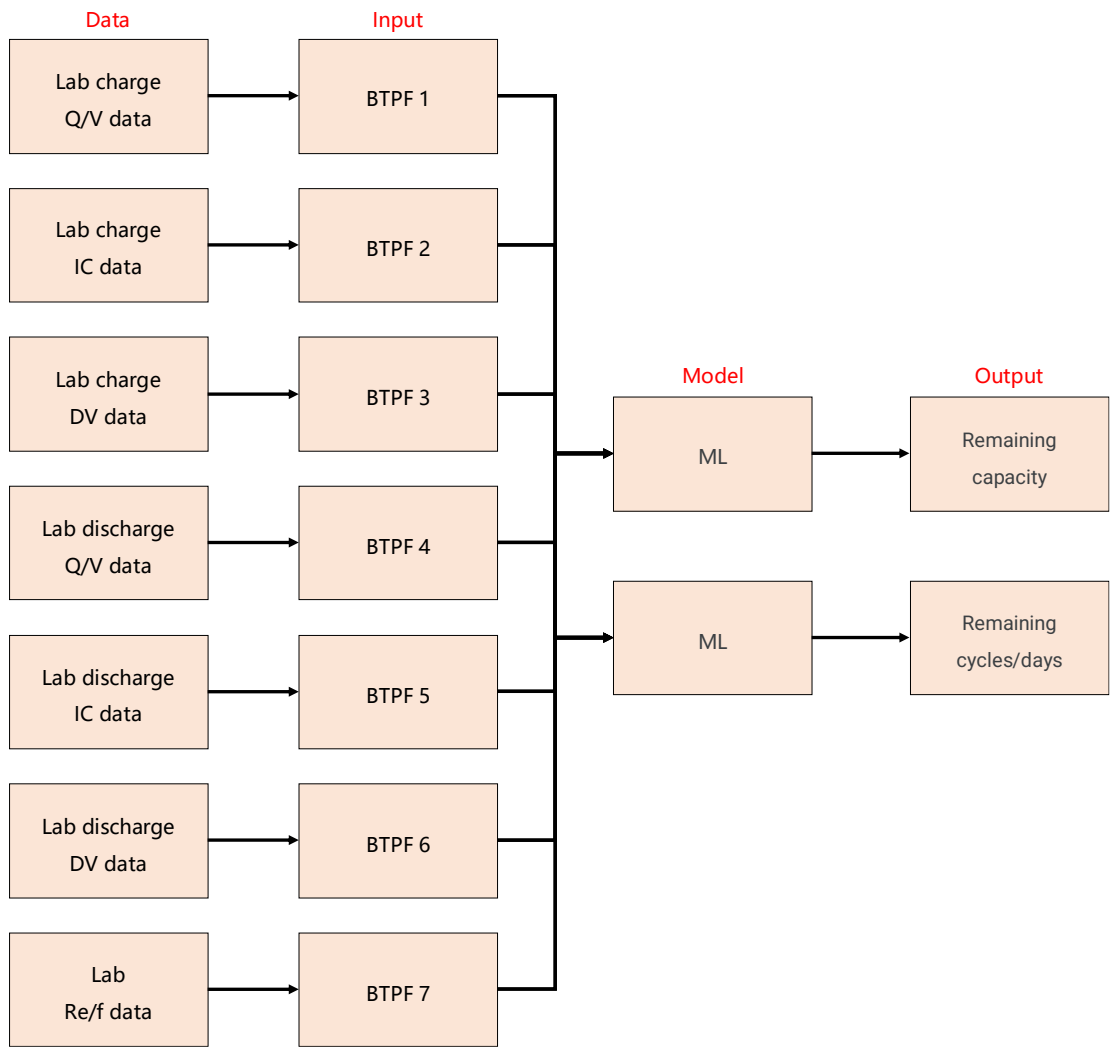

**Supplementary Figure 5** The diagnostic and prognostic methods.

(a) Selection of the BTPF for diagnosis extracted from the laboratory Re/f curve in the mid-high frequency range in the training set with 152 cells. F01 to F16 in axes represent 16 values of frequencies from 2.08 Hz to 1000 Hz. The colors are determined based on the Pearson correlation coefficient values. BTPF are selected from $(16^2-16)/2=120$ candidate two-point features. The maximum Pearson correlation coefficient value corresponding to BTPF is 0.83. BTPF is obtained by subtracting the △Re/f values corresponding to 10 Hz and 312.5 Hz, and taking the absolute value.

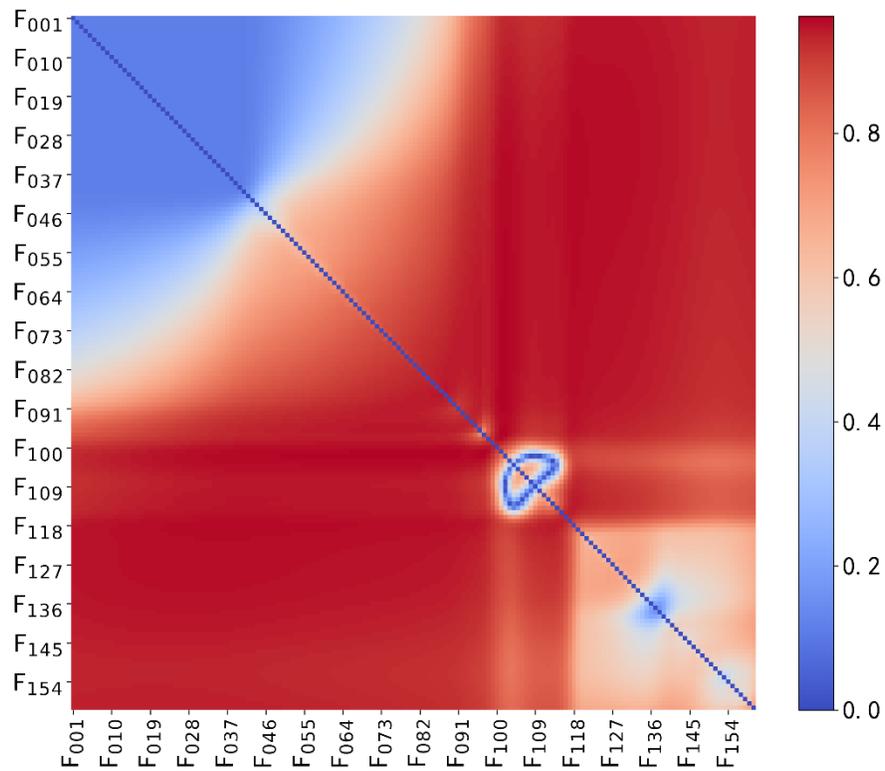

(b) Selection of the BTPF for diagnosis extracted from the laboratory charge Q/V curve in the training set with 152 cells. $F_{001}$ to $F_{160}$ in axes represent 160 values of voltage from 2.5 V to 4.1 V, with the interval of 0.01 V. The colors are determined based on the Pearson correlation coefficient values. BTPF are selected from $(160^2-160)/2=12720$ candidate two-point features. The maximum Pearson correlation coefficient value corresponding to BTPF is 0.96. BTPF is obtained by subtracting the $\triangle Q/V$ values corresponding to 3.49 V and 3.45 V, and taking the absolute value.

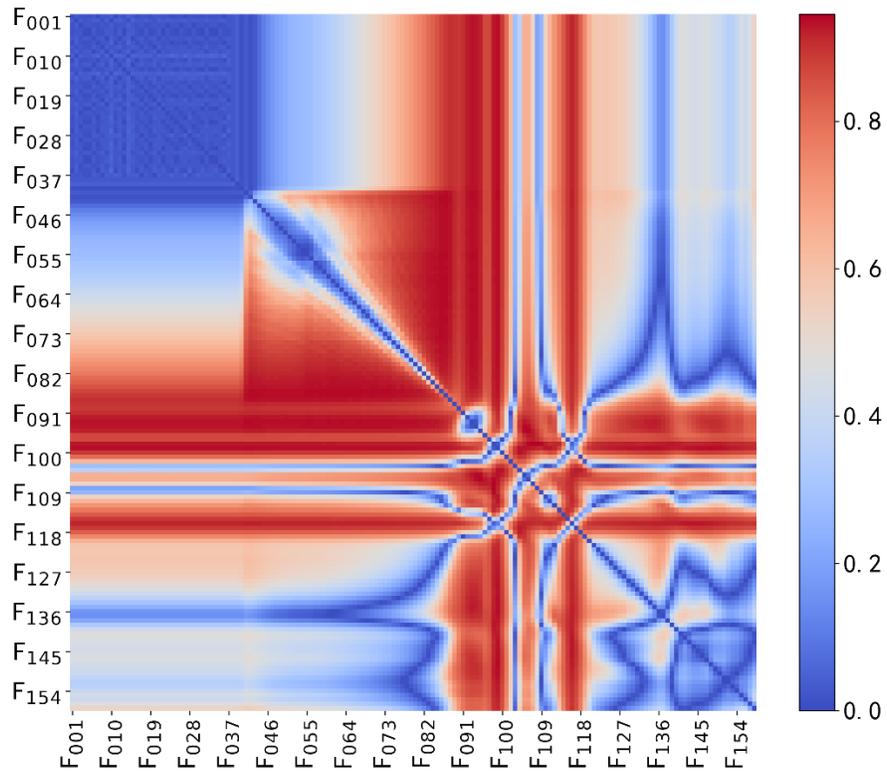

(c) Selection of the BTPF for diagnosis extracted from the laboratory charge IC curve in the training set with 152 cells. F01 to F159 in axes represent 159 values of voltage from 2.51 V to 4.1 V, with the interval of 0.01 V. The colors are determined based on the Pearson correlation coefficient values. BTPF are selected from $(159^2-159)/2=12561$ candidate two-point features. The maximum Pearson correlation coefficient value corresponding to BTPF is 0.95. BTPF is obtained by subtracting the $\triangle dQ/dV$ values corresponding to 3.54 V and 3.43 V, and taking the absolute value.

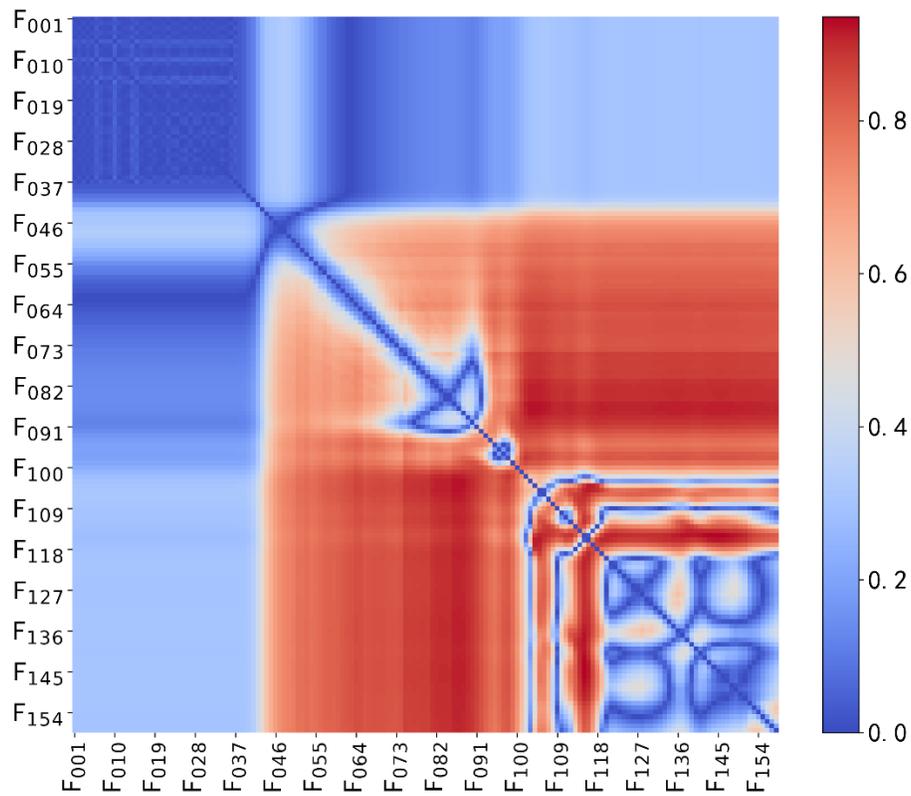

(d) Selection of the BTPF for diagnosis extracted from the laboratory charge DV curve in the training set with 152 cells. F01 to F159 represent 159 values of voltage from 2.51 V to 4.1 V on the charge Q/V curve, with the interval of 0.01 V. The colors are determined based on the Pearson correlation coefficient values. BTPF are selected from $(159^2-159)/2=12561$ candidate two-point features. The maximum Pearson correlation coefficient value corresponding to BTPF is 0.94. BTPF is obtained by subtracting the △dV/dQ values corresponding to 3.93 V and 3.63 V on the charge Q/V curve, and taking the absolute value.

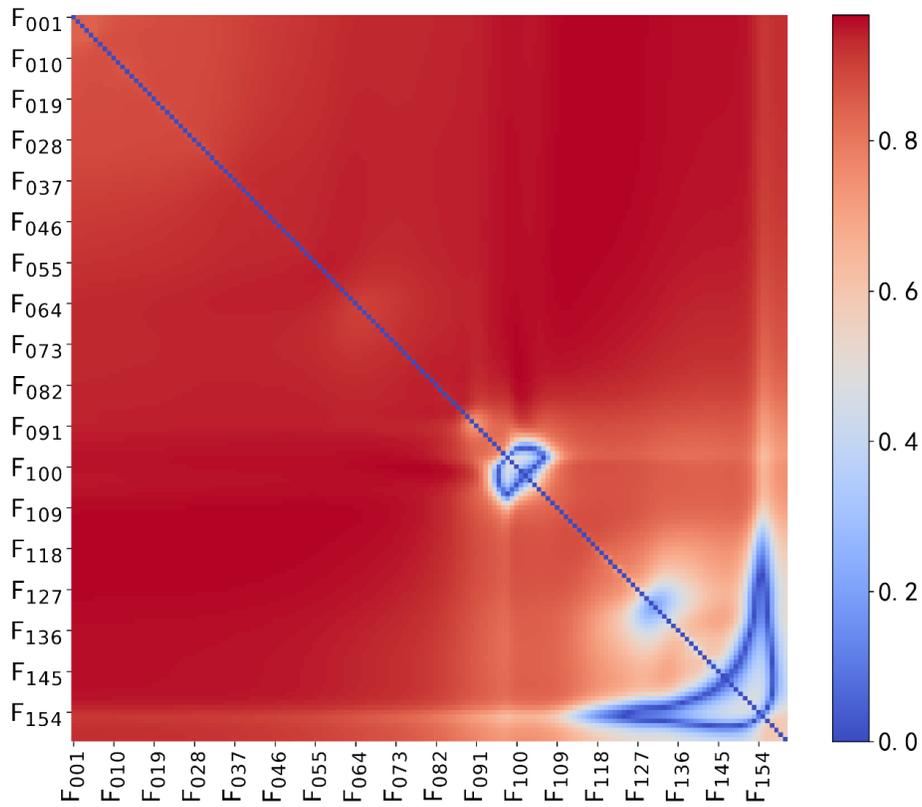

(e) Selection of the BTPF for diagnosis extracted from the laboratory discharge Q/V curve in the training set with 152 cells. $F01$ to $F160$ represent 160 values of voltage from 2.5 V to 4.1 V, with the interval of 0.01 V. The colors are determined based on the Pearson correlation coefficient values. BTPF are selected from $(160^2-160)/2=12720$ candidate two-point features. The maximum Pearson correlation coefficient value corresponding to BTPF is 0.97. BTPF is obtained by subtracting the $\triangle Q/V$ values corresponding to 3.63 V and 2.79 V, and taking the absolute value.

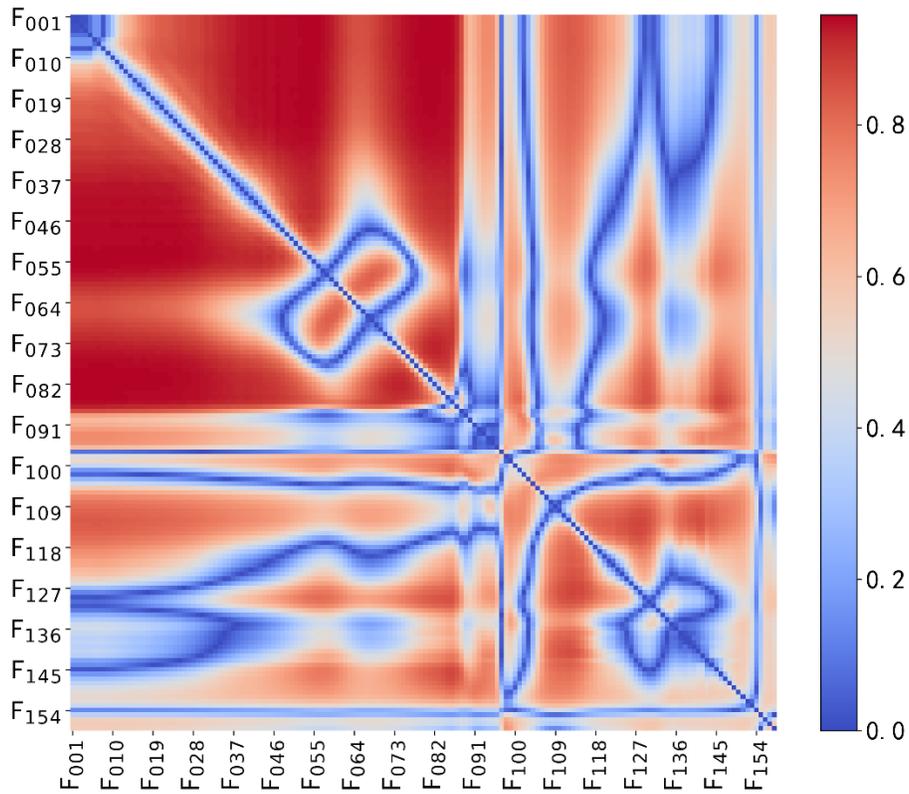

(f) Selection of the BTPF for diagnosis extracted from the laboratory discharge IC curve in the training set with 152 cells. F01 to F159 represent 159 values of voltage from 2.51 V to 4.1 V, with the interval of 0.01 V. The colors are determined based on the Pearson correlation coefficient values. BTPF are selected from $(159^2-159)/2=12561$ candidate two-point features. The maximum Pearson correlation coefficient value corresponding to BTPF is 0.95. BTPF is obtained by subtracting the $\triangle dQ/dV$ values corresponding to 3.29 V and 2.55 V, and taking the absolute value.

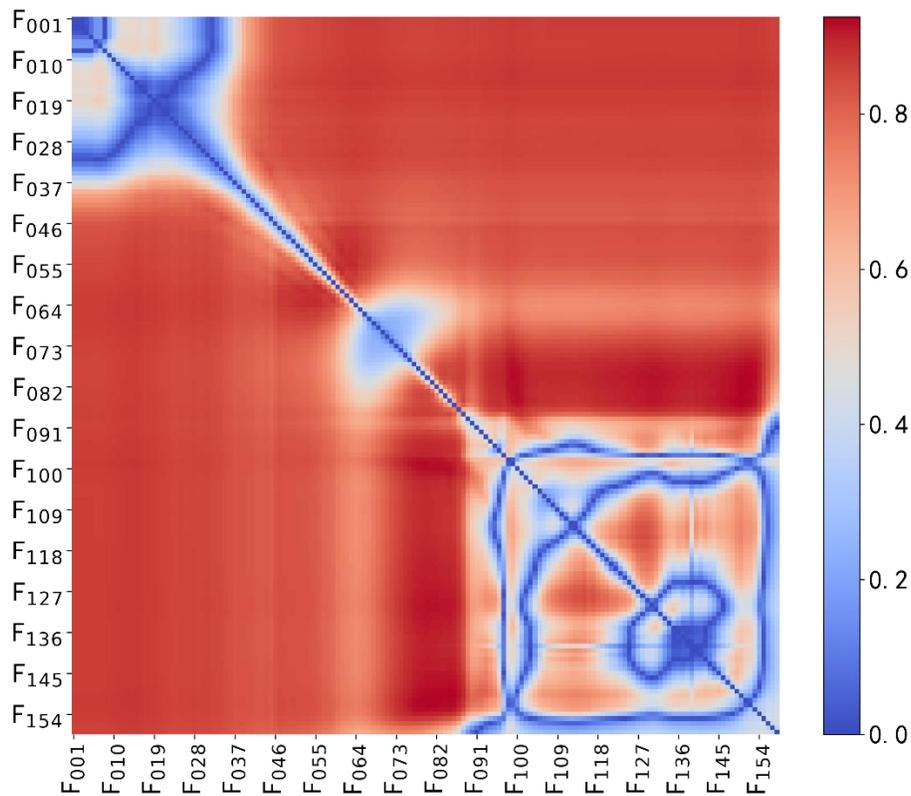

(g) Selection of the BTPF for diagnosis extracted from the laboratory discharge DV data in the training set with 152 cells. F01 to F159 represent 159 values of voltage from 2.51 V to 4.1 V on the discharge Q/V curve, with the interval of 0.01 V. The colors are determined based on the Pearson correlation coefficient values. BTPF are selected from $(159^2-159)/2=12561$ candidate two-point features. The maximum Pearson correlation coefficient value corresponding to BTPF is 0.92. BTPF is obtained by subtracting the △dV/dQ values corresponding to 3.98 V and 3.33 V on the discharge Q/V curve, and taking the absolute value.

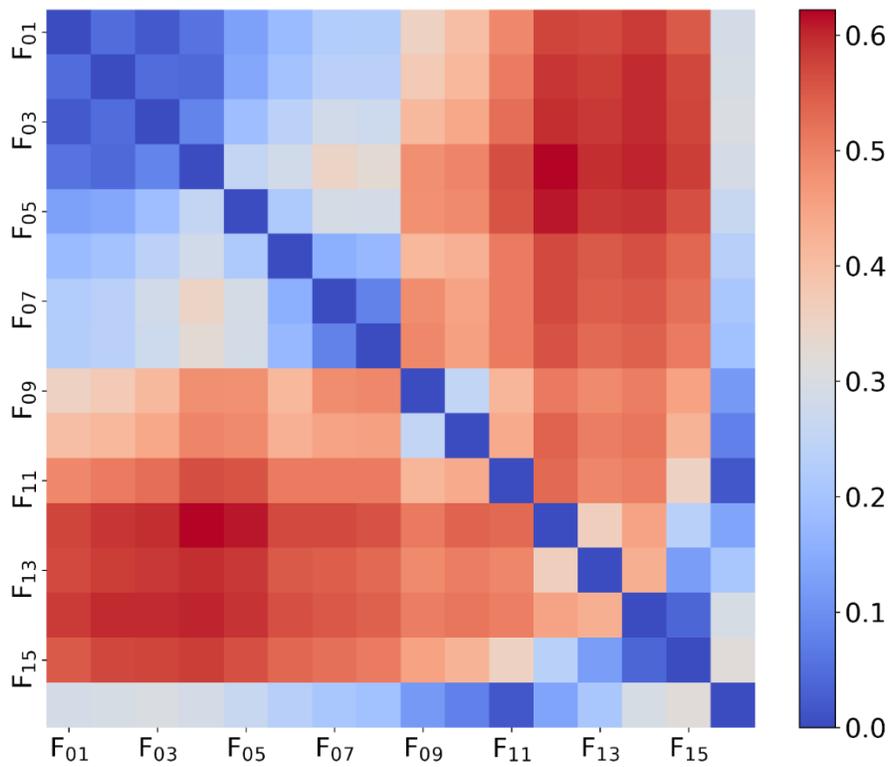

(h) Selection of the BTPF for prognosis extracted from the laboratory Re/f curve in the mid-high frequency range in the training set with 152 cells. F01 to F16 represent 16 values of frequencies from 2.08 Hz to 1000 Hz. The colors are determined based on the Pearson correlation coefficient values. BTPF are selected from $(16^2-16)/2=120$ candidate two-point features. The maximum Pearson correlation coefficient value corresponding to BTPF is 0.63. BTPF is obtained by subtracting the △Re/f values corresponding to 500 Hz and 14.71 Hz, and taking the absolute value.

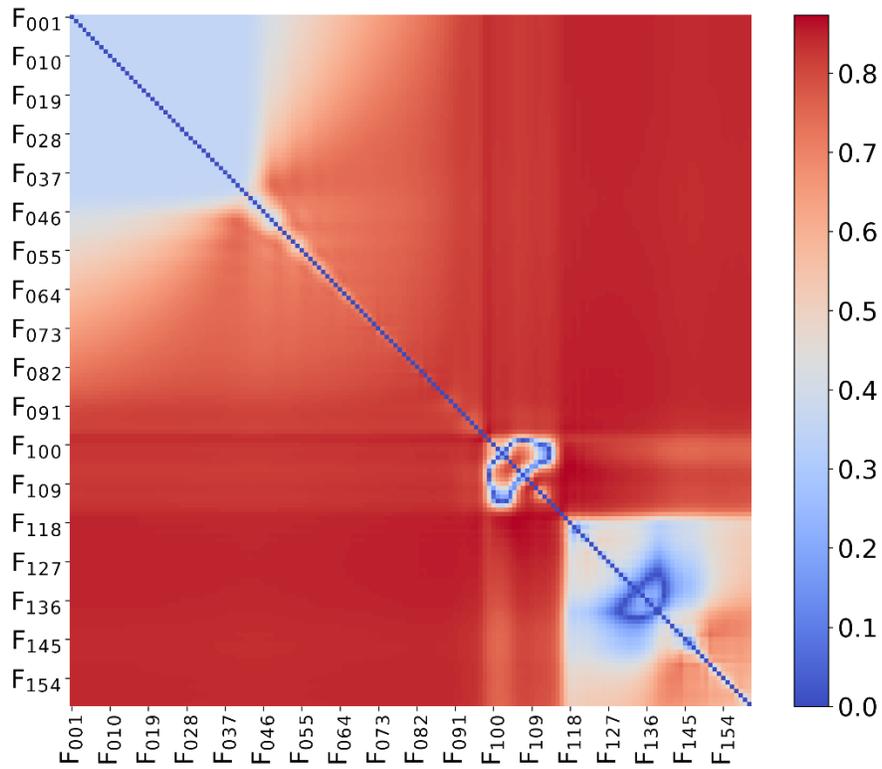

(i) Selection of the BTPF for prognosis extracted from the laboratory charge Q/V curve in the training set with 152 cells. F01 to F160 represent 160 values of voltage from 2.5 V to 4.1 V, with the interval of 0.01 V. The colors are determined based on the Pearson correlation coefficient values. BTPF are selected from (160^2-160)/2=12720 candidate two-point features. The maximum Pearson correlation coefficient value corresponding to BTPF is 0.87. BTPF is obtained by subtracting the △Q/V values corresponding to 3.65 V and 3.54 V, and taking the absolute value.

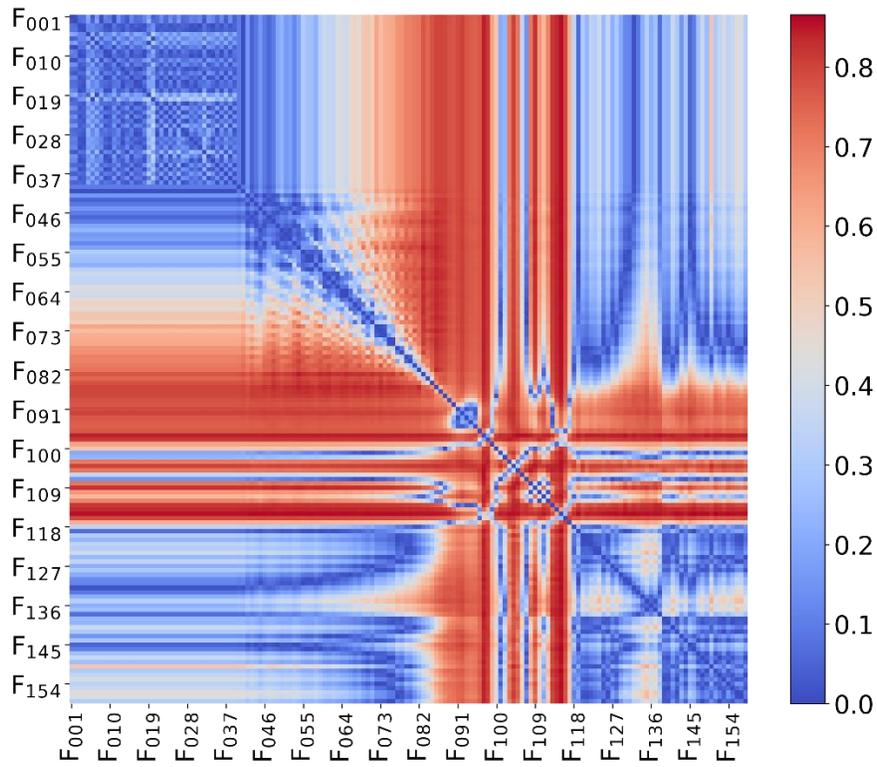

(j) Selection of the BTPF for prognosis extracted from the laboratory charge IC curve in the training set with 152 cells. $F01$ to $F159$ represent 159 values of voltage from 2.51 V to 4.1 V, with the interval of 0.01 V. The colors are determined based on the Pearson correlation coefficient values. BTPF are selected from $(159^2-159)/2=12561$ candidate two-point features. The maximum Pearson correlation coefficient value corresponding to BTPF is 0.87. BTPF is obtained by subtracting the $\triangle dQ/dV$ values corresponding to 3.99 V and 3.63 V, and taking the absolute value.

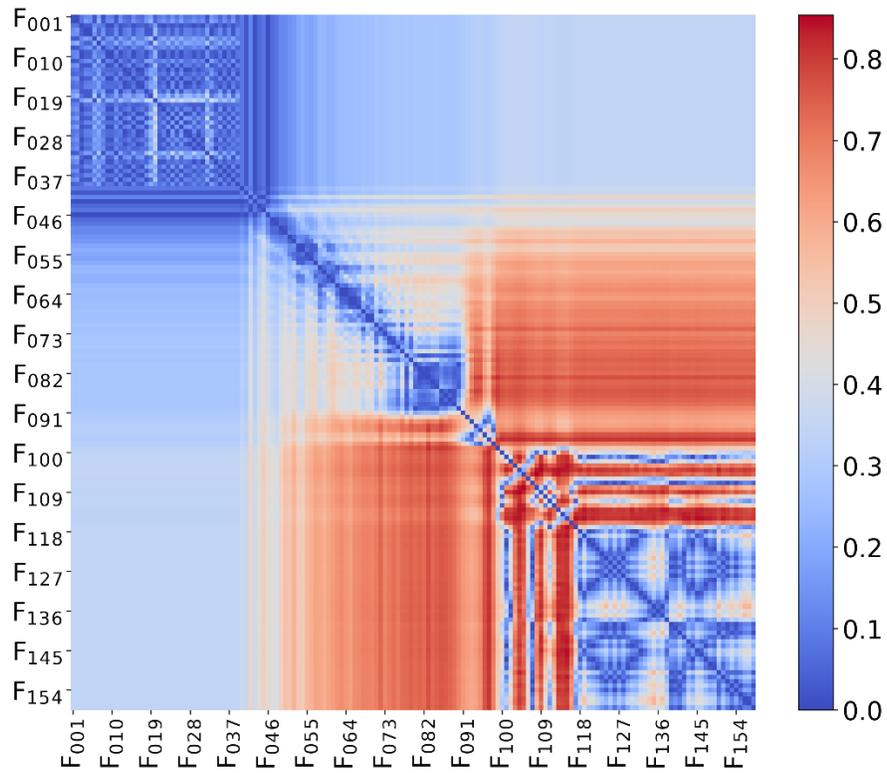

(k) Selection of the BTPF for prognosis extracted from the laboratory charge DV curve in the training set with 152 cells. F01 to F159 represent 159 values of voltage from 2.51 V to 4.1 V on the charge Q/V curve, with the interval of 0.01 V. The colors are determined based on the Pearson correlation coefficient values. BTPF are selected from $(159^2-159)/2=12561$ candidate two-point features. The maximum Pearson correlation coefficient value corresponding to BTPF is 0.85. BTPF is obtained by subtracting the $\triangle dV/dQ$ values corresponding to 3.56 V and 3.53 V on the charge Q/V curve, and taking the absolute value.

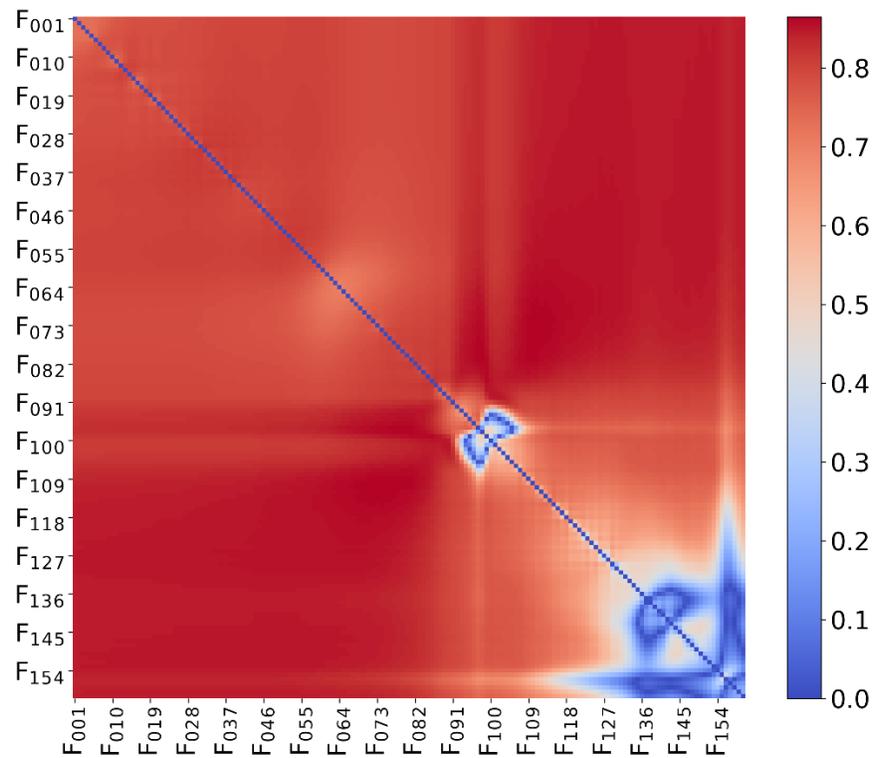

(l) Selection of the BTPF for prognosis extracted from the laboratory discharge Q/V curve in the training set with 152 cells. F01 to F160 represent 160 values of voltage from 2.5 V to 4.1 V, with the interval of 0.01 V. The colors are determined based on the Pearson correlation coefficient values. BTPF are selected from $(160^2-160)/2=12720$ candidate two-point features. The maximum Pearson correlation coefficient value corresponding to BTPF is 0.86. BTPF is obtained by subtracting the $\triangle Q/V$ values corresponding to 3.59 V and 3.23 V, and taking the absolute value.

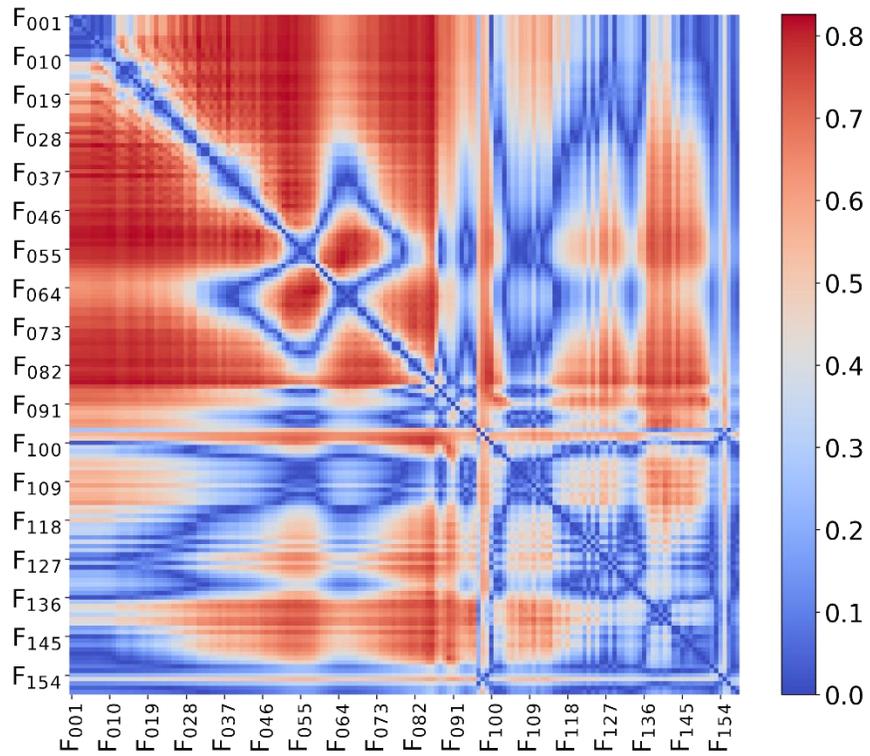

(m) Selection of the BTPF for prognosis extracted from the laboratory discharge IC curve in the training set with 152 cells. F01 to F159 represent 159 values of voltage from 2.51 V to 4.1 V, with the interval of 0.01 V. The colors are determined based on the Pearson correlation coefficient values. BTPF are selected from $(159^2-159)/2=12561$ candidate two-point features. The maximum Pearson correlation coefficient value corresponding to BTPF is 0.83. BTPF is obtained by subtracting the △dQ/dV values corresponding to 2.79 V and 2.56 V, and taking the absolute value.

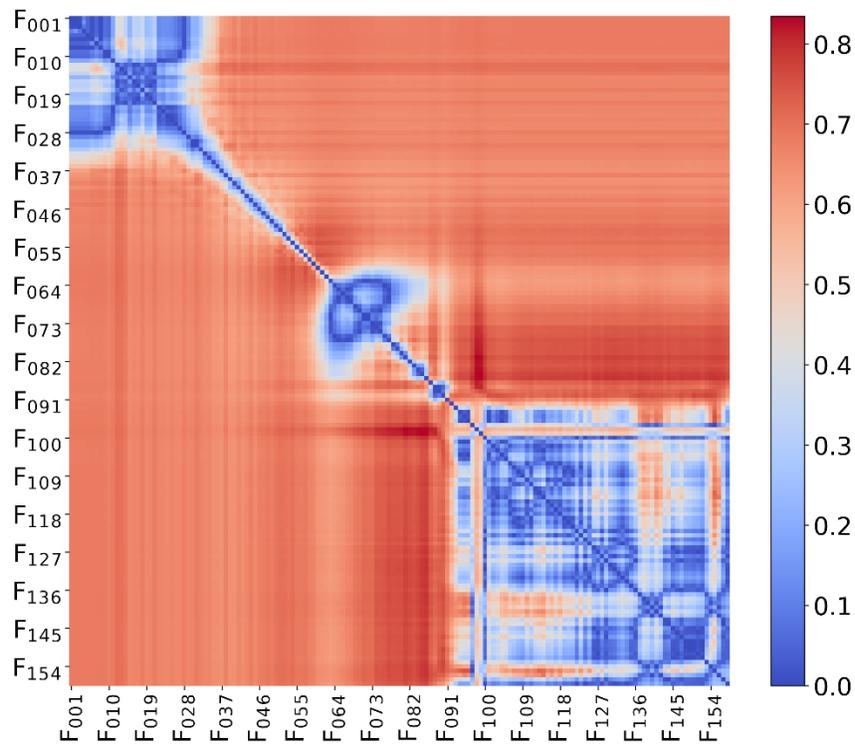

(n) Selection of the BTPF for prognosis extracted from the laboratory discharge DV curve in the training set with 152 cells. F01 to F159 represent 159 values of voltage from 2.51 V to 4.1 V on the discharge Q/V curve, with the interval of 0.01 V. The colors are determined based on the Pearson correlation coefficient values. BTPF are selected from $(159^2-159)/2=12561$ candidate two-point features. The maximum Pearson correlation coefficient value corresponding to BTPF is 0.83. BTPF is obtained by subtracting the $\triangle dV/dQ$ values corresponding to 3.47 V and 3.34 V on the discharge Q/V curve, and taking the absolute value.

**Supplementary Figure 6** Selection of BTPFs for diagnosis and prognosis from laboratory data in the training set of Dataset 1.

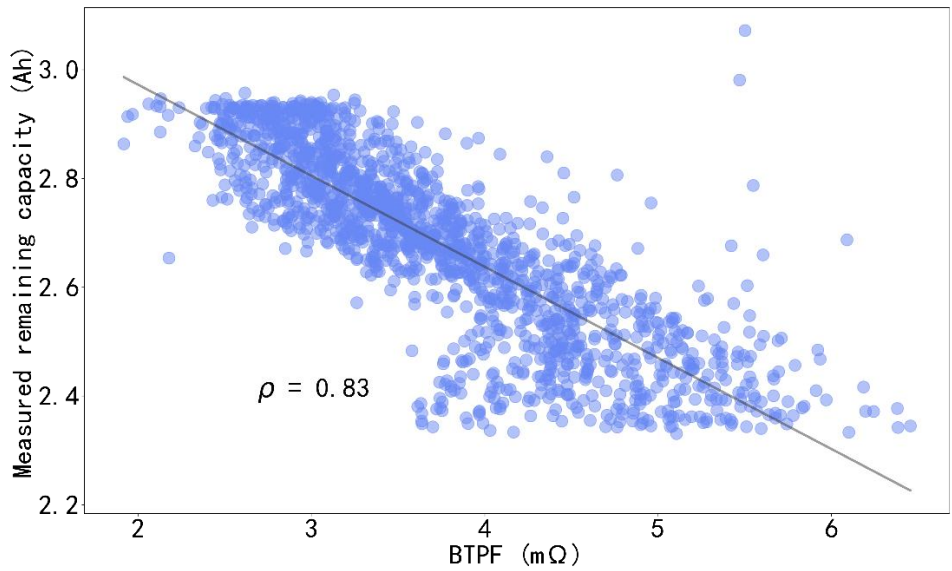

(a) Measured remaining capacity of 152 cells in the training set of Dataset 1 plotted as a function of BTPF in **Supplementary Figure 6 (a)**, with a Pearson correlation coefficient of 0.83.

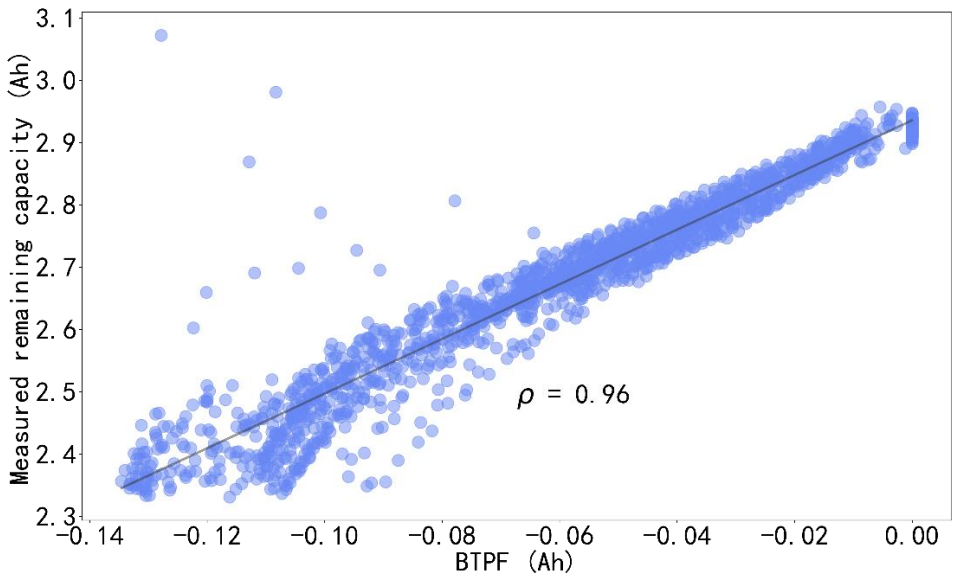

(b) Measured remaining capacity of 152 cells in the training set of Dataset 1 plotted as a function of BTPF in **Supplementary Figure 6 (b)**, with a Pearson correlation coefficient of 0.96.

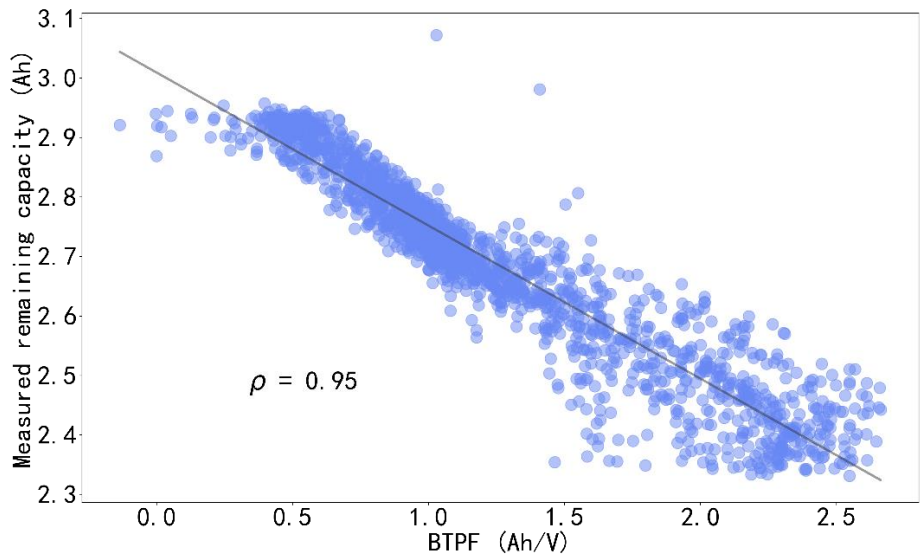

(c) Measured remaining capacity of 152 cells in the training set of Dataset 1 plotted as a function of BTPF in **Supplementary Figure 6 (c)**, with a Pearson correlation coefficient of 0.95.

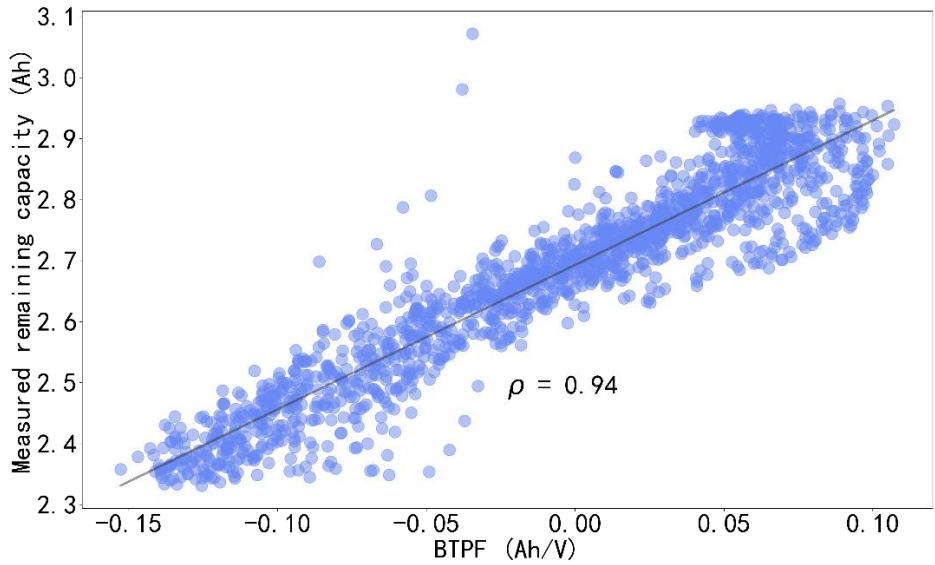

(d) Measured remaining capacity of 152 cells in the training set of Dataset 1 plotted as a function of BTPF in **Supplementary Figure 6 (d)**, with a Pearson correlation coefficient of 0.94.

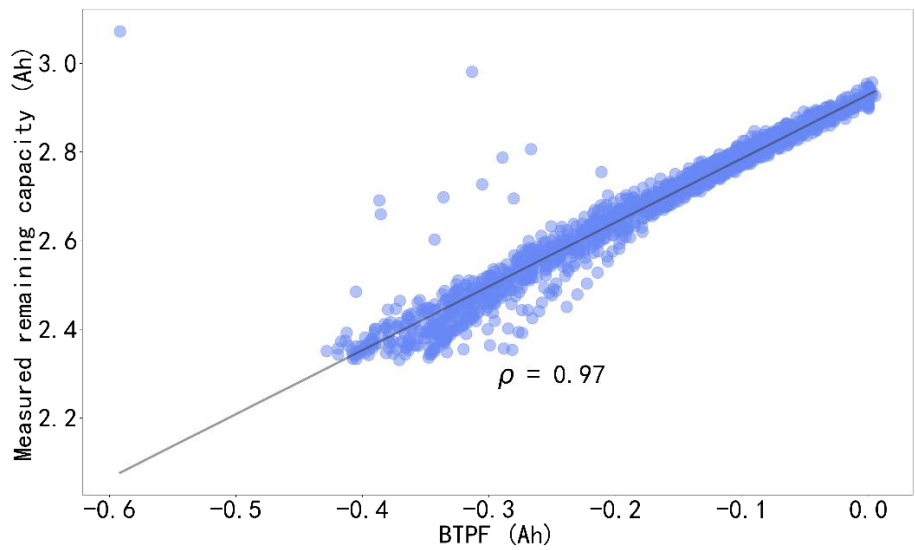

(e) Measured remaining capacity of 152 cells in the training set of Dataset 1 plotted as a function of BTPF in **Supplementary Figure 6 (e)**, with a Pearson correlation coefficient of 0.97.

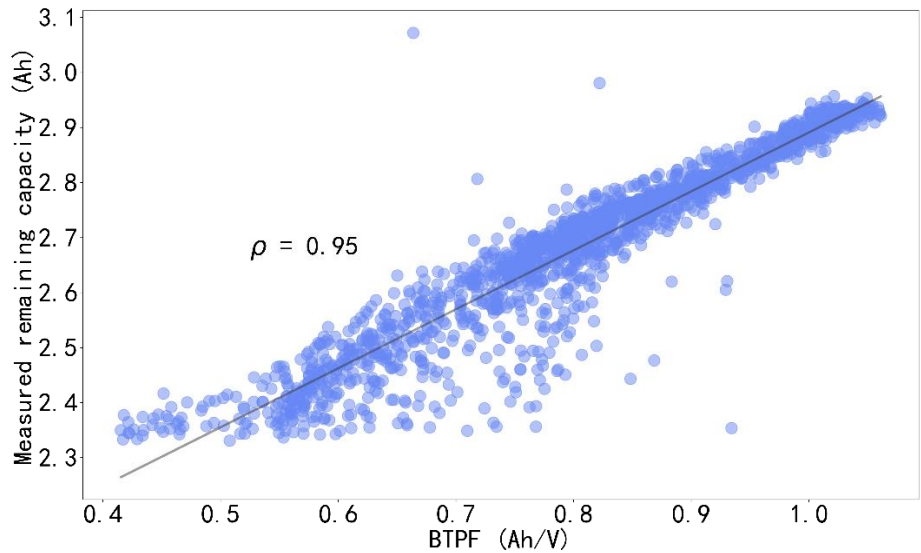

(f) Measured remaining capacity of 152 cells in the training set of Dataset 1 plotted as a function of BTPF in **Supplementary Figure 6 (f)**, with a Pearson correlation coefficient of 0.95.

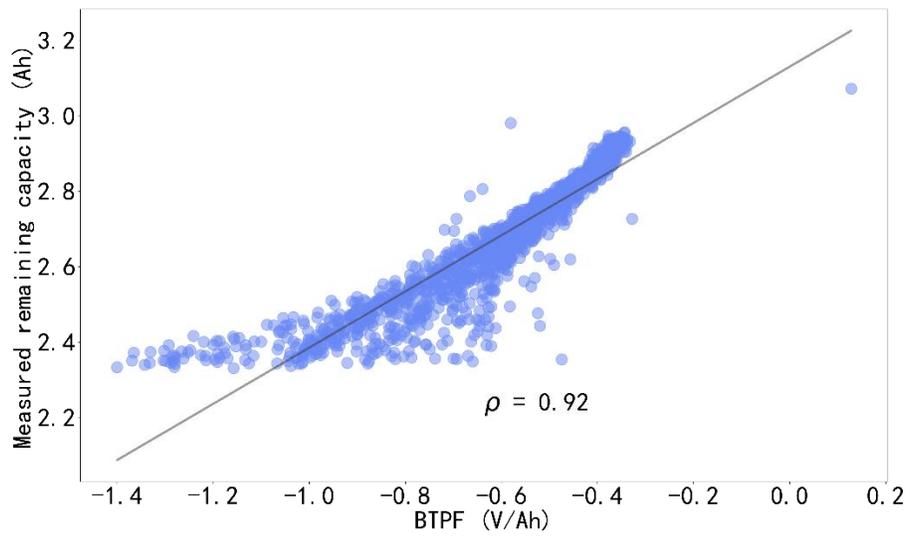

(g) Measured remaining capacity of 152 cells in the training set of Dataset 1 plotted as a function of BTPF in **Supplementary Figure 6 (g)**, with a Pearson correlation coefficient of 0.92.

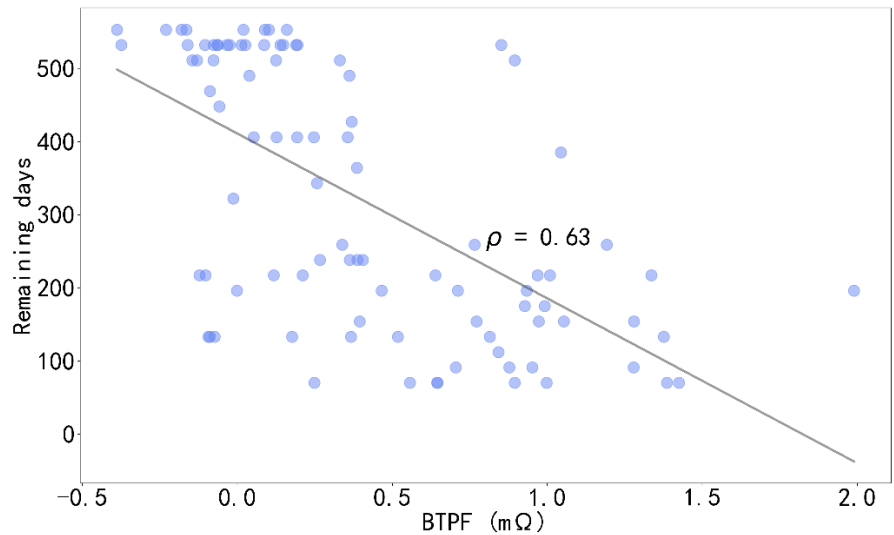

(h) Measured remaining days of 152 cells in the training set of Dataset 1 plotted as a function of BTPF in **Supplementary Figure 6 (h)**, with a Pearson correlation coefficient of 0.63.

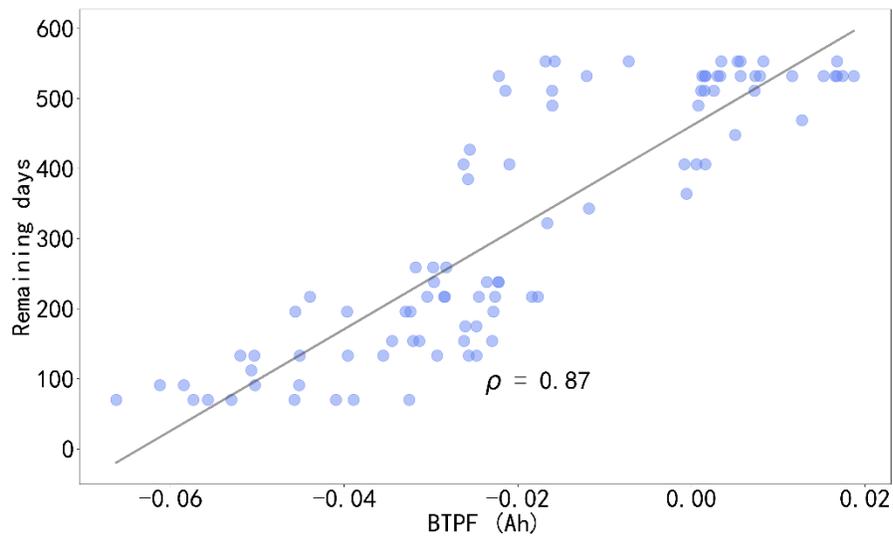

(i) Measured remaining days of 152 cells in the training set of Dataset 1 plotted as a function of BTPF in **Supplementary Figure 6 (i)**, with a Pearson correlation coefficient of 0.87.

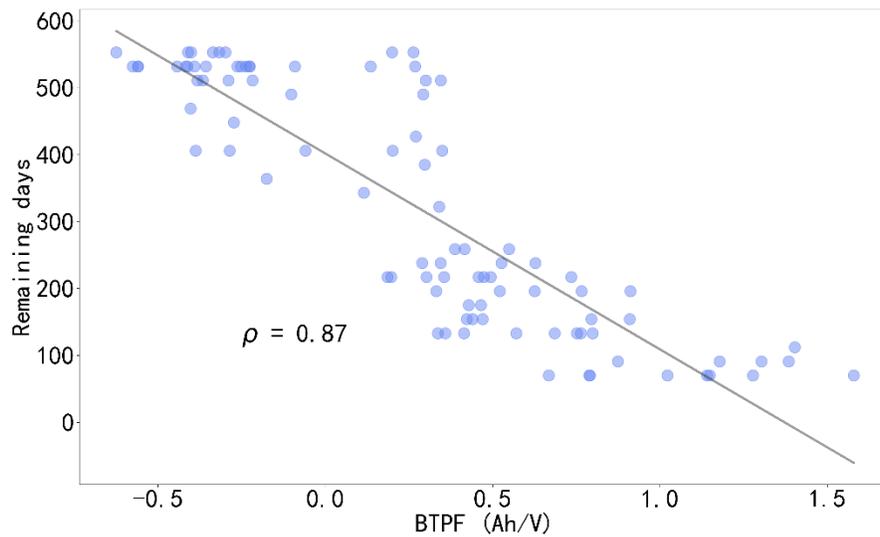

(j) Measured remaining days of 152 cells in the training set of Dataset 1 plotted as a function of BTPF in **Supplementary Figure 6 (j)**, with a Pearson correlation coefficient of 0.87.

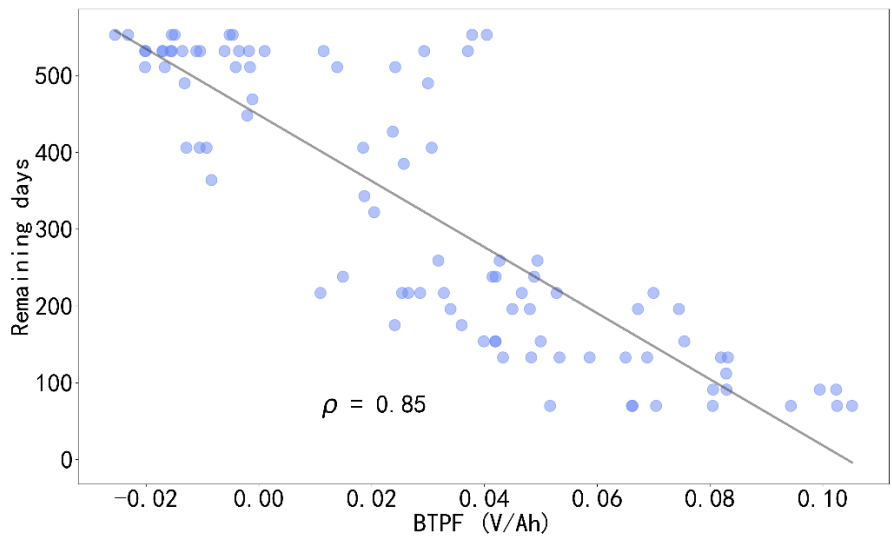

(k) Measured remaining days of 152 cells in the training set of Dataset 1 plotted as a function of BTPF in **Supplementary Figure 6 (k)**, with a Pearson correlation coefficient of 0.85.

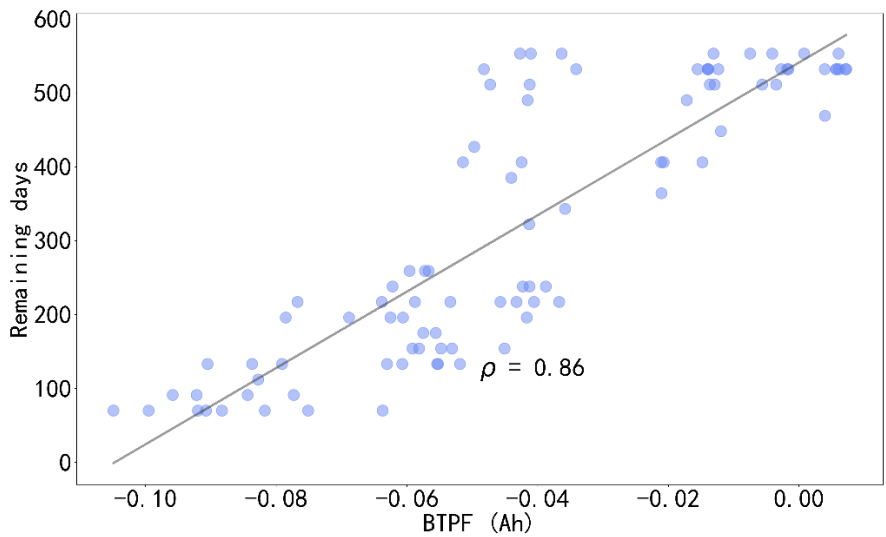

(l) Measured remaining days of 152 cells in the training set of Dataset 1 plotted as a function of BTPF in **Supplementary Figure 6 (l)**, with a Pearson correlation coefficient of 0.86.

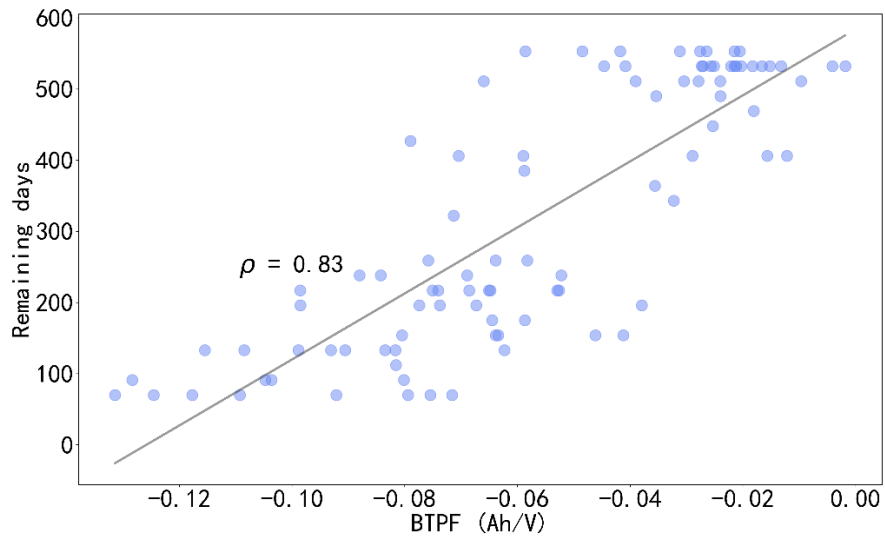

(m) Measured remaining days of 152 cells in the training set of Dataset 1 plotted as a function of BTPF in **Supplementary Figure 6 (m)**, with a Pearson correlation coefficient of 0.83.

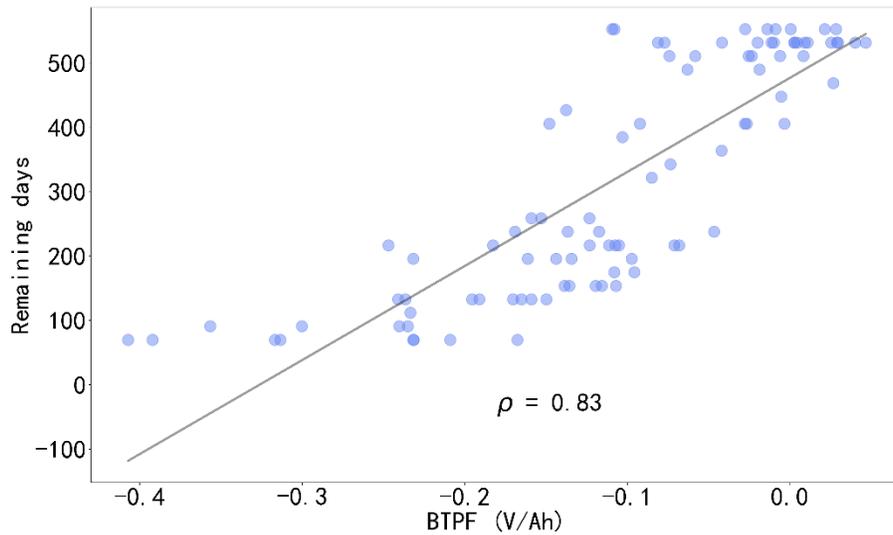

(n) Measured remaining days of 152 cells in the training set of Dataset 1 plotted as a function of BTPF in **Supplementary Figure 6 (n)**, with a Pearson correlation coefficient of 0.83.

**Supplementary Figure 7** Measured remaining capacity or measured life of cells in the training set of Dataset 1 plotted as functions of BTPFs.

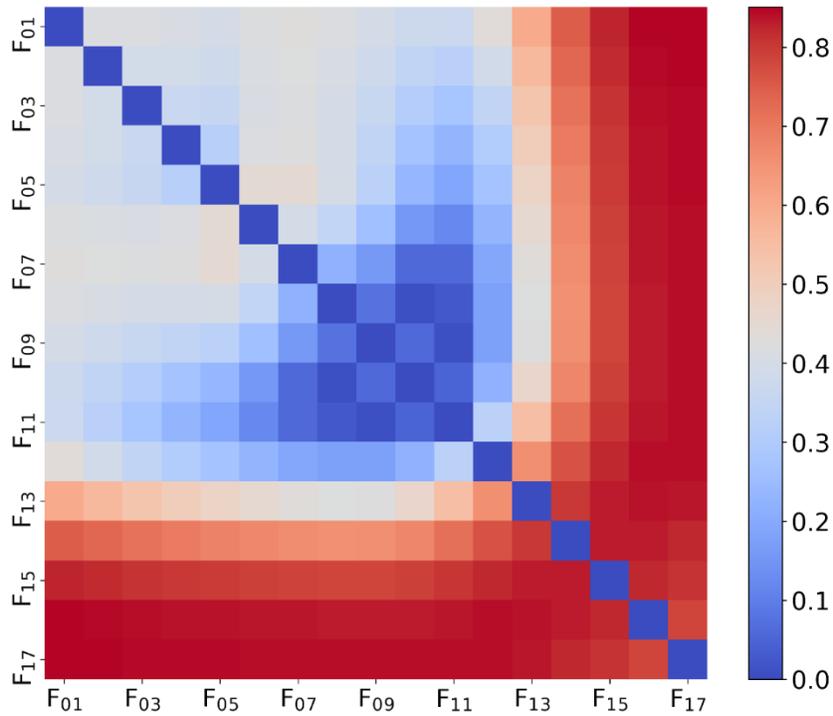

(a) Selection of the BTPF for diagnosis extracted from the laboratory Re/f curve in the mid-high frequency range in the training set with 11 cells. F01 to F17 represent 17 values of frequencies from 1.69 Hz to 936.06 Hz. The colors are determined based on the Pearson correlation coefficient values. BTPF are selected from $(17^2-17)/2=136$ candidate two-point features. The maximum Pearson correlation coefficient value corresponding to BTPF is 0.85. BTPF is obtained by subtracting the △Re/f values corresponding to 936.06 Hz and 1.69 Hz, and taking the absolute value.

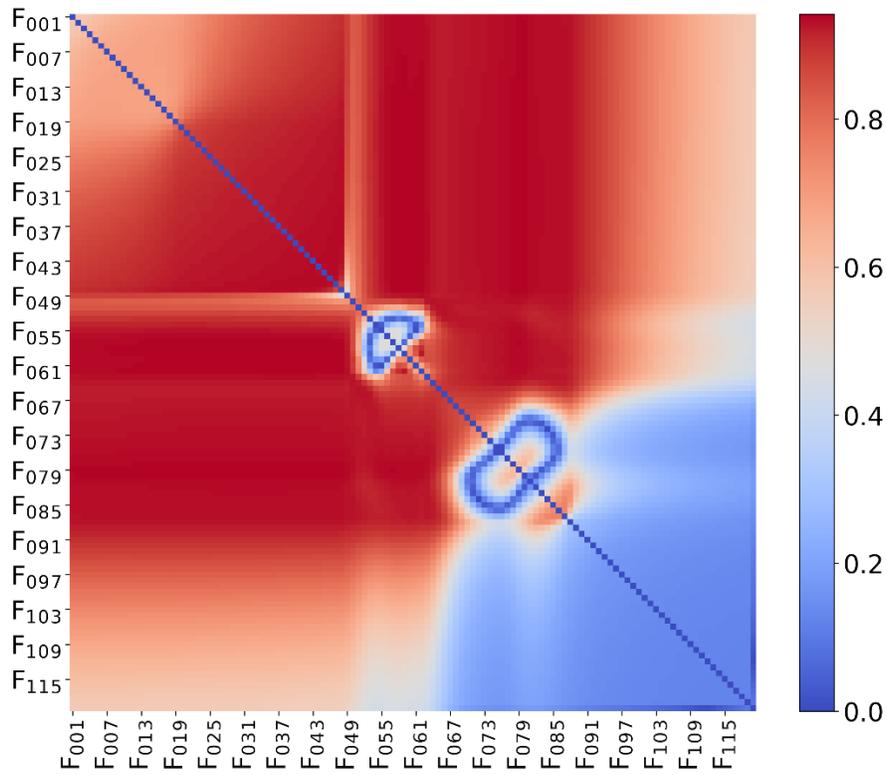

(b) Selection of the BTPF for diagnosis extracted from the laboratory charge Q/V curve in the training set with 11 cells. F01 to F120 represent 120 values of voltage from 3.01 V to 4.2 V, with the interval of 0.01 V. The colors are determined based on the Pearson correlation coefficient values. BTPF are selected from $(120^2-120)/2=7140$ candidate two-point features. The maximum Pearson correlation coefficient value corresponding to BTPF is 0.94. BTPF is obtained by subtracting the $\triangle Q/V$ values corresponding to 3.58 V and 3.42 V, and taking the absolute value.

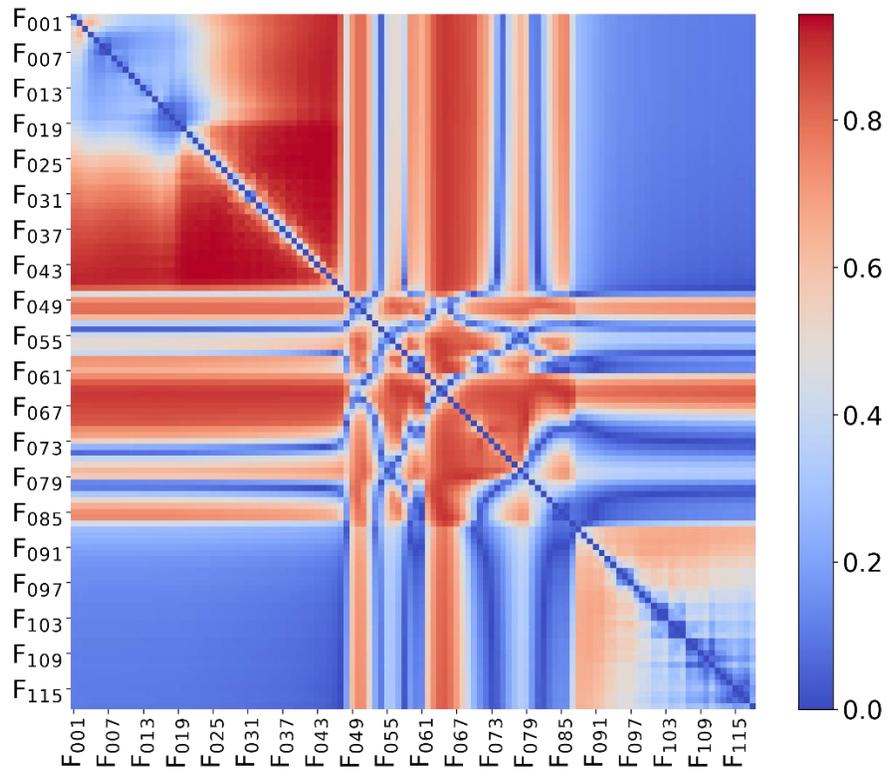

(c) Selection of the BTPF for diagnosis extracted from the laboratory charge IC curve in the training set with 11 cells. F01 to F119 represent 119 values of voltage from 3.02 V to 4.2 V, with the interval of 0.01 V. The colors are determined based on the Pearson correlation coefficient values. BTPF are selected from $(119^2-119)/2=7021$ candidate two-point features. The maximum Pearson correlation coefficient value corresponding to BTPF is 0.94. BTPF is obtained by subtracting the $\triangle dQ/dV$ values corresponding to 3.42 V and 3.22 V, and taking the absolute value.

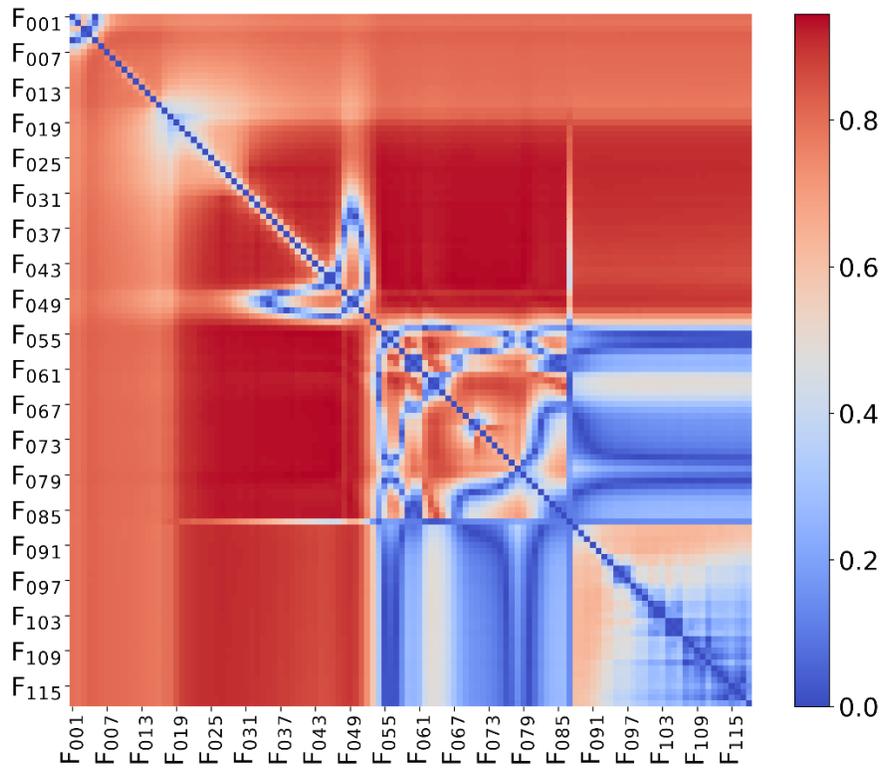

(d) Selection of the BTPF for diagnosis extracted from the laboratory charge DV curve in the training set with 11 cells. F01 to F119 represent 119 values of voltage from 3.02 V to 4.2 V on the charge Q/V curve, with the interval of 0.01 V. The colors are determined based on the Pearson correlation coefficient values. BTPF are selected from $(119^2-119)/2=7021$ candidate two-point features. The maximum Pearson correlation coefficient value corresponding to BTPF is 0.94. BTPF is obtained by subtracting the $\triangle dV/dQ$ values corresponding to 3.54 V and 3.45 V on the charge Q/V curve, and taking the absolute value.

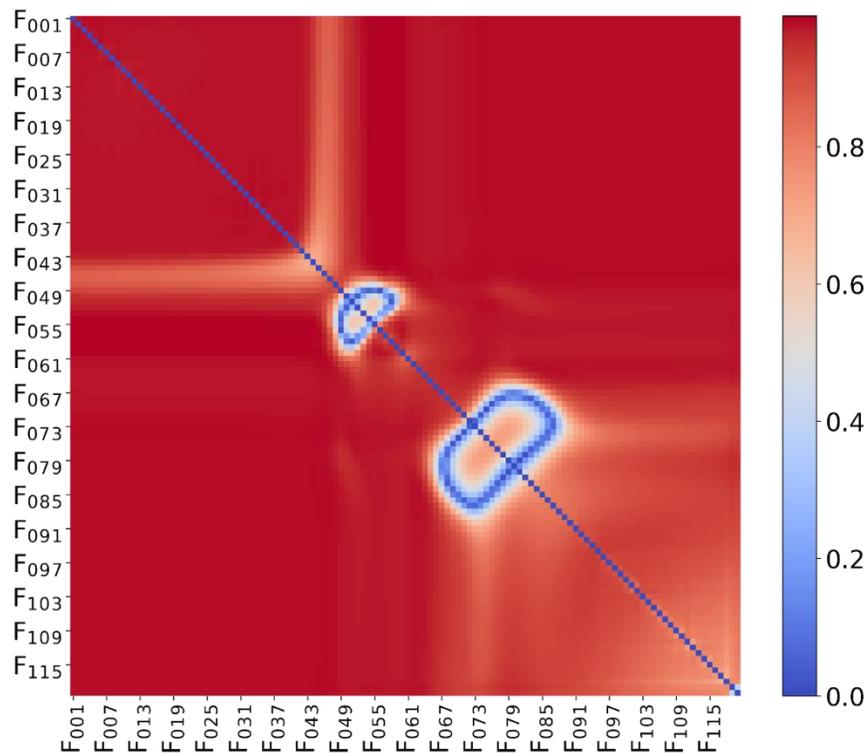

(e) Selection of the BTPF for diagnosis extracted from the laboratory discharge Q/V curve in the training set with 11 cells. F01 to F120 represent 120 values of voltage from 3.01 V to 4.2 V, with the interval of 0.01 V. The colors are determined based on the Pearson correlation coefficient values. BTPF are selected from $(120^2-120)/2=7140$ candidate two-point features. The maximum Pearson correlation coefficient value corresponding to BTPF is 0.99. BTPF is obtained by subtracting the $\triangle Q/V$ values corresponding to 3.55 V and 3.43 V, and taking the absolute value.

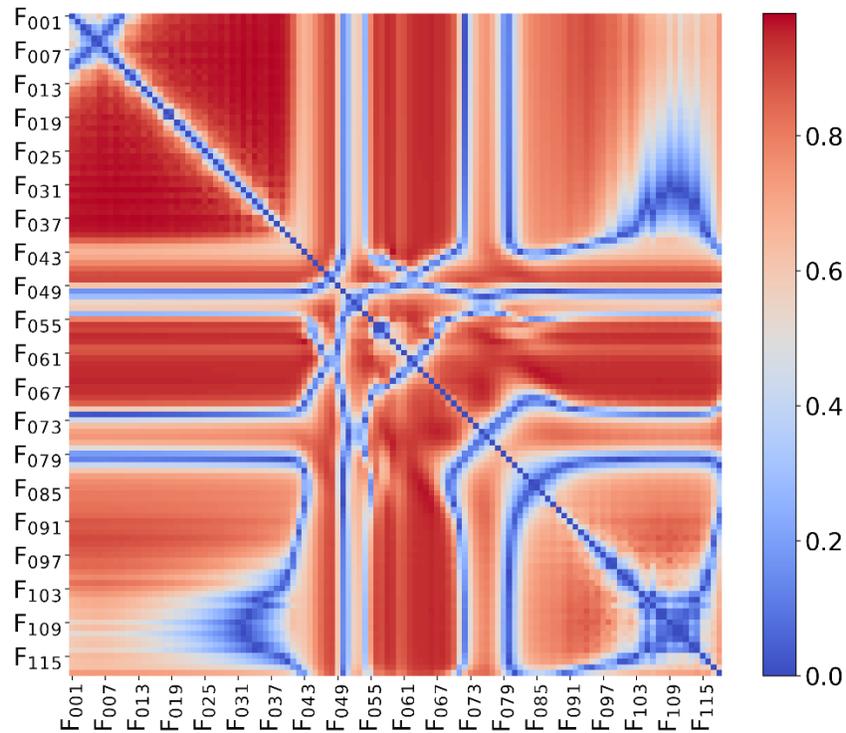

(f) Selection of the BTPF for diagnosis extracted from the laboratory discharge IC curve in the training set with 11 cells. F01 to F119 represent 119 values of voltage from 3.02 V to 4.2 V, with the interval of 0.01 V. The colors are determined based on the Pearson correlation coefficient values. BTPF are selected from (119^2-119)/2=7021 candidate two-point features. The maximum Pearson correlation coefficient value corresponding to BTPF is 0.98. BTPF is obtained by subtracting the △dQ/dV values corresponding to 3.31 V and 3.08 V, and taking the absolute value.

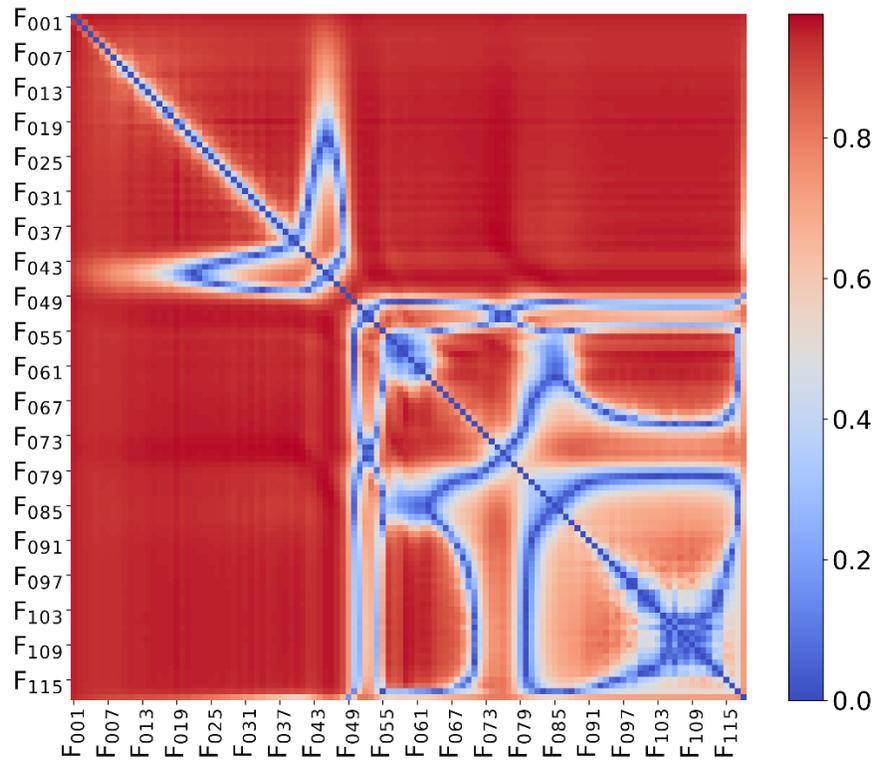

(g) Selection of the BTPF for diagnosis extracted from the laboratory discharge DV curve in the training set with 11 cells. F01 to F119 represent 119 values of voltage from 3.02 V to 4.2 V on the discharge Q/V curve, with the interval of 0.01 V. The colors are determined based on the Pearson correlation coefficient values. BTPF are selected from $(119^2-119)/2=7021$ candidate two-point features. The maximum Pearson correlation coefficient value corresponding to BTPF is 0.98. BTPF is obtained by subtracting the $\triangle dV/dQ$ values corresponding to 3.75 V and 3.39 V on the discharge Q/V curve, and taking the absolute value.

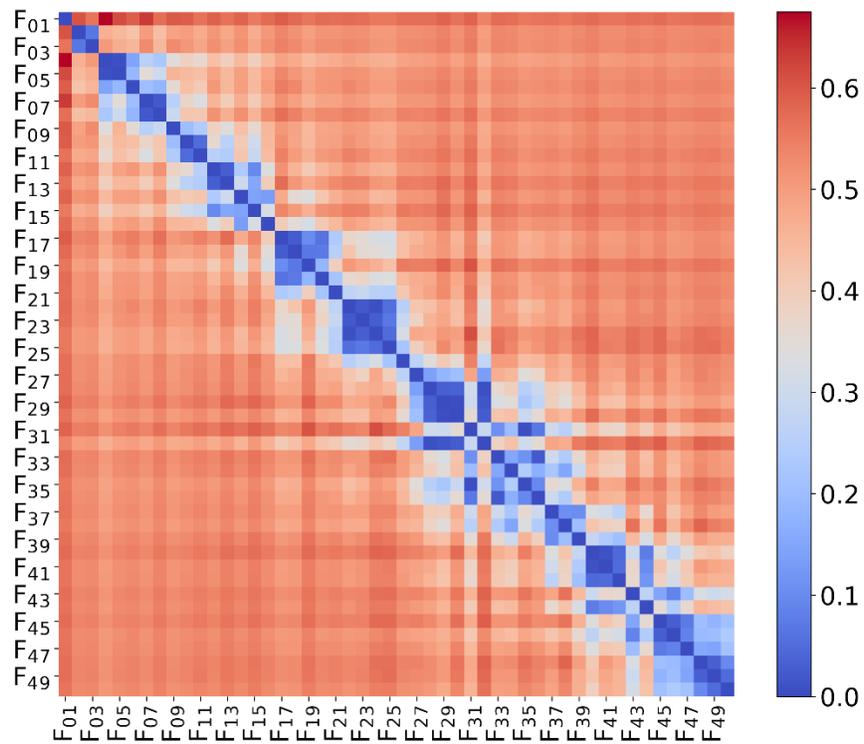

(h) Selection of the BTPF for diagnosis extracted from the laboratory relaxation V/t curve after full charge in the training set with 11 cells. F01 to F50 represent 50 values of times from 0 s to 3000 s, with the interval of 60 s. The colors are determined based on the Pearson correlation coefficient values. BTPF are selected from (50^2-50)/2=1225 candidate two-point features. The maximum Pearson correlation coefficient value corresponding to BTPF is 0.67. BTPF is obtained by subtracting the △V/t values corresponding to 180 s and 0 s, and taking the absolute value.

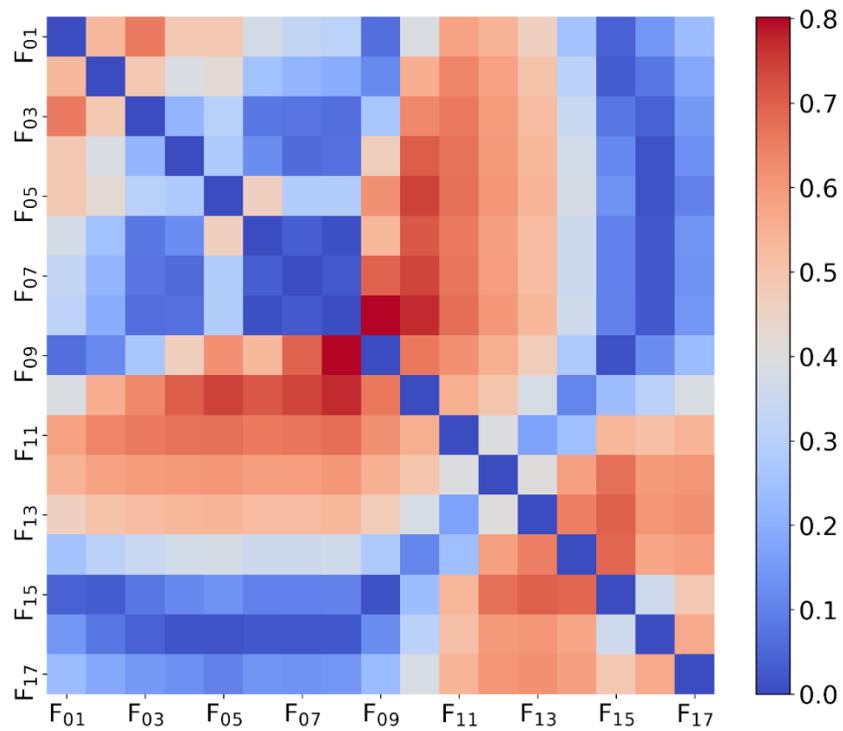

(i) Selection of the BTPF for prognosis extracted from the laboratory Re/f curve in the mid-high frequency range in the training set with 11 cells. F01 to F17 represent 17 values of frequencies from 1.69 Hz to 936.06 Hz. The colors are determined based on the Pearson correlation coefficient values. BTPF are selected from $(17^2-17)/2=136$ candidate two-point features. The maximum Pearson correlation coefficient value corresponding to BTPF is 0.80. BTPF is obtained by subtracting the △Re/f values corresponding to 59.06 Hz and 39.80 Hz, and taking the absolute value.

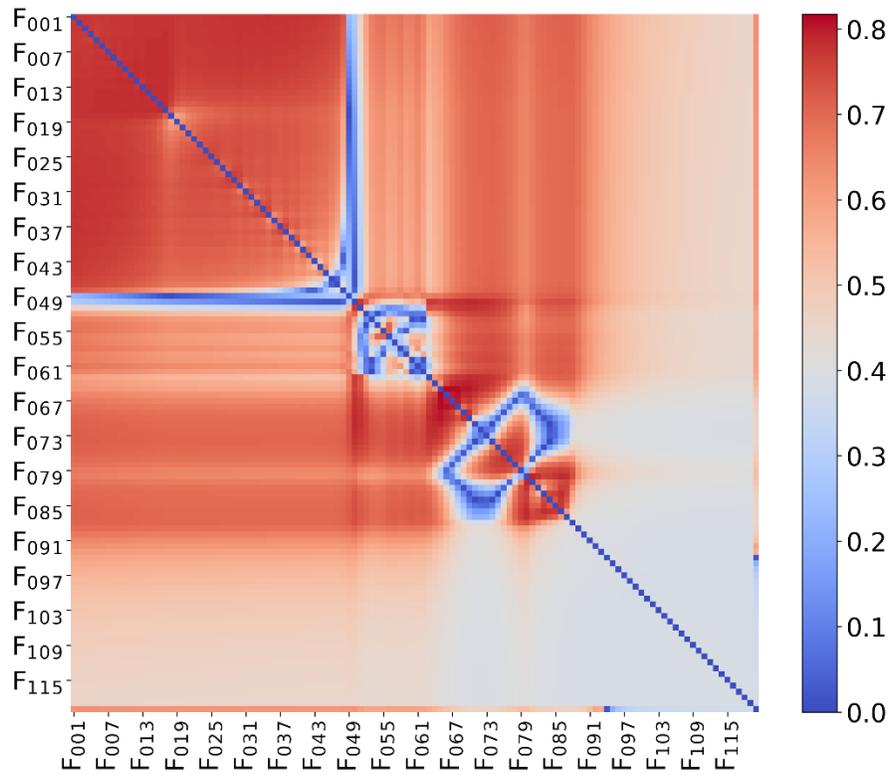

(j) Selection of the BTPF for prognosis extracted from the laboratory charge Q/V curve in the training set with 11 cells. F01 to F120 represent 120 values of voltage from 3.01 V to 4.2 V, with the interval of 0.01 V. The colors are determined based on the Pearson correlation coefficient values. BTPF are selected from $(120^2-120)/2=7140$ candidate two-point features. The maximum Pearson correlation coefficient value corresponding to BTPF is 0.82. BTPF is obtained by subtracting the △Q/V values corresponding to 3.65 V and 3.64 V, and taking the absolute value.

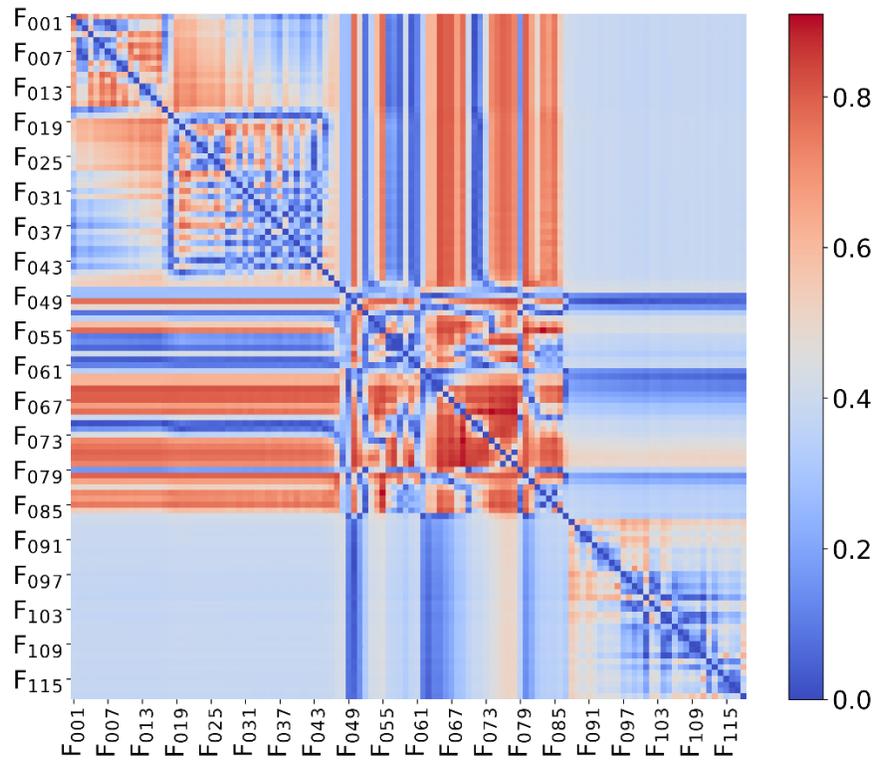

(k) Selection of the BTPF for prognosis extracted from the laboratory charge IC curve in the training set with 11 cells. F01 to F119 represent 119 values of voltage from 3.02 V to 4.2 V, with the interval of 0.01 V. The colors are determined based on the Pearson correlation coefficient values. BTPF are selected from $(119^2-119)/2=7021$ candidate two-point features. The maximum Pearson correlation coefficient value corresponding to BTPF is 0.91. BTPF is obtained by subtracting the $\triangle dQ/dV$ values corresponding to 3.82 V and 3.54 V, and taking the absolute value.

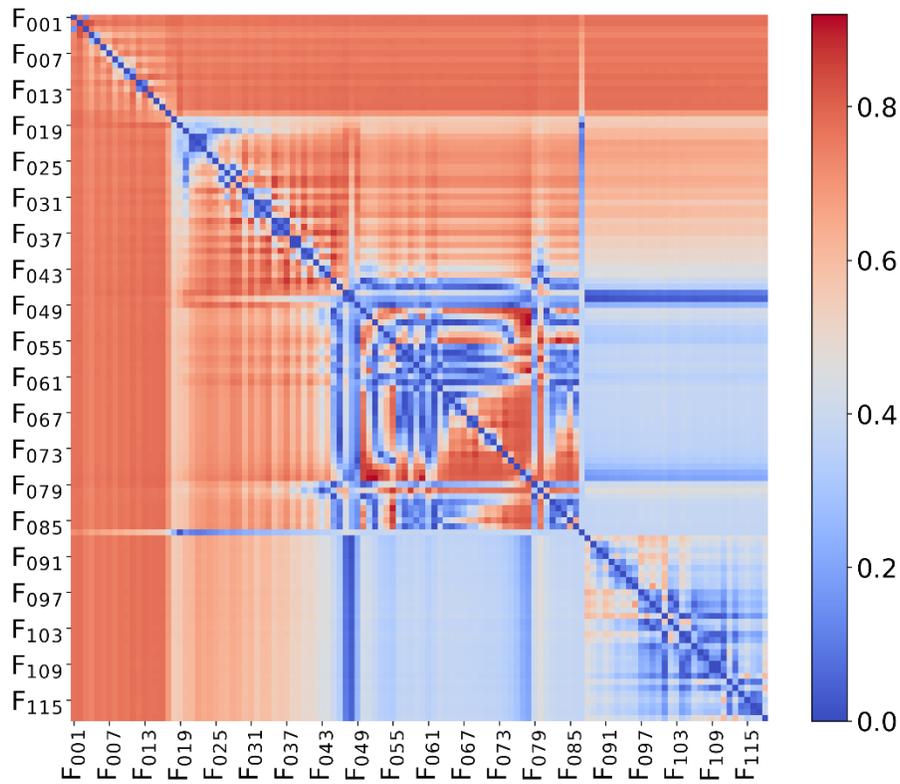

(l) Selection of the BTPF for prognosis extracted from the laboratory charge DV curve in the training set with 11 cells. F01 to F119 represent 119 values of voltage from 3.02 V to 4.2 V on the charge Q/V curve, with the interval of 0.01 V. The colors are determined based on the Pearson correlation coefficient values. BTPF are selected from $(119^2-119)/2=7021$ candidate two-point features. The maximum Pearson correlation coefficient value corresponding to BTPF is 0.92. BTPF is obtained by subtracting the $\triangle dV/dQ$ values corresponding to 3.77 V and 3.51 V on the charge Q/V curve, and taking the absolute value.

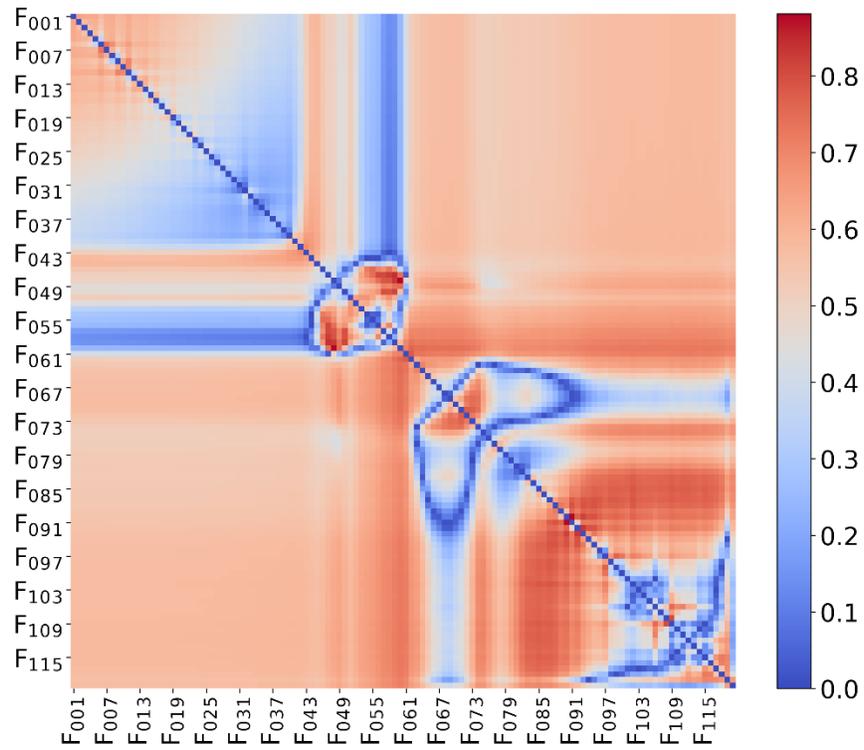

(m) Selection of the BTPF for prognosis extracted from the laboratory discharge Q/V curve in the training set with 11 cells. F01 to F120 represent 120 values of voltage from 3.01 V to 4.2 V, with the interval of 0.01 V. The colors are determined based on the Pearson correlation coefficient values. BTPF are selected from $(120^2-120)/2=7140$ candidate two-point features. The maximum Pearson correlation coefficient value corresponding to BTPF is 0.88. BTPF is obtained by subtracting the $\triangle Q/V$ values corresponding to 3.59 V and 3.47 V, and taking the absolute value.

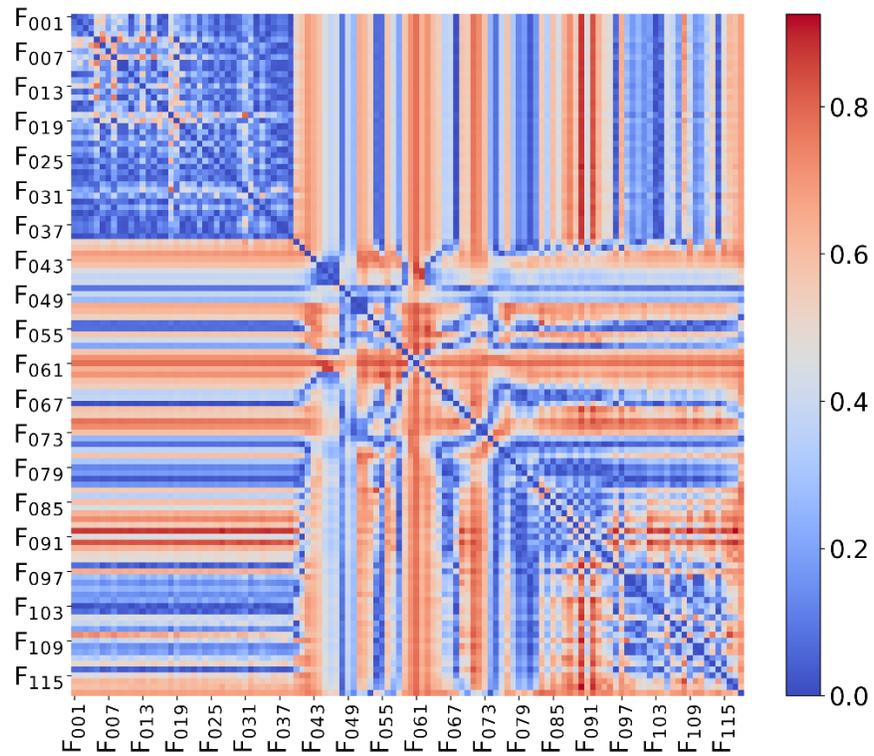

(n) Selection of the BTPF for prognosis extracted from the laboratory discharge IC curve in the training set with 11 cells. F01 to F119 represent 119 values of voltage from 3.02 V to 4.2 V. The colors are determined based on the Pearson correlation coefficient values. BTPF are selected from $(119^2-119)/2=7021$ candidate two-point features. The maximum Pearson correlation coefficient value corresponding to BTPF is 0.93. BTPF is obtained by subtracting the △Q/U values corresponding to 4.16V and 3.89V, and taking the absolute value.

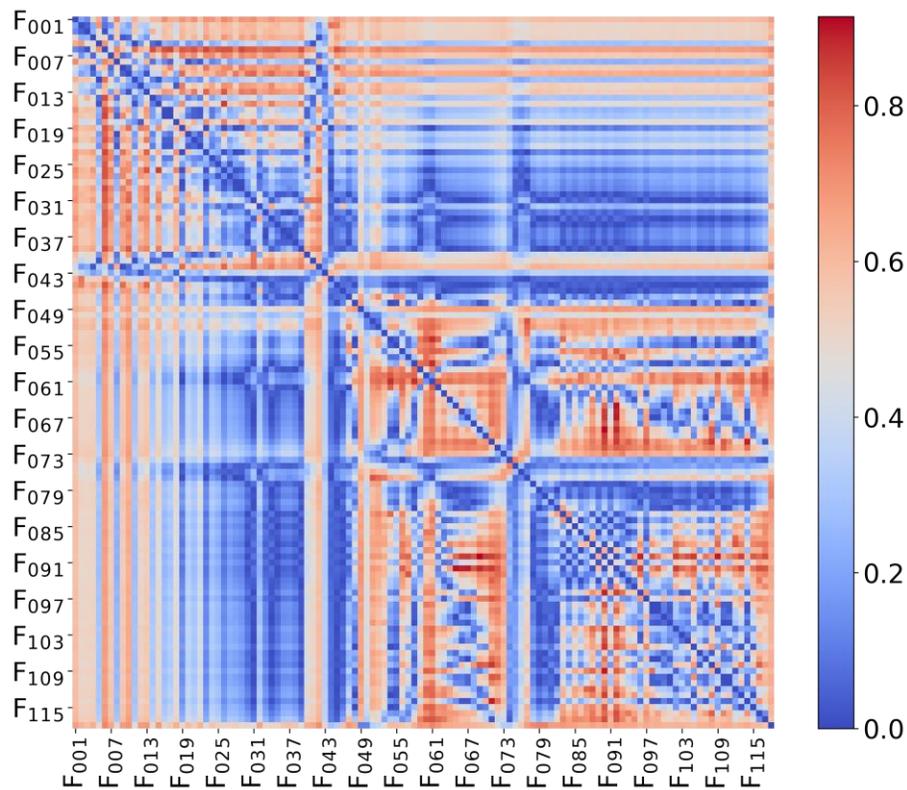

(o) Selection of the BTPF for prognosis extracted from the laboratory discharge DV curve in the training set with 11 cells. F01 to F119 represent 119 values of voltage from 3.02 V to 4.2 V on the discharge Q/V curve, with the interval of 0.01 V. The colors are determined based on the Pearson correlation coefficient values. BTPF are selected from $(119^2-119)/2=7021$ candidate two-point features. The maximum Pearson correlation coefficient value corresponding to BTPF is 0.91. BTPF is obtained by subtracting the $\triangle dV/dQ$ values corresponding to 3.89 V and 3.68 V on the discharge Q/V curve, and taking the absolute value.

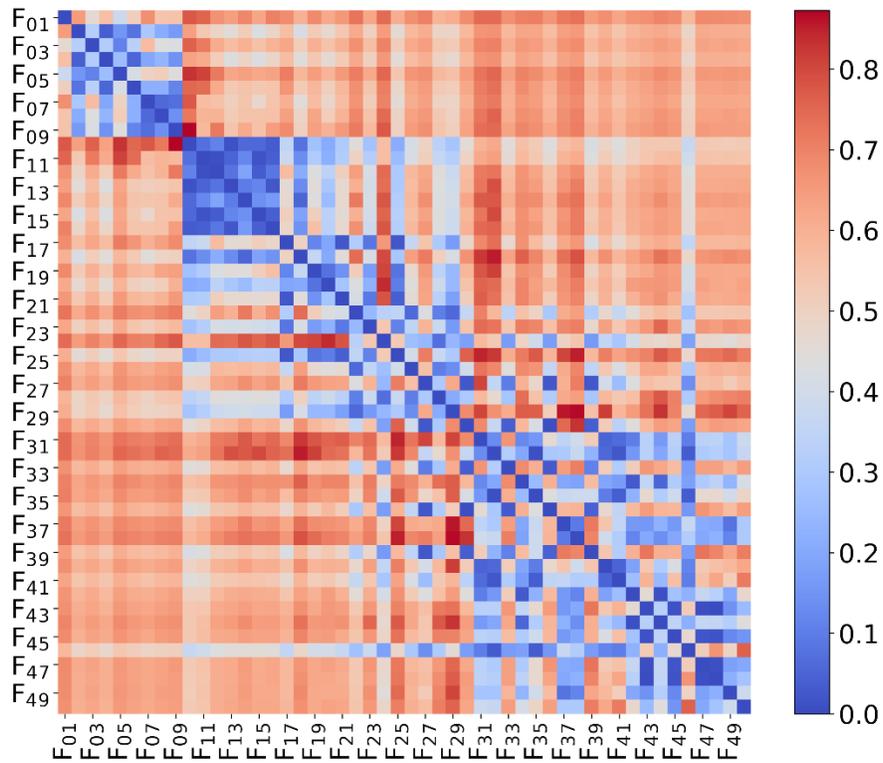

(p) Selection of the BTPF for prognosis extracted from the laboratory relaxation V/t curve after full charge in the training set with 11 cells. F01 to F50 represent 50 values of times from 0s to 3000s, with the interval of 60 s. The colors are determined based on the Pearson correlation coefficient values. BTPF are selected from (50^2-50)/2=1225 candidate two-point features. The maximum Pearson correlation coefficient value corresponding to BTPF is 0.87. BTPF is obtained by subtracting the △V/t values corresponding to 540 s and 480 s, and taking the absolute value.

**Supplementary Figure 8** Selection of BTPFs for diagnosis and prognosis from laboratory data in the training set of Dataset 2.

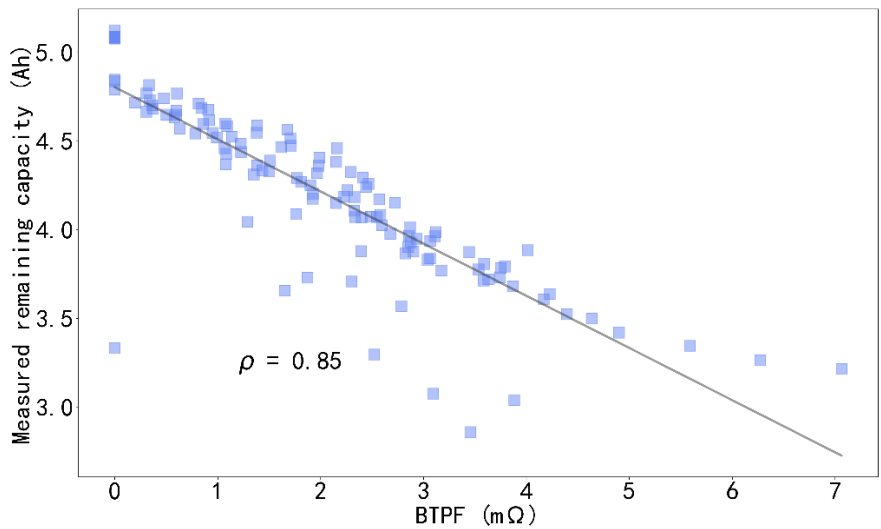

(a) Measured remaining capacity of 11 cells in the training set of Dataset 2 plotted as a function of BTPF in **Supplementary Figure 8 (a)**, with a Pearson correlation coefficient of 0.85.

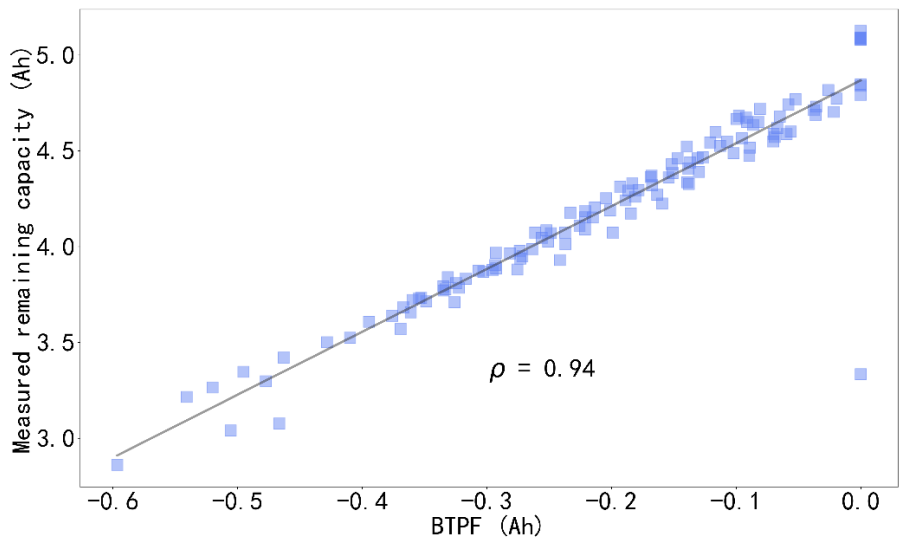

(b) Measured remaining capacity of 11 cells in the training set of Dataset 2 plotted as a function of BTPF in **Supplementary Figure 8 (b)**, with a Pearson correlation coefficient of 0.94.

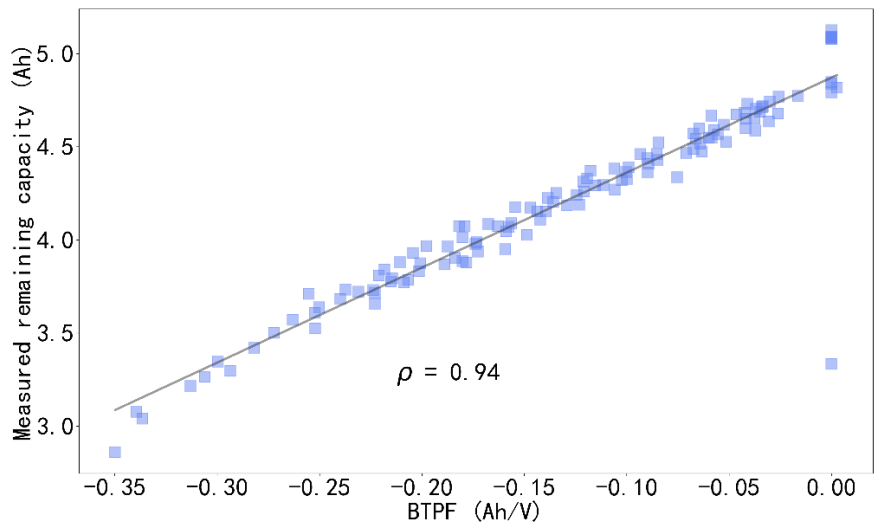

(c) Measured remaining capacity of 11 cells in the training set of Dataset 1 plotted as a function of BTPF in **Supplementary Figure 8 (c)**, with a Pearson correlation coefficient of 0.94.

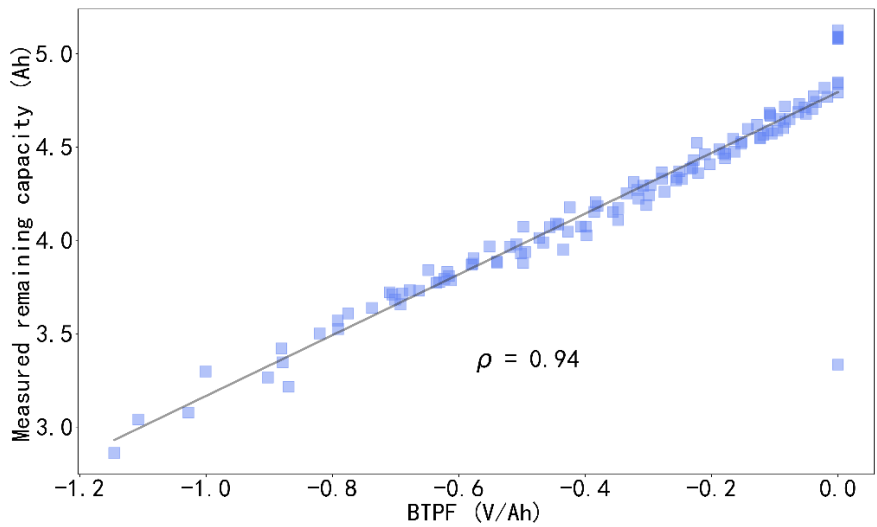

(d) Measured remaining capacity of 11 cells in the training set of Dataset 1 plotted as a function of BTPF in **Supplementary Figure 8 (d)**, with a Pearson correlation coefficient of 0.94.

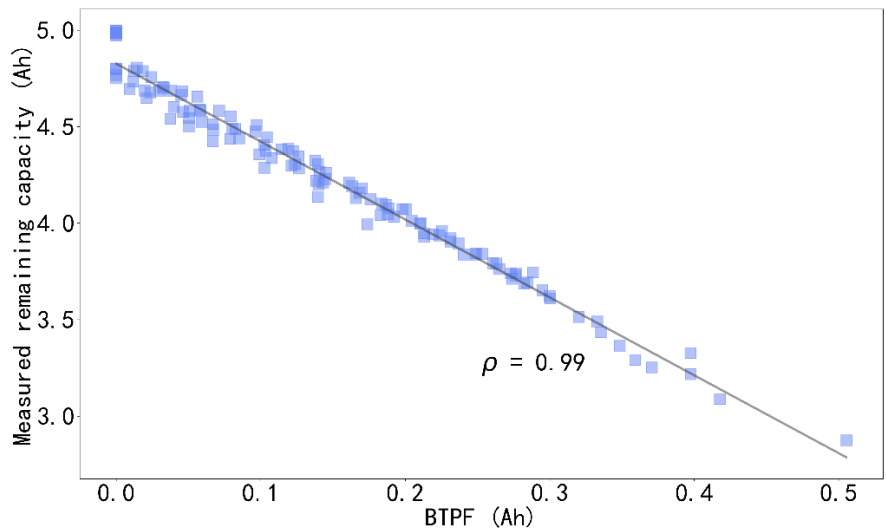

(e) Measured remaining capacity of 11 cells in the training set of Dataset 1 plotted as a function of BTPF in **Supplementary Figure 8 (e)**, with a Pearson correlation coefficient of 0.99.

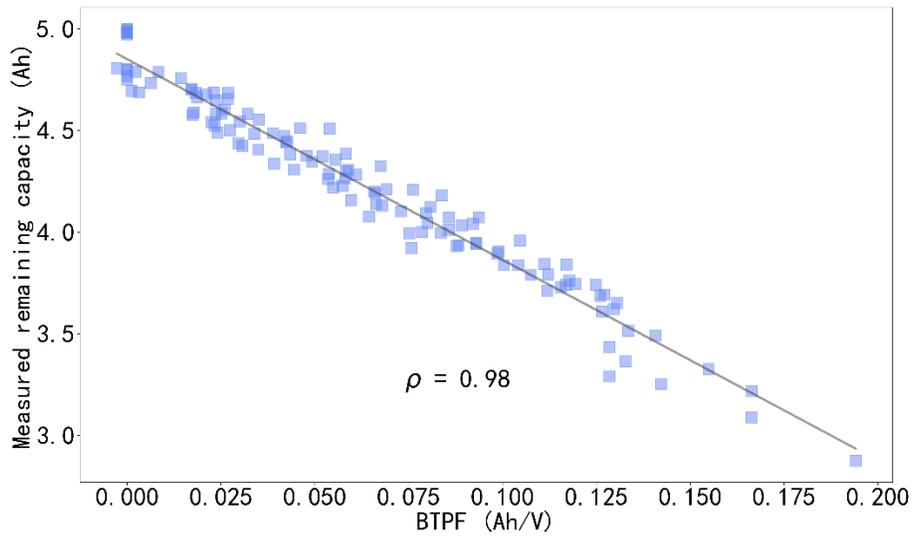

(f) Measured remaining capacity of 11 cells in the training set of Dataset 1 plotted as a function of BTPF in **Supplementary Figure 8 (f)**, with a Pearson correlation coefficient of 0.98.

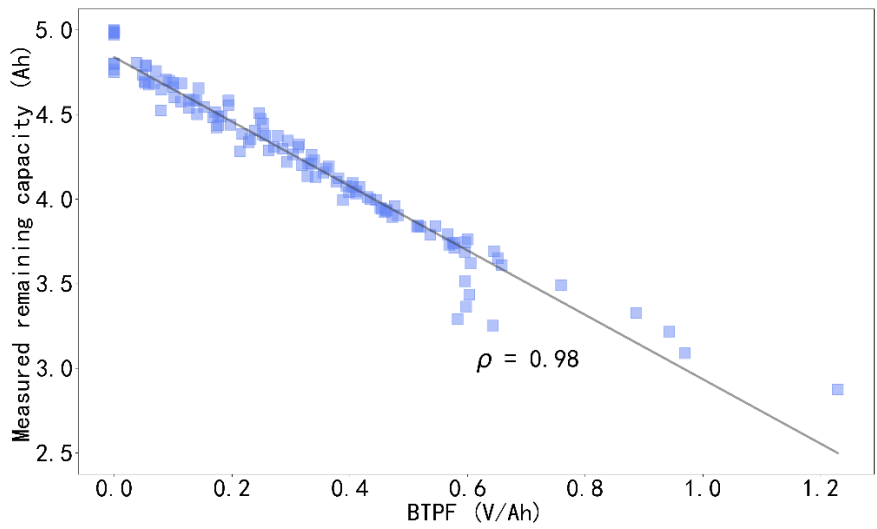

(g) Measured remaining capacity of 11 cells in the training set of Dataset 1 plotted as a function of BTPF in **Supplementary Figure 8 (g)**, with a Pearson correlation coefficient of 0.98.

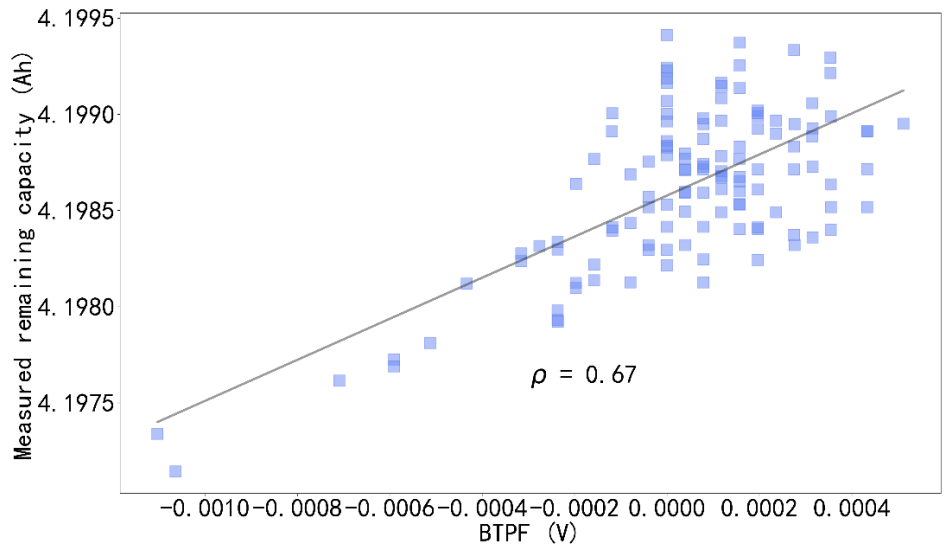

(h) Measured remaining capacity of 11 cells in the training set of Dataset 1 plotted as a function of BTPF in **Supplementary Figure 8 (h)**, with a Pearson correlation coefficient of 0.67.

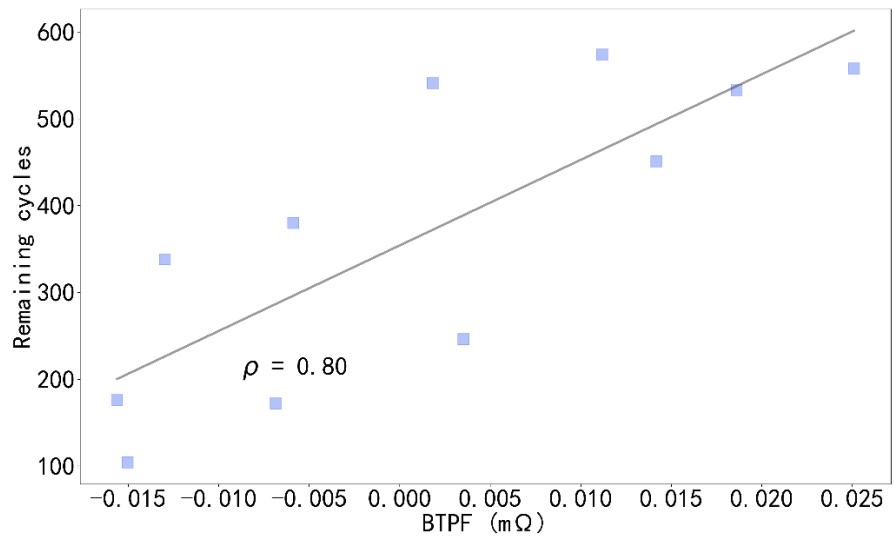

(i) Measured remaining cycles of 11 cells in the training set of Dataset 1 plotted as a function of BTPF in **Supplementary Figure 8 (i)**, with a Pearson correlation coefficient of 0.80.

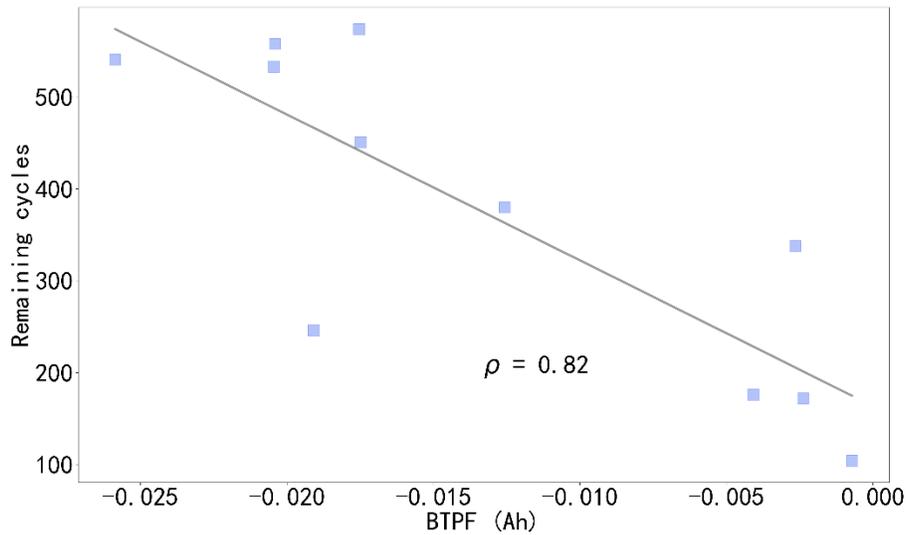

(j) Measured remaining cycles of 11 cells in the training set of Dataset 1 plotted as a function of BTPF in **Supplementary Figure 8 (j)**, with a Pearson correlation coefficient of 0.82.

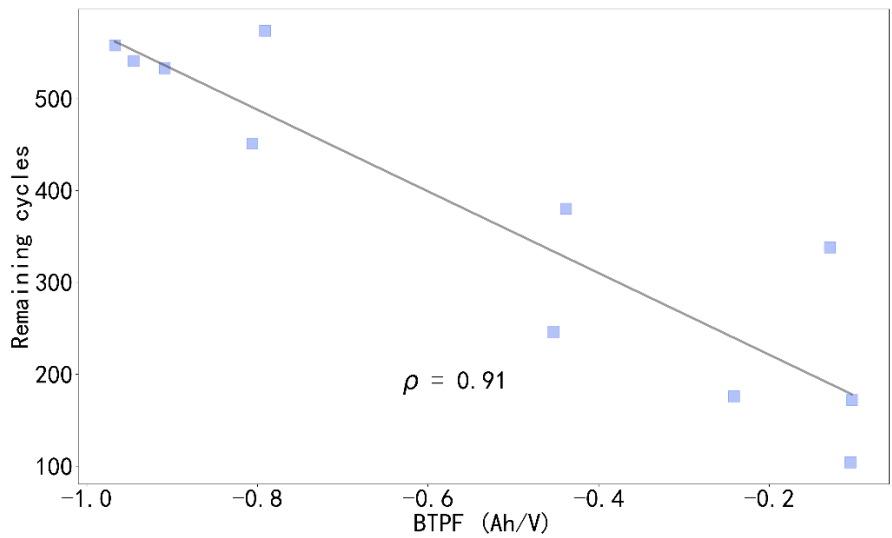

(k) Measured remaining cycles of 11 cells in the training set of Dataset 1 plotted as a function of BTPF in **Supplementary Figure 8 (k)**, with a Pearson correlation coefficient of 0.91.

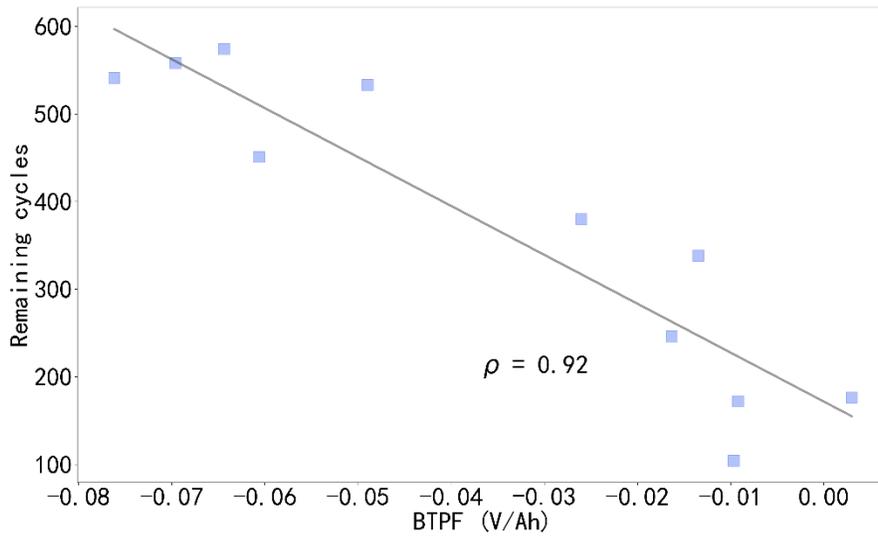

(l) Measured remaining cycles of 11 cells in the training set of Dataset 1 plotted as a function of BTPF in **Supplementary Figure 8 (l)**, with a Pearson correlation coefficient of 0.92.

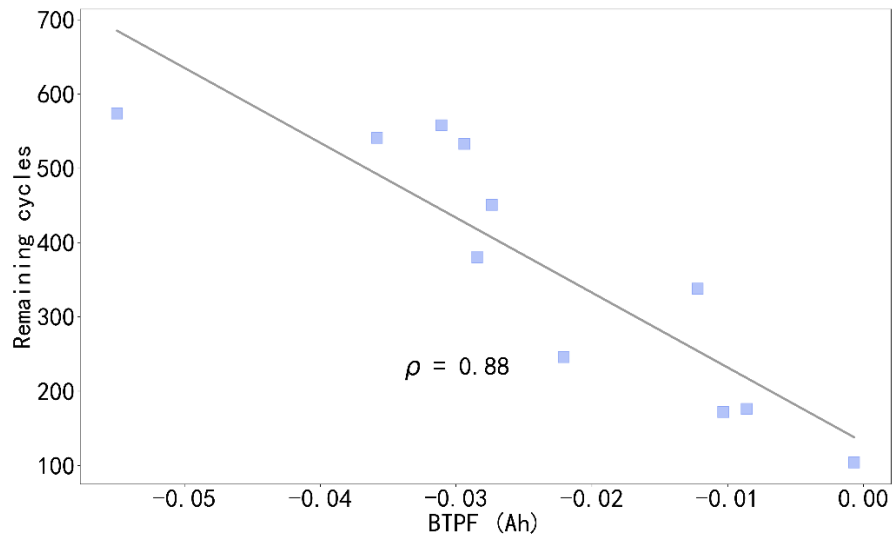

(m) Measured remaining cycles of 11 cells in the training set of Dataset 1 plotted as a function of BTPF in **Supplementary Figure 8 (m)**, with a Pearson correlation coefficient of 0.88.

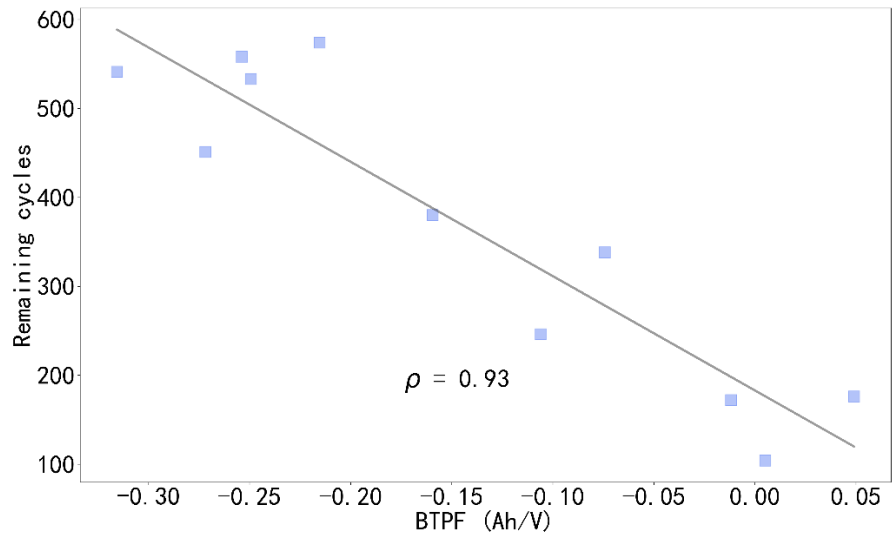

(n) Measured remaining cycles of 11 cells in the training set of Dataset 1 plotted as a function of BTPF in **Supplementary Figure 8 (n)**, with a Pearson correlation coefficient of 0.93.

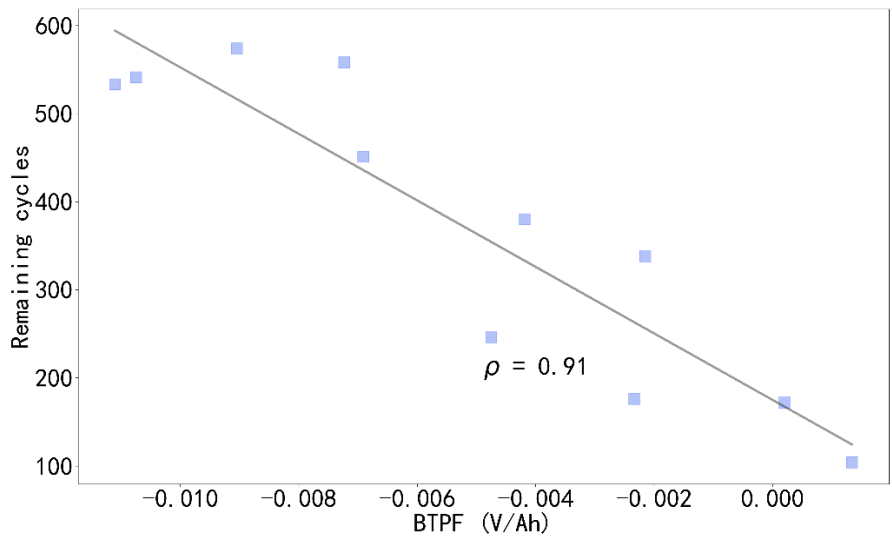

(o) Measured remaining cycles of 11 cells in the training set of Dataset 1 plotted as a function of BTPF in **Supplementary Figure 8 (o)**, with a Pearson correlation coefficient of 0.91.

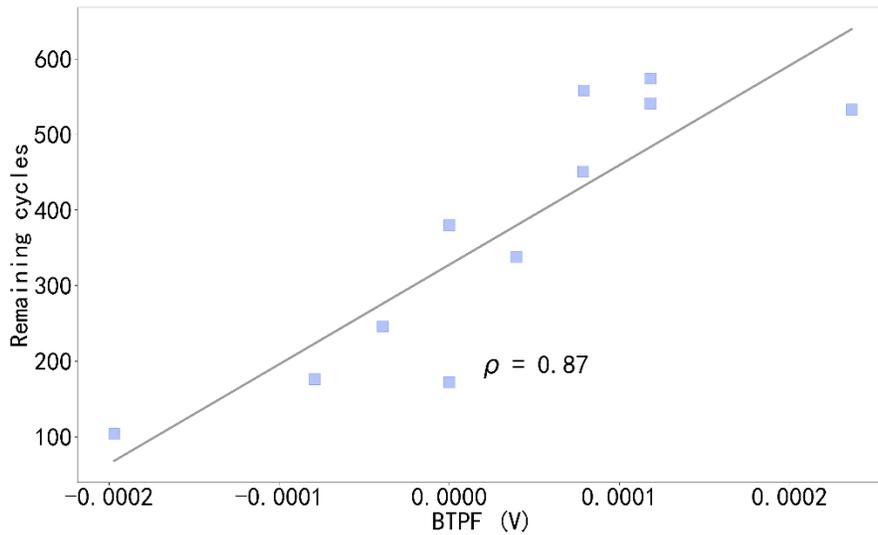

(p) Measured remaining cycles of 11 cells in the training set of Dataset 1 plotted as a function of BTPF in **Supplementary Figure 8 (p)**, with a Pearson correlation coefficient of 0.87.

**Supplementary Figure 9** Measured remaining capacity or measured life of cells in the training set of Dataset 1 plotted as functions of BTPFs.

## Supplementary Note 6

The grid search method [2] is selected for hyperparameter optimization of the RF regression model [3]. The grid search method is a systematic hyperparameter optimization method that exhaustively searches all possible parameter combinations to find the best hyperparameters by defining a set of parameter grids. In this paper, the hyperparameter optimization of the RF regression model includes the number of trees (*n_estimators*), max depth (*max_depth*), minimum number of samples required to be at a leaf node (*min_samples_leaf*), the number of features to consider when looking for the best split (*max_features*), and the fraction of samples to be used for training each tree (*subsample*) [4].

## Supplementary Note 7

Mean absolute error (MAE), mean absolute percentage error (MAPE), and root mean square error (RMSE) are chosen to evaluate diagnostic and prognostic results.

MAE is defined as

$$\text{MAE} = \frac{1}{n}\sum_{i=1}^{n} |y_i - \hat{y}_i| \tag{S1}$$

where $n$ is the number of samples, $y_i$ is the actual value of the $i$th sample, and $\hat{y}_i$ is the predicted value of the $i$th sample.

MAPE is defined as

$$\text{MAPE} = \frac{100\%}{n}\sum_{i=1}^{n} \left|\frac{y_i - \hat{y}_i}{y_i}\right| \tag{S2}$$

RMSE is defined as

$$\text{RMSE} = \sqrt{\frac{1}{n}\sum_{i=1}^{n} (y_i - \hat{y}_i)^2} \tag{S3}$$

**Supplementary Table 1** Test results of Step 2 to Step 4 when $SOC^L$=10% on the test set of Dataset 1.

| Steps | Lab data | MAE | RMSE | MAPE |
|---|---|---|---|---|
| Step 2 | $Re_1^L$ | 0.26 mΩ | 0.39 | 1.21% |
|  | $Re_2^L$ | 0.12 mΩ | 0.19 | 0.69% |
| Step 3 | Re/f curve | 0.21 mΩ | 0.37 | 0.87% |
|  | DRT curve | 0.13 mΩ | 0.04 | 0.56% |
| Step 4 | Charge Q/V curve | 0.02 Ah | 0.04 | 4.86% |
|  | Charge IC curve | 0.06 Ah/V | 0.14 | 9.91% |
|  | Charge DV curve | 0.86 V/Ah | 1.65 | 9.97% |
|  | Discharge Q/V curve | 0.04 Ah | 0.05 | 2.86% |
|  | Discharge IC curve | 0.07 Ah/V | 0.13 | 9.96% |
|  | Discharge DV curve | 0.48 V/Ah | 1.16 | 8.59% |

**Supplementary Table 2** Test results of Step 2 to Step 4 when $SOC^L$=30% on the test set of Dataset 1.

| Steps | Lab data | MAE | RMSE | MAPE |
|---|---|---|---|---|
| Step 2 | $Re_1^L$ | 0.24 mΩ | 0.40 | 1.51% |
|  | $Re_2^L$ | 0.11 mΩ | 0.17 | 0.65% |
| Step 3 | Re/f curve | 0.15 mΩ | 0.21 | 0.70% |
|  | DRT curve | 0.02 mΩ | 0.03 | 0.10% |
| Step 4 | Charge Q/V curve | 0.02 Ah | 0.03 | 5.01% |
|  | Charge IC curve | 0.05 Ah/V | 0.12 | 9.86% |
|  | Charge DV curve | 0.82 V/Ah | 1.58 | 9.19% |
|  | Discharge Q/V curve | 0.03 Ah | 0.05 | 2.68% |
|  | Discharge IC curve | 0.07 Ah/V | 0.13 | 8.68% |
|  | Discharge DV curve | 0.47 V/Ah | 1.12 | 8.30% |

**Supplementary Table 3** Test results of Step 2 to Step 4 when $SOC^L$=50% on the test set of Dataset 1.

| Steps | Lab data | MAE | RMSE | MAPE |
|---|---|---|---|---|
| Step 2 | $Re_1^L$ | 0.24 mΩ | 0.37 | 1.16% |
|  | $Re_2^L$ | 0.12 mΩ | 0.19 | 0.67% |
| Step 3 | Re/f curve | 0.16 mΩ | 0.25 | 0.75% |
|  | DRT curve | 0.17 mΩ | 0.29 | 1.25% |
| Step 4 | Charge Q/V curve | 0.02 Ah | 0.03 | 4.83% |
|  | Charge IC curve | 0.05 Ah/V | 0.12 | 9.21% |
|  | Charge DV curve | 0.82 V/Ah | 1.58 | 9.21% |
|  | Discharge Q/V curve | 0.03 Ah | 0.05 | 2.77% |
|  | Discharge IC curve | 0.07 Ah/V | 0.13 | 8.71% |
|  | Discharge DV curve | 0.47 V/Ah | 1.12 | 8.29% |

**Supplementary Table 4** Test results of Step 2 to Step 4 when $SOC^L$=70% on the test set of Dataset 1.

| Steps | Lab data | MAE | RMSE | MAPE |
|---|---|---|---|---|
| Step 2 | $Re_1^L$ | 0.26 mΩ | 0.45 | 1.25% |
| | $Re_2^L$ | 0.14 mΩ | 0.24 | 0.77% |
| Step 3 | Re/f curve | 0.26 mΩ | 0.35 | 0.95% |
| | DRT curve | 0.17 mΩ | 0.02 | 0.25% |
| Step 4 | Charge Q/V curve | 0.02 Ah | 0.03 | 4.43% |
| | Charge IC curve | 0.05 Ah/V | 0.12 | 9.32% |
| | Charge DV curve | 0.84 V/Ah | 1.63 | 9.33% |
| | Discharge Q/V curve | 0.03 Ah | 0.05 | 2.69% |
| | Discharge IC curve | 0.06 Ah/V | 0.13 | 9.60% |
| | Discharge DV curve | 0.57 V/Ah | 1.15 | 8.88% |

**Supplementary Table 5** Test results of Step 2 to Step 4 when $SOC^L$=90% on the test set of Dataset 1.

| Steps | Lab data | MAE | RMSE | MAPE |
|---|---|---|---|---|
| Step 2 | $Re_1^L$ | 0.24 mΩ | 0.37 | 1.19% |
| | $Re_2^L$ | 0.12 mΩ | 0.19 | 0.68% |
| Step 3 | Re/f curve | 0.18 mΩ | 0.32 | 0.85% |
| | DRT curve | 0.07 mΩ | 0.01 | 0.19% |
| Step 4 | Charge Q/V curve | 0.02 Ah | 0.03 | 4.72% |
| | Charge IC curve | 0.05 Ah/V | 0.12 | 10.55% |
| | Charge DV curve | 0.86V/A | 1.65 | 9.42% |
| | Discharge Q/V curve | 0.03 Ah | 0.04 | 2.69% |
| | Discharge IC curve | 0.06 Ah/V | 0.13 | 9.83% |
| | Discharge DV curve | 0.51V/A | 1.03 | 8.80% |

## References

[1] Hao Liu, Yanbin Zhao, Huarong Zheng, Xiulin Fan, Zhihua Deng, Mengchi Chen, Xingkai Wang, Zhiyang Liu, Jianguo Lu & Jian Chen. Two points are enough, *arXiv*, 2408.11872 (2024).

[2] Petro Liashchynskyi & Pavlo Liashchynskyi. Grid search, random search, genetic algorithm: a big comparison for NAS, *arXiv*, 1912.06059 (2019).

[3] Leo Breiman. Random forests, *Machine learning*, **45**, 5-32 (2001).

[4] RandomForestClassifier-scikit-learn 1.5.2 documentation, https://scikit-learn.org/1.5/modules/generated/sklearn.ensemble.RandomForestClassifier.html.